# A Comprehensive Review of Coastal Compound Flooding Literature

Joshua Green[1,2], Ivan D. Haigh[1,2], Niall Quinn[2], Jeff Neal[2,3], Thomas Wahl[4], Melissa Wood[1], Dirk Eilander[5,6], Marleen de Ruiter[5], Philip Ward[5,6], Paula Camus[1,7]

---

1 School of Ocean and Earth Science, National Oceanography Centre, University of Southampton, European Way, Southampton SO14 3ZH, UK
2 Fathom, Square Works, 17-18 Berkeley Square, Bristol BS8 1HB, UK
3 School of Geographical Sciences, University of Bristol, University Road, Bristol BS8 1SS, UK
4 Department of Civil, Environmental, and Construction Engineering and National Center for Integrated Coastal Research, University of Central Florida, Orlando, FL 32816, USA
5 Institute for Environmental Studies, Vrije Universiteit Amsterdam, 1081 HV Amsterdam, The Netherlands
6 Deltares, 2629 HV Delft, The Netherlands
7 Geomatics and Ocean Engineering Group. Departamento de Ciencias y Técnicas del Agua y del Medio Ambiente. E.T.S.I.C.C.P. Universidad de Cantabria, Santander, Spain

Correspondence to: Joshua Green, J.Green@soton.ac.uk

**Abstract**

Compound flooding, where the combination or successive occurrence of two or more flood drivers leads to an extreme impact, can greatly exacerbate the adverse consequences associated with flooding in coastal regions. This paper reviews the practices and trends in coastal compound flood research methodologies and applications, as well as synthesizes key findings at regional and global scales. Systematic review is employed to construct a literature database of 271 studies relevant to compound flood hazards in a coastal context. This review explores the types of compound flood events, their mechanistic processes, and synthesizes the definitions and terms exhibited throughout the literature. Considered in the review are six flood drivers (fluvial, pluvial, coastal, groundwater, damming/dam failure, and tsunami) and five precursor events and environmental conditions (soil moisture, snow, temp/heat, fire, and drought). Furthermore, this review summarizes the trends in research methodology, examines the wide range of study applications, and considers the influences of climate change and urban environments. Finally, this review highlights the knowledge gaps in compound flood research and discusses the implications of review findings on future practices. Our five recommendations for future compound flood research are to: 1) adopt consistent definitions, terminology, and approaches; 2) expand the geographic coverage of research; 3) pursue more inter-comparison projects; 4) develop modelling frameworks that better couple dynamic earth systems; and 5) design urban and coastal infrastructure with compound flooding in mind. We hope this review will help to enhance understanding of compound flooding, guide areas for future research focus, and close knowledge gaps.

## 1) Introduction

Flooding is the costliest and most common hazard worldwide [1-4], and can lead to a wide range of environmental, economic, and social repercussions. Over 1.8 billion people, almost a quarter (23%) of the world's population, are exposed to 1-in-100 year flooding [4]. The vast majority (89%) of these people live in low- and middle-income countries, and socially vulnerable communities are disproportionately at risk [4]. Since 1980, global floods have caused over 250,000 fatalities and $1 trillion USD in losses [5,6]. In 2021 alone there were more than 50 severe flood disasters recorded worldwide, causing economic losses totaling 82 billion USD [3].

A large proportion of deaths and the economic losses associated with flooding have historically occurred in densely populated coastal regions. Today, near-coastal zones and low-elevation coastal zones, subject to flooding from a range of drivers, are home to 2.15 billion and ~900 million people globally [7]. In the past decade, floods associated with strong onshore wind and pressure fields (e.g., 2013/2014 UK Winter Floods, 2017 Atlantic Hurricane Season, 2019 Atlantic Hurricane Dorian, 2019 East Africa Tropical Cyclone Idai, 2019 Pacific Typhoon Season, and 2022 Eastern Australia Floods) have showcased the ever-present threat of extreme flood impacts in coastal settings. Even in regions where coastal defence standards are among the highest in the world (e.g., Europe, Japan, Netherlands), potential defence failure during events that exceed the standard of protection (e.g., major overtopping or a breach) still pose considerable risk to populations and development in coastal floodplains. Moreover, flooding is a rapidly growing threat to most coastal regions and their communities due to: (i) sea-level rise, changes in storminess, and rainfall patterns driven by climate change [8,9]; (ii) population growth, urbanisation, and continued development in floodplains [10]; and (iii) the continued decline in the extent of shorelines and habitats which act as natural buffers to flooding [11,12]. Average global flood losses in large coastal cities are estimated to increase approximately tenfold by 2050 due to socio-economic change alone, reaching up to US$1 trillion or more per year when considering sea-level rise and land subsidence [10]. Thus, there is clear importance in advancing our understanding of flooding in coastal regions.

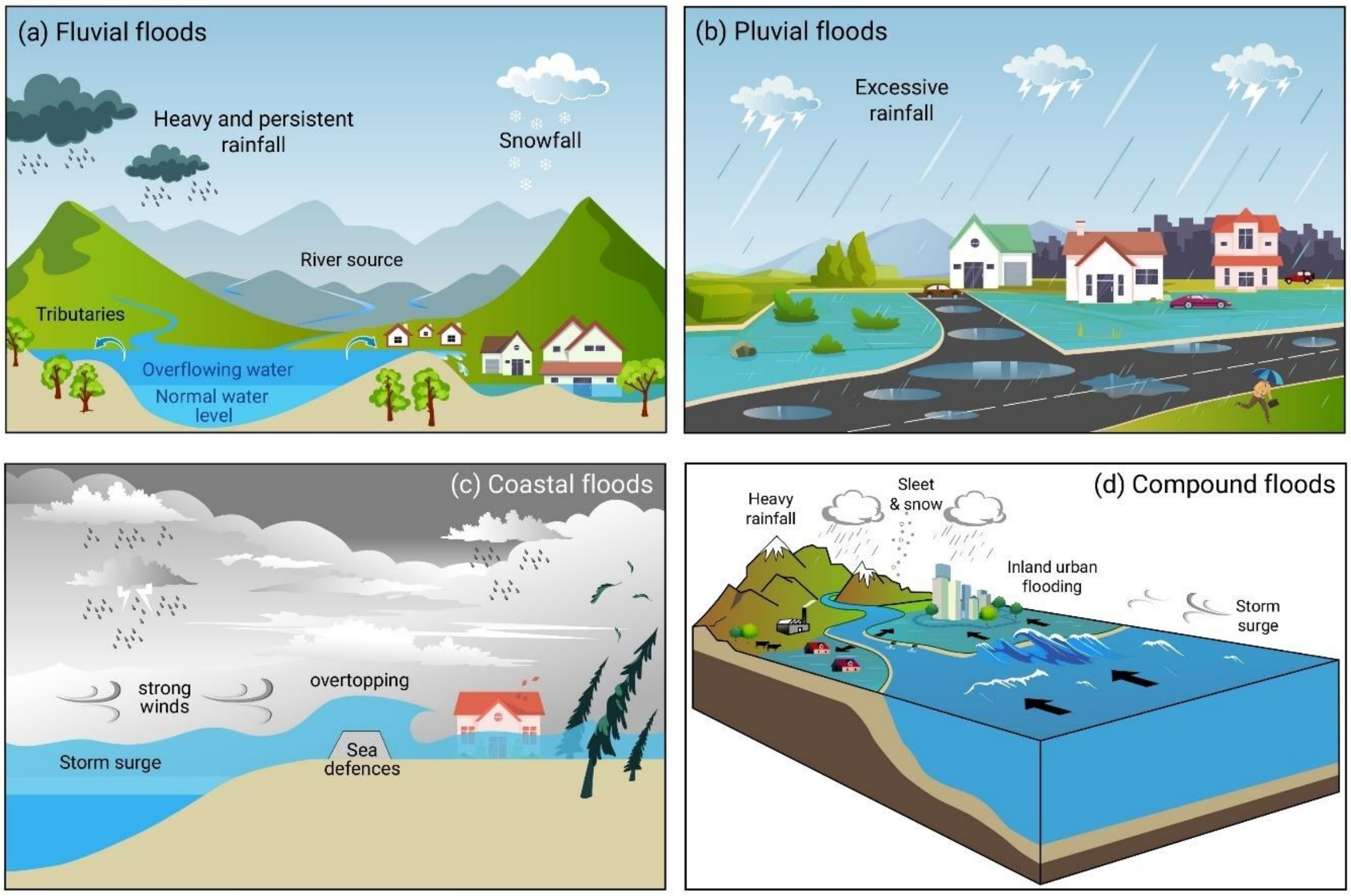

*Figure 1. Schematic diagram of flood drivers showing (a) fluvial (river discharge), (b) pluvial (rainfall runoff), and (c) coastal (surge, tide, waves, and total sea level) components, as well as their (d) compound flood interactions.*

This review focuses on compound flooding in coastal and estuarine regions, which primarily arises from three main sources: (1a) river discharge (**fluvial**); (1b) precipitation surface runoff (**pluvial**); and (1c) coastal oceanographic processes including storm surge, astronomical tides, wave action, and relative sea level rise (SLR) (**coastal**) as shown in Figure 1. Traditionally, most existing flood risk assessments consider these main drivers of flooding separately; and many oversimplify or ignore key interactions all together. However, in many coastal regions, floods are often caused by more than one driver as the processes are naturally correlated. For example, intense tropical/extratropical cyclones (TCs/ETCs) can generate heavy precipitation that enhances river discharges, while at the same time strong winds and low pressures cause large storm surges and waves. When fluvial, pluvial, and/or coastal drivers occur at the same time, or within a few hours or days, the adverse effects of flooding can be measurably exacerbated [13,14]. The synergy of multiple hazard drivers can result in disproportionately extreme events, even if individual flood drivers are not extreme themselves. This is often referred to as 'compound events' [15-19]. It is only in the last decade that we are beginning to recognize the necessity of compound event-based approaches to flood risk assessment, as traditional univariate methods of analysis fail to capture the non-linear impacts of multiple flood drivers [17,20-27].

In recent decades our knowledge of individual flood drivers has improved tremendously, as a result of better in-situ and remote sensed datasets, and advances in statistical and numerical modelling techniques. However, our understanding of compound flood events is still limited, from the synergetic processes to the spatiotemporal trends and scales of interacting drivers. Compound event-based research is relatively new [28,29], having only gained notable attention in 2012 when it was formally defined in the Intergovernmental Panel on Climate Change's (IPCC) Special Report on Climate Extremes (SREX) [18], and as a key guiding principle of the 2015 UN Sendai Framework on Disaster Risk Reduction [30] . Additionally, there has been growing public awareness of extreme compound flooding following a decade of increasingly frequent extreme weather events, where catastrophic disasters arose from multiple interacting flood drivers. For example, in 2017 Hurricane Harvey resulted in record-breaking rainfall, river discharge, and runoff, which when combined with long-lasting storm surge resulted in catastrophic flooding in Houston, Texas [31-33]. This was the second costliest ($152.5B) natural hazard in US history [34]. As a result of this event, it has been recognised that by failing to consider compound flooding, the risk to Houston and elsewhere had been, and currently remains, greatly underestimated.

Compound flood research at local, regional, and recently global scales has experienced growing recognition and substantial advancements over the past decade, with rapid increases in the number of academic publications (particularly since 2020). However, to date there have only been a handful of published reviews that have synthesized current understanding of compound flooding. Moreover, the reviews that do exist have only focused on specific elements of the broader compound flood subject. Bensi, et al. [35] reviewed the drivers and mechanisms of compound flooding, the methods of joint distribution analysis regarding probability hazard assessment, and the key findings of various bivariate coastal-fluvial and coastal-pluvial flood studies. To the best of our knowledge, three publications have reviewed compound flood modelling approaches in coastal regions [36-38] . Santiago-Collazo, et al. [36] summarized practices of numerical compound flood modelling methodologies including different frameworks for linking (or coupling) multiple hydrologic, hydrodynamic, and ocean circulation models. Xu, et al. [37] examined the advancements, benefits, limitations, and uncertainties of varying numerical and statistical (joint probability and dependence) models and frameworks for compound flood inundation. Lastly, Jafarzadegan, et al. [38] provided an overview of advancements in compound riverine and coastal modelling including hybrid methods (i.e., both process-based and data-driven) including linked statistical-hydrodynamic models and physics-informed machine learning (ML) approaches. More broadly, two additional papers by Hao, et al. [39] and Zhang, et al. [40] reviewed the advancing work on compound flood extremes in the realm of hydrometeorology, evaluating the physical drivers and underlying mechanisms [39] plus analytical and modelling research methods [40]. Hao, et al. [39] outlined the characteristics and key statistical tools for assessing compound flood and other compound hydroclimatic extremes (drought, heatwave, cold wave, extreme rainfall). Zhang, et al. [40] discussed these same statistical approaches when reviewing drivers, mechanisms, and means of quantifying risk for compound flooding and four other compound extremes (drought, hot-wet, cold-wet, cold-dry). In addition, they reflected on methods of numerical modelling and collate findings on pluvial-surge, fluvial-surge, sea level-tide, and fluvial-tide compound flood studies. Regarding compound events and driver dependence, Hao and Singh [21] and Zscheischler and Seneviratne [41] reviewed standard methods of measuring dependence (using copulas) as well as approaches for quantifying the likelihood of compound floods. Abbaszadeh, et al. [42] reviewed the sources and challenges of uncertainty in flood modelling and forecasting and offer guidance on reducing uncertainty in the context of compound floods. In addition to these aforementioned papers that reviewed specific aspects of compound flooding, there are a number of articles (e.g., Leonard, et al. [17,]Zscheischler, et al. [19,]Ridder, et al. [26,]Bevacqua, et al. [28,]Simmonds, et al. [43,]AghaKouchak, et al. [44,]Van den Hurk, et al. [45]) that have reviewed broader compound event research involving a wider range of hazards beyond just flooding. These papers have discussed compound flooding and provide a diversity of detailed case examples, but largely focus on the frameworks, typologies, theories, and perspectives of compound event-based research and disaster risk reduction as a whole [17,19,26,28,43,44]. Overall, these previous reviews have provided an excellent synthesis of specific aspects of compound flooding, however, they have each only focused on a narrow area within the much broader compound flooding discipline. To date, a detailed state-of-the-art review of the entire body of compound flood literature is yet to be done.

Therefore, the overall aim of this paper is to carry out a comprehensive systematic review and synthesis of compound flood literature, with a focus on coastal regions. To address this aim we have six objectives around which the paper is structured:

1. To survey the range of compound event definitions and terminologies, and examine how they pertain to the scope of compound flooding (Section 2);
2. To briefly discuss the key physical processes contributing to flood events from individual drivers (Section 3);
3. To develop an extensive literature database on compound flood research (Section 4);
4. To identify trends in the characteristics of compound flood research (Section 5);
5. To synthesize the key findings (dependence hotspots and driver dominance), considerations (costal urban infrastructure and climate change), and standard practices (application cases and analytical methods) of compound flood research (Section 6); and
6. To reflect on the knowledge gaps in multivariate flood hazard research and suggest potential directions for research going forward (Section 7).

Finally, overall conclusions are given (Section 8). Compound flood research is a nascent field of science. As well as providing a comprehensive review, identifying knowledge gaps, and suggesting potential areas for future research, one of our secondary goals of this paper is to provide an initial starting point to better inform researchers and decision-makers new to the emerging field of compound flooding.

## 2) Definitions and Types of Compound Events & Multi-hazard Events

Our first objective is to survey the range of compound event terminologies observed in literature, and to establish the scope of compound flooding considered in this review. First, we do this broadly, reflecting on the definitions of compound events across different types of hazards (and risks) that have been defined in the literature, and then we examine how the various definitions pertain specifically to compound flood types and accompanying drivers. After this, we seek to champion a unifying definition framework (i.e., encompasses a diversity of perspectives and use-cases around compound events) for this review.

Throughout natural hazard literature, terminology around 'compound event, 'compound hazard', and 'multi-hazard' are highly inconsistent. In the past, these terms have sometimes been applied interchangeably. Some instead referenced compound hazards as a type of multi-hazard event within the larger umbrella of the multi-hazard framework. We believe each of these terms are distinct from one another, and thus for the purposes of this review we use the phrase 'compound event'. Examples of different compound event (and related) terminologies are listed in Table 1 (general disaster and hazard definitions are also provided for context). Several terms have been used to describe similar concepts that all broadly involve the consideration of multiple hazards, drivers, mechanisms, variables, and extremes in a multivariate and non-linear assessment of risk (i.e., hazard exposure x vulnerability x capacity) and impact as defined by the IPCC [46,47].

Use of the term 'compound event' (and similar phrases) has been observed in older academic publications [16], however it was only formally defined in an official context in the 2012 IPCC SREX (Seneviratne, et al. [18]). As of present, the most widely accepted definitions of compound events are those from the IPCC SREX [18], Leonard, et al. [17], and Zscheischler, et al. [19], which we briefly discuss below.

| Term Category | Reference | Term | Definition |
|---|---|---|---|
| **General** | UNDRR [48] | Disaster | A serious disruption of the functioning of a community or a society at any scale due to **hazardous events interacting with conditions of exposure, vulnerability, and capacity**, leading to one or more of the following: human, material, economic and environmental **losses and impacts**. |
| **General** | IPCC [47] | Disaster | Severe alterations in the normal functioning of a community or a society due to **hazardous physical events interacting with vulnerable social conditions**, leading to widespread adverse human, material, economic, or environmental effects that require immediate emergency response to satisfy critical human needs and that may require external support for recovery. |
| **General** | UNDRR [48] | Hazard | **A process, phenomenon or human activity** that may cause loss of life, injury or other health impacts, property damage, social and economic disruption, or environmental degradation. |
| **General** | IPCC [47] | Hazard | The potential occurrence of a **natural or human-induced physical event** that may cause loss of life, injury, or other health impacts, as well as damage and loss to property, infrastructure, livelihoods, service provision, and environmental resources. |
| **General** | IPCC [47] | Disaster Risk | The likelihood over a specified time period of severe alterations in the normal functioning of a community or a society due to **hazardous physical events interacting with vulnerable social conditions**, leading to widespread adverse human, material, economic, or environmental effects that require immediate emergency response to satisfy critical human needs and that may require external support for recovery. |
| **General** | UNDRR [48] | Disaster Risk | The potential loss of life, injury, or destroyed or damaged assets which could occur to a system, society or a community in a specific period of time, determined probabilistically as a function of **hazard, exposure, vulnerability, and capacity**. |
| **General** | IPCC [47] | Impacts | The **effects** on natural and human systems of **physical events, of disasters, and of climate change**. |
| **General** | UNDRR [48] | Disaster Impact | The **total effect**, including negative effects (e.g., economic losses) and positive effects (e.g., economic gains), of a **hazardous event or a disaster**. The term includes economic, human and environmental impacts, and may include death, injuries, disease and other negative effects on human physical, mental and social well-being. |
| **General** | Herring [49] | Extreme Event | A time and place in which **weather, climate, or environmental conditions**—such as temperature, precipitation, drought, or flooding—statistically **rank above a threshold value** near the upper or lower ends of the range of historical measurements. Though the threshold is subjective, some scientists define extreme events as those that occur in the highest or lowest 5% or 10% of historical measurements. Other times they describe events by how far they are from the mean, or by their recurrence interval or probability. |
| **General** | Sarewitz and Pielke [50] | Extreme Event | An occurrence that, with respect to some class of occurrences, is either **notable, rare, unique, profound, or otherwise significant in terms of its impacts, effects or outcomes**. An extreme event is not simply 'something big and rare and different'. 'Eventness' demands some type of temporal and spatial boundaries, while 'extremeness' reflects an event's potential to cause change. |
| **General** | IPCC [46] | Extreme Weather Event | An **extreme weather event** is an event that is **rare at a particular place and time of year**. Definitions of rare vary, but an extreme weather event would normally be as rare as or rarer than the $10^{th}$ or $90^{th}$ percentile of a probability density function estimated from observations. The characteristics of what is called extreme weather may vary from place to place in an absolute sense. When a pattern of extreme weather persists for some time, such as a season, it may be classed as an extreme climate event, especially if it yields an average or total that is itself extreme (e.g., drought or heavy rainfall over a season). |
| **Multi-** | UNDRR [48] | Multi-hazard | 1) The selection of **multiple major hazards** that the country faces, and<br>2) The specific contexts where **hazardous events may occur simultaneously, cascadingly, or cumulatively** over time, and taking into account the potential interrelated effects |
| **Multi-** | Zschau [51] | Multi-hazard | **More than one hazard** where hazard interactions are considered |
| **Multi-** | Komendantova, et al. [52] | Multi-hazard | The analysis of different relevant **hazards, triggering, and cascade effects** threatening the same exposed elements **with or without temporal concurrence** |
| **Multi-** | Tilloy, et al. [53] | Multi-hazard | **More than one natural hazard with interrelationships** between the hazards that impact the **same location and time period**. |
| **Multi-** | Gill and Malamud [54] | Multihazards | **All possible and relevant hazards, and their interactions**, in a given **spatial region and/or temporal period** |
| **Multi-** | Hewitt and Burton [16] | Multiple Hazards | **Elements** of quite different kinds **coinciding accidentally, or more often, following one another** with damaging force, for instance floods in the midst of drought, or hurricane followed by landslides and floods. |
| **Multi-** | Zschau [51] | Multi-hazard Risk | Risk in a **multihazard** framework where **no hazard interactions are considered** on the vulnerability level |
| **Multi-** | Eshrati, et al. [20] | Multi-hazards Risk | **The consideration of multiple** (if possible all relevant) **hazards** posing risk to a certain area under observation. |
| **Multi-** | Kappes, et al. [22] | Multi-hazard Risk | The **totality of relevant hazards in a defined area**. Hazards are, as natural processes, part of the same overall system, influence each other and interact. Thus, **multi-hazard risk** contains emergent properties: It is not just the sum of single-hazard risks since their relations would not be considered and this would lead to unexpected effects. |
| **Multi-** | Kappes, et al. [55] | Multi-hazard Risk | A first definition of the term '**multi-hazard**' in a risk reduction context could read as follows: the **totality of relevant hazards in a defined area** (Kappes 2011). However, whether a hazardous process is relevant has to be defined according to the specific setting of the respective area and to the objective of the study. Additionally, not all studies on multiple hazards share the aim of involving '**all relevant processes of a defined area**' but can rather be described as **'more-than-one-hazard**' approaches. In summary, two approaches to multi-hazard can be distinguished: 1) **primarily spatially oriented** and aims at including all relevant hazards, and 2) **primarily thematically defined**. |
| **Multi-** | Eshrati, et al. [20] | Multi-hazards Interaction Types | Hazards relationship refers to many different types of influence of hazards to each other.<br>1) **Triggering** of a hazard by another<br>2) **Simultaneous impact** of several hazards due to the same triggering event<br>3) **Disposition alteration** of a hazard after another hazard occurrence<br>4) **Multiple effects** of a hazard phenomenon |
| **Multi-** | Tilloy, et al. [53] | Multi-hazards Interaction Types | 1) **Independence** where spatial and temporal overlapping of the impact of two hazards without any dependence or triggering relationship<br>2) **Triggering/Cascading** where a primary hazard that triggers and a secondary hazard<br>3) **Change Conditions**: one hazard altering the disposition of a second hazard by changing environmental conditions<br>4) **Compound hazard** (association) where different hazards are the result of the same "primary event", or large-scale processes which are not necessarily hazard<br>5) **Mutual exclusion (negative dependence)** where two hazards can also exhibit negative dependence or be mutually exclusive |
| **Multi-** | Kappes, et al. [22] | Multi-hazard Interaction Types | 1) **Disposition Altering** where modification of environmental characteristics, whether long-term basic disposition (e.g., relief, climate, vegetation cover) or faster variable disposition (e.g. daily to seasonal weather, water balance, vegetation period) causes the exceedance of a threshold and resulting hazard<br>2) **Triggering/Cascading** where one hazards is directly triggered or provoked by another hazard, or a chain of two or more hazards are induced as a result of a shared external event |
| **Multi-** | Gill and Malamud [54] | Multihazard Interaction Types | Multiple hazard interaction types are divided into four categories:<br>1) **Coincidence relationship** involving the spatial and temporal coincidence of natural hazards. |
| | | | 2) **Triggering relationship** where a hazard is triggered. (e.g., lightning triggering a wildfire, groundwater abstraction triggering regional subsidence, a flood triggering a landslide which then triggers a further flood)<br>3) **Increased probability relationship** where the probability of a hazard in increased. (e.g., a wildfire increasing the probability of landslides, regional subsidence increasing the probability of flooding)<br>4) **Decreased probability relationship** where the probability of a hazard is decreased. (e.g., urbanisation catalysing storm-triggered flooding, storms impeding urban fire-triggered structural collapse) |
| **Multi-** | Zschau [51] | Multi-risk | Risk in a **multi-hazard** framework where **hazard interactions are considered** on the vulnerability level. |
| **Multi-** | Komendantova, et al. [52] | Multi-risk | A comprehensive risk defined from **interactions between all possible hazards and vulnerabilities**. |
| **Compound / Other** | IPCC SREX (Seneviratne, et al. [18]) IPCC [47] | Compound Event | In climate science, compound events can be:<br>1) **Two or more extreme events occurring simultaneously or successively**,<br>2) **Combinations of extreme events** with underlying **conditions that amplify the impacts** of the events, or<br>3) **Combinations of events** that are not themselves extreme but **lead to an extreme event or impact** when combined. The contributing events can be of similar (clustered multiple events) or different types. Examples of compound events resulting from events of different types are varied – for instance, high sea level coinciding with tropical cyclone landfall, or cold and dry conditions (e.g., the Mongolian Dzud), or the impact of hot events and droughts n wildfire, or a combined risk of flooding from sea level surges and precipitation-induced high river discharge (Svensson and Jones, 2002; Van den Brink et al., 2005). Compound events can even result from 'contrasting extremes', for example, the projected occurrence of both droughts and heavy precipitation events in future climate in some regions. |
| **Compound / Other** | Hewitt and Burton [16] | Compound Event | **Several elements acting together above their respective damage threshold,** for instance wind, hail, and lightning damage in a severe storm. Many of the most severe meteorological hazards are **compound**, or become disastrous through involvement in a **multiple hazard situation** |
| **Compound / Other** | Leonard, et al. [17] | Compound Event | Emphasizes three key characteristics of a **compound event**: (1) the **extremeness of the impact** rather than variables or events it depends on; (2) the requirement of **multiple variables or events** on which the impact depends; and (3) the role of **statistical dependence**. Consider a coastal flood where the flood level depends on a rainfall event and an elevated ocean level. The coastal flood is a compound event because (1) the impact metric, a flood level, is considered to be extreme; (2) the impact depends on multiple variables, the rainfall and ocean boundary; and (3) the ocean level can have a statistical dependence with rainfall due to influences such as storm surge, wind setup, or seasonality. |
| **Compound / Other** | Zscheischler, et al. [27] | Compound Event | **Compound weather and climate events** are the **combination of multiple drivers and/or hazards** that contributes to societal or environmental risk. Drivers include processes, variables and phenomena in the climate and weather domain that may span over multiple spatial and temporal scales. Hazards are usually the immediate physical precursors to negative impacts (such as floods, heatwaves, wildfire), but can occasionally have positive outcomes (for example, greening in the Alps during the 2003 heatwave in Europe). |
| **Compound / Other** | Zscheischler, et al. [19] | Compound Event Interaction Types | **Compound weather and climate events** have been organized into four type classes:<br>1) **Preconditioned**: where a hazard causes or leads to an amplified impact because of a precondition<br>2) **Multivariate**: co-occurrence of multiple climate drivers and/or hazards in the same geographical region causing an impact<br>3) **Temporally Compounding (sequential)**: succession of hazards that affect a given geographical region, leading to, or amplifying, an impact compared with a single hazard<br>4) **Spatially Compounding**: events where spatially co-occurring hazards cause an impact |
| **Compound / Other** | Raymond, et al. [56] | Connected Extreme Event | The concept of **connected extreme weather and climate events** further recognizes that **compound event impacts** are often substantially and nonlinearly influenced by non-physical factors such as exposure and vulnerability, cutting across sectors and scales (from personal to society wide). These 'societal mechanisms' can tie together the impacts **from two or more climate extremes**. It is the creation or strengthening of the connections between events, in the impacts space and involving anthropogenic systems, that leads to our **terminology** |
| | | | **of 'connected' events as being distinct from 'compound' events**, and also from **interacting-risk** or **multi-risk** frameworks that focus on **combinations of physical hazards**. |
| **Compound / Other** | Pescaroli and Alexander [57] | Compound Risk | Risk from:<br>1) **Extremes that occur simultaneously or successively**;<br>2) **Extremes combined with background conditions** that amplify their overall impact; or<br>3) **Extremes** that result from **combinations of "average" events**. |
| **Compound / Other** | De Ruiter, et al. [58] | Dependent Hazards (Triggering / Cascading) | Include **triggering and cascading disasters**, such as landslides triggered by a flood, or fires caused in the aftermath of an earthquake (Daniell et al., 2017). Cascading events are commonly defined as a primary hazard triggering a secondary hazard (Pescaroli & Alexander, 2015) |
| **Compound / Other** | Kappes, et al. [22], Kappes, et al. [55] | Cascading / Triggering Hazards | The **triggering of one hazard by another**, eventually leading to subsequent hazard events. This is referred to as **cascade, domino effect, follow-on event, knock-on effect, or triggering effect**. |
| **Compound / Other** | UNDRR [59] | Cascading Hazard | **Cascading hazard** processes refer to a primary impact (trigger) such as heavy rainfall, seismic activity or unexpectedly rapid snow melt, followed by a **chain of consequences** that can cause secondary impacts |
| **Compound / Other** | Mishra, et al. [60] | Cascading / Compound Extreme Event | A **cascading (compound) event** occurs due to the **combination of two or more individual extreme events occurring successively (simultaneously)**. Examples of cascading events are: (a) a severe drought event followed by an extreme flood (drought-flood regime), and (b) extreme drought followed by wildfire (drought-wildfire regimes), which can be further compounded by flooding events. The compound event can also be a combination of human and natural related disasters (Mishra et al., 2021). |
| **Compound / Other** | Cutter [61] | Compound / Cascading / Triggering Hazard | Natural scientists working in the hazards arena inherently understand the **compounding physical processes and interactions that trigger a natural hazard event** such as an earthquake and follow on sequences of other events that occur as a direct or indirect result of the initial triggering event. **Compounding interactions** can trigger a secondary hazard (e.g., lightning causing a wildfire) or increase the probability of a hazard (e.g.,wildfire destroying slope vegetation and when rain events occur mudflows ensue). **Compounding interactions are both spatially and temporally coincident and can amplify the effects**, especially if they occur over relatively short time periods and overlap geographically. **Compounding processes, compounding events, or compounding hazards** are synonyms for describing these types of processes or outcomes. **Cascading hazards** occur as a direct or indirect result of an initial hazard. One characteristic feature of cascading natural events is proximity in time and space, suggesting that there are sufficient forces or energy in the initial event to trigger the subsequent events in the physical system. |
| **Compound / Other** | Pescaroli and Alexander [62] | Cascading Disasters | **Extreme events**, in which **cascading effects** increase in progression over time and generate unexpected secondary events of strong impact. These tend to be at least as serious as the original event, and to contribute significantly to the overall duration of the disaster's effects. In cascading disasters **one or more secondary events can be identified and distinguished** from the original source of disaster. |
| **Compound / Other** | De Ruiter, et al. [58] | Consecutive Disasters | **Two or more disasters that occur in succession**, and whose direct impacts overlap spatially before recovery from a previous event is considered to be completed. This can include a broad range of **multi-hazard types**, such as **compound events** (Zscheischler et al., 2018) and **cascading events** (Pescaroli & Alexander, 2015). **Consecutive disasters** can occur due to dependency between natural hazards (e.g., **triggering events**) or when **independent hazards** occur in the same space-time window |
| **Compound / Other** | Pescaroli and Alexander [57] | Interacting / Interconnected Risk | Risk from **physical dynamics** that develop through the existence of a widespread network of causes and effects, tends to overlap with **compound risk** in the hazard domain. Focus on the area in which hazard interacts with vulnerability to create disaster risk |
| **Compound / Other** | Pescaroli and Alexander [57] | Cascading Risk | Risk from '**toppling dominoes**' or '**systematic accidents'**. Associated mostly with the anthropogenic domain and the vulnerability component of risk. |

*Table 1. Examples of different compound event (and related) terminologies, types, and definitions in scientific literature. Unique aspects of varying definitions are emphasized in bold.*

The IPCC SREX [18] defines compound events as a ‘combination of multiple divers or hazards with adverse environmental or social risk/impact’. A more detailed explanation is as follow:

> *“(1) two or more extreme events occurring simultaneously or successively, (2) combinations of extreme events with underlying conditions that amplify the impact of the events, or (3) combinations of events that are not themselves extremes but lead to an extreme event or impact when combined. The contributing events can be of similar (clustered multiple events) or different type(s)”*

According to this definition, compound flooding could, for instance, describe the occurrence of a moderate rainfall event that causes surface runoff and discharges at the coast, in addition to elevated coastal water level from storm surge and wave action (whether simultaneous or a few days later). None, one, or both of the two events may be considered extreme according to threshold or probability-based approaches, but together they lead to extreme coastal water levels. This definition also emphasizes the potential for compounding from the temporal clustering of the same (or different) types of events (e.g., storm clustering involving quick succession of storm events and associated coastal hazards [63]).

Leonard, et al. [17] argue that the IPCC SREX [18] definition is unable to capture extreme event edge cases (i.e., unexpected or outlier situations) and is not founded on the physical systems at play. They instead propose a definition that focuses on the variable interactions and event impact, as follows:

> *“Our definition emphasizes three characteristics: (1) the extremeness of the impact rather than the climate or weather event; (2) the multivariate nature of the event; and (3) statistical dependence between variables or events that cause the impact.”*

Thus, according to this definition, classification of compound flood events necessitates an extreme impact. In the context of flooding, the IPCC SREX may recognize, for example, the simultaneous overtopping of riverine channels and surfacing of groundwater as compounding. However, unless the impact is extreme, it would not pass as a compound flood according to Leonard, et al. [17]. This interpretation also requires definitive dependence between the extremes in question. Therefore, a fluke spatiotemporal overlap of extreme rainfall due to an atmospheric river in a region with elevated river levels from recent snowmelt would not be considered a compound flood as the two events are fully independent.

More recently, Zscheischler, et al. [27] proposed a broader definition that is specific to compound weather/climate events, as follows:

> "*The combination of multiple drivers and/or hazards that contributes to societal or environmental risk*."

Under this definition, the extremeness of individual drivers and/or hazards is not considered, however their combination must still exhibit some extent of impact to contribute to overall risk. Furthermore, compound events are strictly limited to the combination of natural (weather/climate) drivers and hazards. Thus, anthropogenic hazards (e.g., dam failure and deforestation) are not included within their scope of compound events. To date, the definition proposed in Zscheischler, et al. [27] appear to offer potential for unified discussion of compound climate events across scientific disciplines. In the past few years numerous compound flood studies have accordingly adopted their definition framework [21,26,28,37,40].

Finally, for the scope of this review, we adopt the IPCC definitions of 'hazard' and 'compound event' [18,47], and thus consider compound events as a combination of two or more co-occurring or consecutive drivers (natural or anthropogenic), that together have a greater impact than either of the individual events. Neither the individual driver nor their combinations must explicitly be considered extreme. Potential driver interaction types within this compound event framework include the temporal and/or spatially overlapping combination of multiple hazards (often from a shared modulators, e.g., storm event prompts simultaneously rainfall and storm surge), the direct triggering or cascading of one hazard by another (e.g., heavy rainfall on top of existing bankfull river discharge), and the random or by-chance spatial/temporal overlapping of independent hazards (e.g., atmospheric river rainfall during peak spring snowmelt).

## 3) Flood Processes and Mechanisms

Having considered the compound event definitions, our second objective is to briefly discuss the key physical processes contributing to flooding and the individual drivers/hazards recognized in this review. In this review we focus on coastal regions. Here, flooding mainly arises from three main flood drivers, namely (i) fluvial, (ii) pluvial and (iii) coastal. In this section we start by discussing these three drivers and their mechanisms individually (Section 3.1). It is these three drivers, in different combinations, that most often result in compound flood events. Schematic diagrams illustrating the varying flood processes associated with these three main drivers are shown in Figure 1. However, flooding can also arise from three less frequent auxiliary flood drivers, that is (iv) groundwater, (v) damming and dam failure, and (vi) tsunamis. These additional flood drivers are also briefly discussed (Section 3.2). Finally, we also highlight several precursor events and environmental conditions that can influence the magnitude and/or occurrence of flooding (Section 3.3).

### 3.1 Main Drivers of Flooding in Coastal Regions

Fluvial flooding (Figure 1a), also known as river (or riverine) flooding is induced by the accumulation of large volumes of rainfall and/or freshwater. Intense precipitation during extreme meteorological events (e.g., TCs/ETCs and atmospheric rivers) and weather seasons (e.g., monsoons) can inundate rivers quickly. Elevated volumes of water cause the level in rivers, creeks, and streams to rise above their channel banks and spill out into the adjacent low-lying area known as the floodplain. Thus, fluvial flooding depends on the hydrometeorological conditions and catchment characteristics (e.g., size, shape, slope, land cover, and soil type). The peak of river flooding can have a time lag of hours to weeks between the rainfall over a catchment and the exceedance of downstream channels [31]. In the spring, fluvial flooding can also be driven by snowmelt (or glacial melt) as large reservoirs of melting freshwater flows into downstream river channels. Freshwater fluvial flooding occurs worldwide but is more frequent in high latitude (e.g., Canada and Northern Europe) and high elevation (e.g., Hindu Kush and Andes Mountains) regions.

Pluvial flooding (Figure 1b) is the result of rapid heavy rainfall (flash flooding) or long sustained rainfall. As the rain reaches the ground, the soil has the potential to become saturated, causing either ponding or surface runoff (overland flooding) that flows down terrain and into rivers (in practice the boundary between pluvial and fluvial flooding is not well defined and is usually based on catchment area rather than physical process). Pluvial flooding is thus closely dependent on surface drainage. Urban flooding is closely linked with pluvial flooding where excessive runoff in areas of human development has insufficient drainage, often due to impervious surfaces such as concrete and asphalt [64]. Urban flooding also ties in with sewer and stormwater flooding in which pluvial surface runoff infiltrate waste management infrastructure and exceed system capacity [64-66].

Coastal flooding (Figure 1c) mainly occurs from the combinations of high astronomical tides, storm surge, wave action (mainly runup and set up), superimposed on relative mean sea level. Each of these components of total sea level contribute differently to flooding, but we have chosen to group them together for simplicity. Coastal flooding primarily refers to flooding at the interface of land and ocean; however, it is sometimes also used when discussing instances of flooding by these mechanisms along the shoreline of lakes (e.g., Great Lakes). Tides are the regular and predictable rise and fall of the sea level caused by the gravitational attraction and rotation of the Earth, Moon, and Sun. Tides exhibit diurnal, semi-diurnal, or mixed diurnal cycles and experience shifts in amplitude on fortnightly, bimonthly, and interannual timescales. Storm surges are driven by storm events with low atmospheric pressure that cause sea levels to rise, and strong winds that force water towards the coastline. Storms also generate waves, locally or remotely (e.g., swell), via the interaction of wind on a water's surface due to boundary friction and energy transfer. Waves mostly contribute to enhanced coastal flooding via setup (the increase in mean water level due to the presence of breaking waves) and runup (the maximum vertical extent of wave uprush on a beach or structure). Mean sea level is the average height of the sea after filtering out the short-term variations associated with tides, storm surges, and waves. Increases in relative mean sea level arise as a result of vertical land movements (i.e., isostatic SLR) and changes in ocean volume (i.e., eustatic SLR) from thermal expansion of water, mass loss from glaciers and polar ice sheets, and changes in terrestrial water storage [12].

### 3.2 Other Drivers of Flooding

In Section 3.1 we considered the three main flood drivers, which most frequently contribute to compound flooding in coastal regions. However, other less frequent drivers can also play an important role in compound floods and are briefly summarised below. Groundwater flooding is the rise of the water table to the ground surface or an elevation above human development [67]. This occurs during an increase in the volume of water entering an underlying aquifer. This can be the result of prolonged rainfall and snowmelt, but in the case of unconfined coastal aquifers can also be driven by SLR and saltwater intrusion [68-70]. Groundwater flooding is often observed along shorelines that are equal to or below sea level [68-70], in regions with high ground-surface connectivity [71], and in areas experiencing ground subsidence (downward vertical shift of Earth's surface from processes such as compaction and groundwater extraction) [72]. As coastal groundwater flooding is the result of long-term changes, it is slow to dissipate and usually persists longer than floods driven by fluvial and pluvial processes [72].

Damming and dam failure (whether occurring naturally or from anthropogenic activities) can result in flooding from a rapid release or build-up of large volumes of water. Natural damming including beaver dams, ice jams, volcanic dams, morainal dams, and landslide dams can inhibit flow and cause backwater flooding (and even lake formation) [73]. Anthropogenic damming is the intentional inundation (via impoundment) of a hydrological network for purposes of resources management [74]. Natural dam failures such as glacial outbursts and landslide dam overtopping can release vast quantities of water that overwhelm and inundate downstream landscapes [73]. The failure of human engineered water reservoirs (e.g., dams, levees, dykes, water supply systems) can also cause substantial downstream flooding; often posing a greater threat due to the close proximity to human development (e.g., 2017 Oroville Dam crisis [75] and 2023 Derna dam collapses [76]).

Tsunamis are a series of impulsive waves generated by the sudden displacement of large volumes of water due to undersea earthquakes and landslides, shifts in the tectonic plates, and underwater volcanic eruptions [77]. While large magnitude tsunami events occur infrequently compared to other flood drivers, they still have the potential to cause catastrophic flooding in coastal regions. Tsunamis are also unique in their potential to drive coastal flooding at oceanic scales, sometimes spanning multiple countries and continents (e.g., 2004 Indonesia tsunami [78,79] and 2022 Tonga tsunami [80,81]).

### 3.3 Precursor Events and Environmental Conditions

In addition to the aforementioned six flood drivers, we also bring to attention five important precursor events and environmental conditions that can strongly influence flooding and whether or not it occurs. First, soil moisture conditions commonly exacerbate surface flooding due to reduced drainage capacity during periods of sustained high antecedent soil moisture [82]. Elevated freshwater volumes from snow and glacial melt may escalate fluvial and groundwater flooding [83-85]. Extreme temp/heat have the potential to increase atmospheric water content and thus intensify pluvial and fluvial flooding [86]. Wildfires can worsen pluvial and fluvial floods by modifying soil properties such that ash deposits and burnt hydrophobic soils cause rapid surface flows and channelization [87-89]. Finally, drought is known to potentially intensify pluvial flooding when long term water deficiencies dry out and harden the soil, in turn reducing ground infiltration and causing rapid surface flows [90].

We note that many of these precursors and conditions have partially overlapping influences on flooding as they are inherently interlinked by shared climatic and meteorologic forcings.

## 4) Literature Database Methodology

Our third objective is to develop a database of the extensive English-written scientific literature on compound flood research. In this section we describe how the database was compiled, and then we review and discuss the database contents in objectives four (Section 5) and five (Section 6).
A combination of *systematic review* and *content analysis* were used to collect scientific literature and filter for publications relevant to the scope and themes of this paper. Published journal articles, academic theses, conference proceedings, and scientific reports up until the end of the year 2022 were sourced using the Web of Science, Semantic Scholar, Google Scholar, and Dimensions AI search engines. Papers were filtered by topic, title, abstract, and full text (when possible) entering different combinations of key search terms as shown in Table 2. Potential valid articles were also identified from the bibliographies of compound flood papers using literature mapping tools, including Connected papers, Citation Gecko, Local Citation Network, Open Knowledge Maps. Research literature was then filtered for relevance based on the set of criteria defined below.

To be include in our review applicable papers must:

1) focus primarily on compound flooding, and not simply mention it fleetingly in the abstract or conclusion when in fact addressing univariate flooding; and
2) involve multivariate statistical analysis, numerical modelling (hydrological and/or hydrodynamic), and/or discussion of two or more flood drivers, precursors events, or environmental conditions, of which at least one being one of the main three flood drivers (fluvial, pluvial, coastal).

Research studies deemed appropriate were added to the literature review database and manually categorized according to:

1) case study geographic scope;

2) case study scenario;

3) flood drivers, precursor events, and/or environmental conditions considered;

4) research approach (numerical modelling, statistical modelling/analysis, or both); and

5) study application (earth system processes, risk assessment, impact assessment, forecasting, planning and management, and methodological advancement).

| **Search Terms** |
|---|
| “compound* flood*” |
| “joint* “flood*” |
| “coincid* flood*” |
| “comb* flood*” |
| “multivariate flood*” |
| “multi* flood*” |
| “multi-hazard” AND “flood*” |
| “cascading” AND “flood*” |
| “trigger*” AND “flood*” |
| “concurrent” AND “flood*” |
| “precondition” AND “flood*” |
| “antecedent” AND “flood*” |
| “*connected” AND “flood*” |
| (“cooccur*” OR “co-occurr*”) AND “flood*” |
| (“interrelated” OR “interacting”) AND “flood*” |
| (“joint probability” OR “joint occurrence”) AND “flood*” |
| (“river” OR “discharge”) AND (“precipitation” OR “rain”) AND “flood*” |
| (“precipitation” OR “rain”) AND (“surge” OR “tide” OR “wave”) AND “flood*” |
| (“river” OR “discharge”) AND (“surge” OR “tide” OR “wave”) AND “flood*” |
| “fluvial” AND “pluvial” AND “flood*” |
| “fluvial” AND “coastal” AND “flood*” |
| “pluvial” AND “coastal” AND “flood*” |
| “fluvial” AND “pluvial” AND “coastal” AND “flood*” |

*Table 2. Literature database keywords and Boolean search terms. Asterisks act as multi-character wildcards used to capture alternative phrasing of truncated root words (e.g., ‘flood*’ returns ‘flood-s’, ‘flood-ed’, and ‘flood-ing’)*

Keeping in line with the compound event definition framework outlined in Section 2, and the individual flood mechanisms detailed in Section 3, this review recognizes compound flooding as a combination of two or more of the six flood drivers (fluvial, pluvial, coastal, groundwater, damming/dam failure, and tsunami) and five precursor events and environmental conditions (soil moisture, snow, temp/heat, fire, and drought). In this paper, the coastal driver category will encapsulate lake coasts in addition to oceanic coasts, as lakes exhibit wind-driven oscillating waves (seiche) that contribute to compound flooding similarly to oceanic tides and storm-surge. Not considered in the review are studies that assess the cooccurrence or consecutive occurrence of flood characteristics that are not specific to a particular flood driver variable (e.g., flow velocity, flood volume, flood duration, flood intensity, flood depth/height). Additionally, this review does not recognize the confluence or convergence of rivers channels within the same river network as compound flooding. While there is considerable literature on this subject (e.g., Bender, et al. [91]), fluvial-fluvial compounding predominantly occurs inland and therefore is not included within the scope of this paper, which focuses on coastal regions. This review does however recognize compounding of like-type flood drivers in the case of pluvial-pluvial temporal clustering as well as coastal-coastal between different coastal components (e.g., tide-surge, surge-waves, tide-waves).

While this review aims to provide an overview of existing research on compound flooding, it is necessary to recognize limitations of the literature review database. Most notably, this review only considers English scientific literature and thus may not fully represent the perspectives and findings of all research communities. Throughout the literature database development process, a small number (<5) of non-English compound flood studies were identified but omitted to preserve consistent methodology. Additionally, the final literature database used in this study is extensive but not exhaustive, as some compound flood literature may have been overlooked or excluded based on the drivers, precursor events, and environmental conditions considered within the review's scope.

From these literature search and database curation methodologies, we identified a total of 271 compound flood publications. A detailed overview of the compound flood literature database is presented in the supplementary material (Appendix 1).

## 5) Review of Literature Database

The fourth objective of the review is to identify and reflect on trends in the characteristics of compound flood research. We discuss general bibliometric characteristics of compound flood literature including: publications over time (Section 5.1), the geographic scope of compound flood case studies (Section 5.2), and the key scientific journals and/or institutions (Section 5.3). We then review the flood drivers considered (Section 5.4), the analytical approaches applied in the studies (Section 5.4), and their various research applications (Section 5.5).

### 5.1) Publications by Year

As mentioned previously, we identified 271 publications on compound flooding up to the end of the year 2022. The number of publications per year, identified in the review, are shown in Figure 2. Up until the year 2000 there were very few compound flood studies (16) [92-107], the earliest being published in 1970 [107]. Since then, there has been an exponential increase in the number of compound flood related papers. The past three years (2020-2022) in particular has spawned a considerable number of compound flood papers (129), accounting for nearly half (48%) of existing publications.

### 5.2) Publications by Geographic Region

The number of compound flood related papers, organized by geographical region on which the study focuses, are displayed in Figure 3a, and spatially mapped in Figure 3b. Although there has been increasing focus on the compound nature of flooding, the spatial scope of compound flood research is largely limited to a few geographic regions. Nearly half the publications are directed at compound flooding along the US coastlines (110, 40%). The spatial distribution of US-related studies is visualized in Figure 3c. Following the US, some of the next most frequently studied regions are the UK (35, 13%), China (19, 7.0%), Global (12, 4.4%), Europe (12, 4.4%), Australia (9, 3.3%), the Netherlands (8, 3.0%), Canada (7, 2.6%), and Taiwan (7, 2.6%). Additional geographic regions assessed in <7 studies are presented in Figure 3a.

### 5.3) Publications by Journals and Institutions

A total of 107 unique scientific journals and institutions (i.e., universities and government agencies) have published compound flood research (i.e., articles, reports, and theses). More than half (140, 52%) of the compound flood literature is published in 15 academic research journals (Figure 4), with the top 5 most frequent journals being Natural Hazards and Earth System Sciences (26, 9.6%), Journal of Hydrology (15, 5.5%), Hydrology and Earth System Sciences (12, 4.4%), Water Resources Research (11, 4.1%), and Water (10, 3.7%). Although a considerable volume of compound flood research is published by a select few journals and institutions, a total of 65 journals and institutions have only published a single compound flood study. We suspect that this will change in the years to come as the field of compound flood hazards gains further attention.

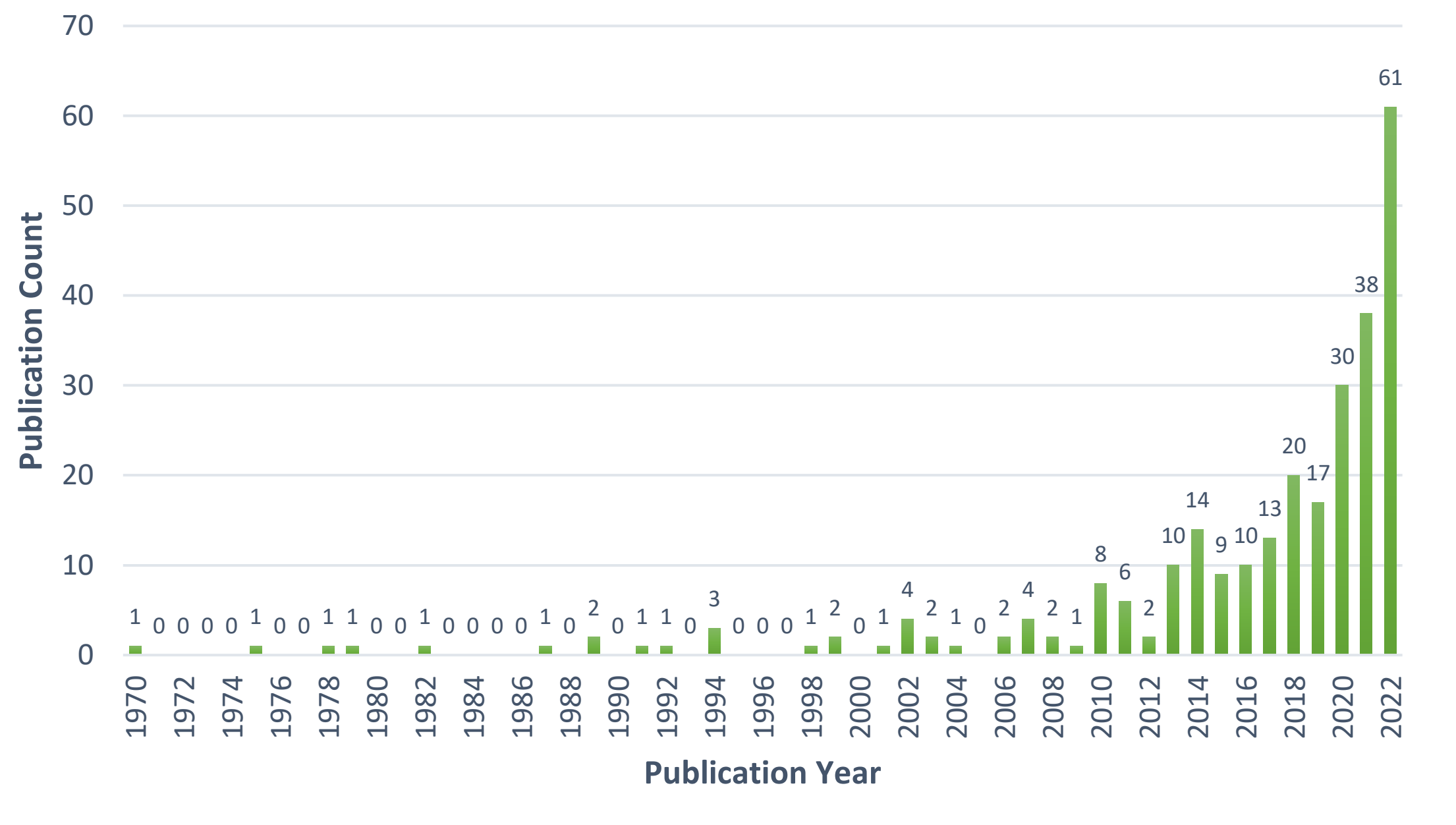

*Figure 2. Histogram showing compound flood literature review database publications over time from 1970 to 2022.*

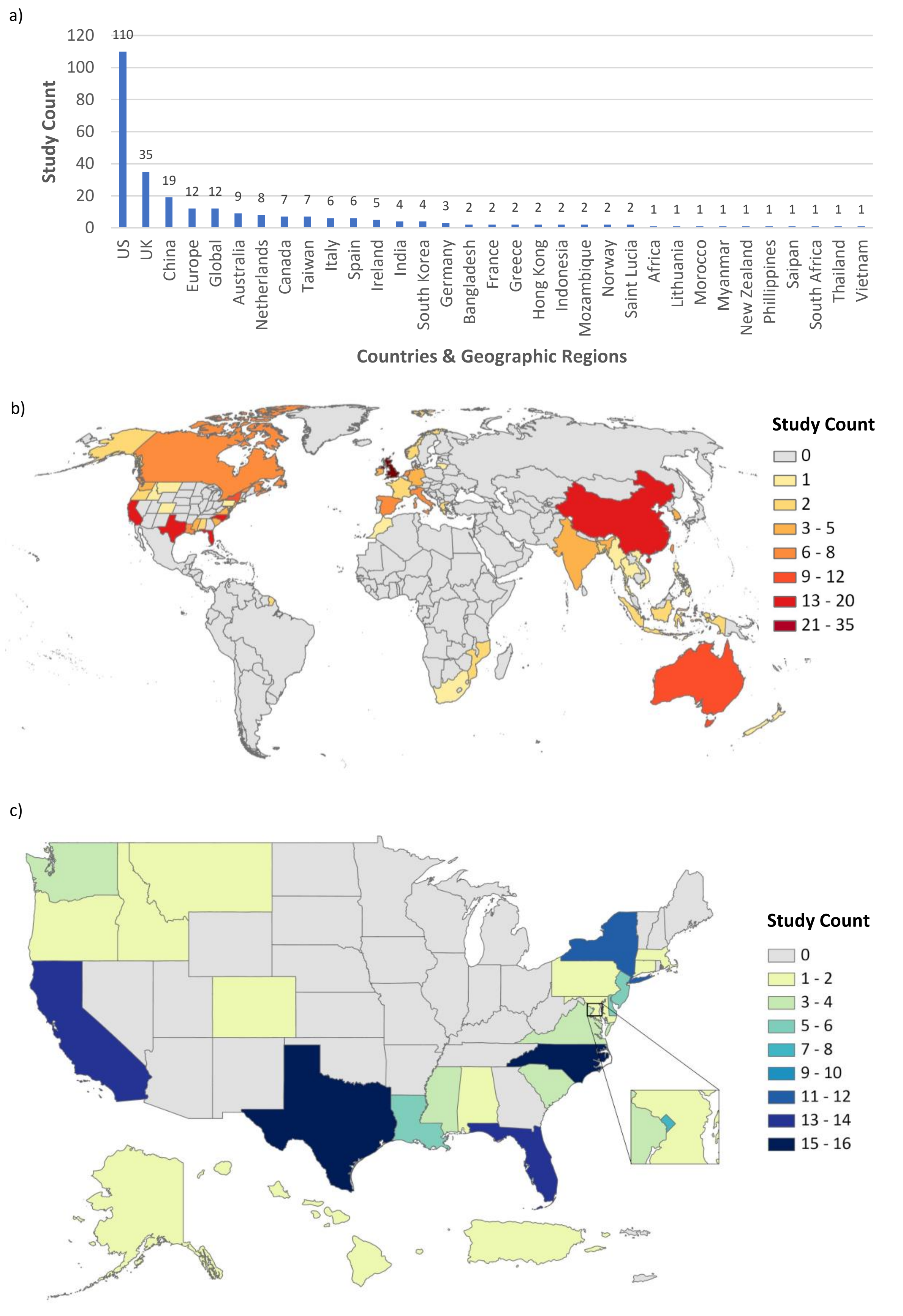

*Figure 3. (a) Histogram showing geographic frequency of compound flood case study regions; and geographic maps showing the frequency of compound flood case study sites (b) across the world and (c) throughout the United States (including Alaska, Hawaii, Puerto Rico, and Washington DC).*

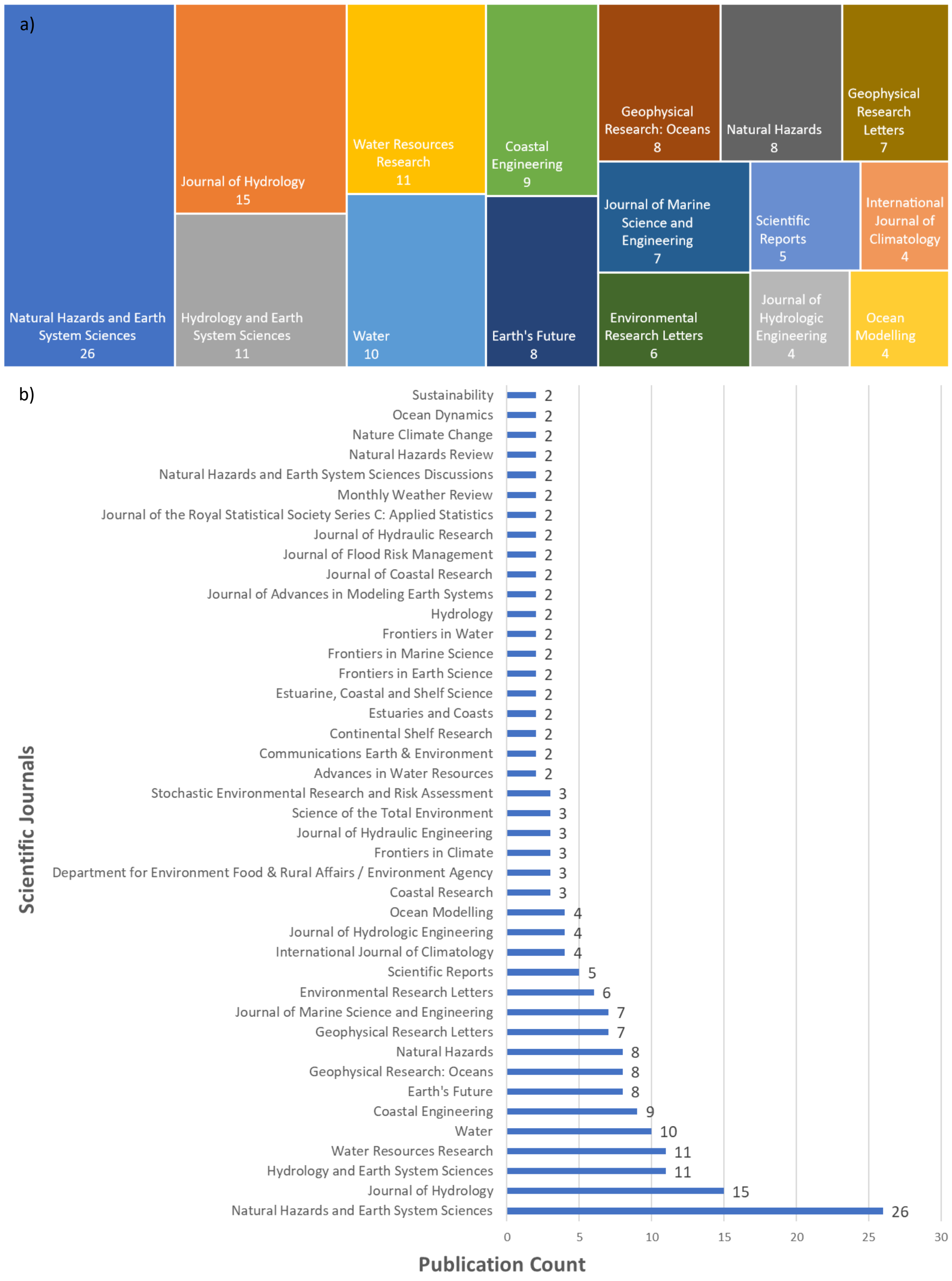

*Figure 4. (a) Treemap of the top 15 most frequent scientific journals and/or institutions that have published compound flood research; and (b) histogram of scientific journals and/or institutions that have published at least two compound flood papers.*

## 5.4) Review of Flood Drivers Considered

Across the 271 studies in the review database, a total of 11 unique compound flood drivers, precursor events, and environmental conditions were identified. These are listed in Table 3 and visualized in Figure 5. Due to the highly complex interactions between terrestrial, oceanic, and atmospheric systems, most studies choose to limit the scope of their research to a select few flood driving mechanisms. For instance, some focus on TC/ETC and extreme precipitation events, while others addressed elevated river discharge in tandem with storm surge. Looking at the combination of drivers analysed, 42 (15%) studies considered exactly the three main components of compound flooding (fluvial, pluvial, coastal); note that analysis of three drivers does not necessarily dictate trivariate analysis (e.g., fluvial-pluvial-coastal), but can also describe two separate bivariate analyses (e.g., fluvial-coastal and pluvial-fluvial) that together include three drivers. The remainder of the studies largely considered combinations of the main drivers (often as bivariate analyses), the most prominent being fluvial-coastal (83, 31%), pluvial-coastal (77, 28%), and coastal-coastal (36, 13%) (e.g., surge and tide) (Figure 5). These results are to be expected as compounding is most prevalent at the coast. Examples of unique and less frequently studied compound flood driver combinations include pluvial-snow [108,109], pluvial-fire [87,110], coastal-tsunami [111,112], pluvial-temp/heat [83], pluvial-drought [26], and fluvial-damming/dam failure [2].

| Flood Drivers, Precursors Events, and Environmental Conditions | Number of Studies in which Considered | Other Corresponding Terms & Variables |
|---|---|---|
| Coastal | 249 (92%) | tide, astronomical tide, storm-tide, surge, storm surge, swell, storm swell, waves, sea surface height, sea level, ocean level, sea water level, total sea level, non-tidal residuals, NTR, H, S, T, W |
| Pluvial | 149 (55%) | precipitation, flash flood, rainfall, rainfall runoff, rainfall anomalies, rainfall extremes, surface runoff, surface inundation, P |
| Fluvial | 141 (52%) | river discharge, riverine discharge, riverine flow, streamflow, streamflow discharge, river level, fluvial discharge, channel discharge, channel flow, Q, R |
| Groundwater | 6 (2.2%) | water table, groundwater level, groundwater head |
| Soil Moisture | 4 (1.5%) | soil saturation, soil moisture extremes, soil moisture anomalies, antecedent soil moisture |
| Snow | 4 (1.5%) | snowmelt, snowfall, glacial melt, freshwater melt |
| Damming/Dam Failure | 2 (0.74%) | dam, levee, barrier, wall, reservoir; dam breach, dam failure, dyke breach, dyke failure, levee breach, levee failure, reservoir breach, reservoir failure |
| Temp/Heat | 2 (0.74%) | temperature extremes, temperature anomalies, extreme heat, |
| Fire | 2 (0.74%) | wildfire |
| Tsunami | 2 (0.74%) | -- |
| Drought | 1 (0.37%) | -- |

*Table 3. List of unique flood drivers, precursor events, and environmental conditions (plus terms and variables) observed in compound flood research from the literature review database*

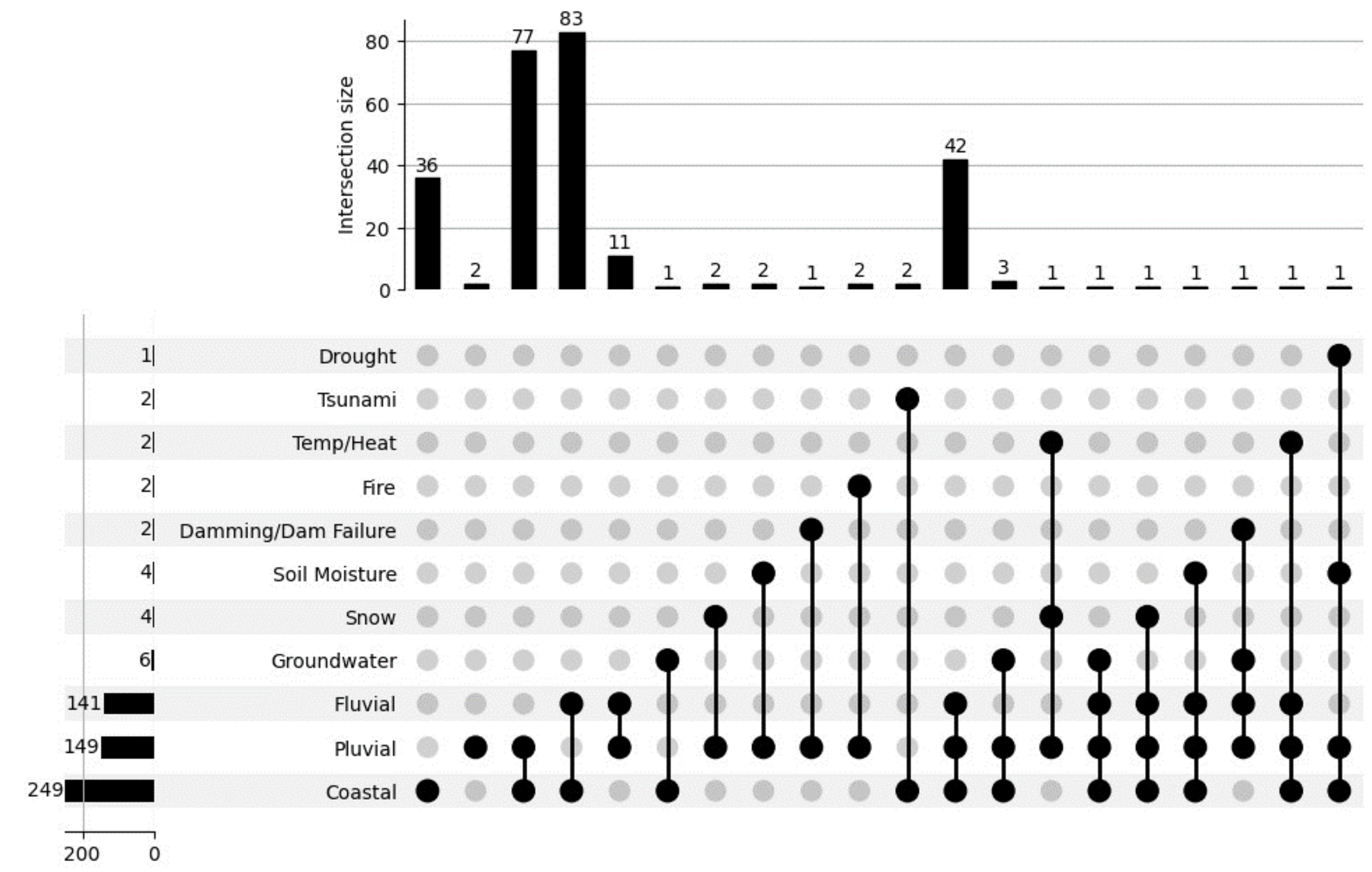

*Figure 5. UpSet plot [113] visualizing the combinations and frequency of driver multi-classifications assigned across the literature. The vertical histogram presents the total count of studies considering each of the eleven drivers (plus precursor events and environmental conditions) categorized nonexclusively, while the horizontal histogram presents the total count for each driver multi-classification combination exclusively. Flood driver classifications for like-type compounding (e.g., pluvial-pluvial and coastal-coastal) are indicated by a non-linked circle. Note that analysis of three drivers does not necessarily dictate trivariate analysis (e.g., fluvial-pluvial-coastal). It may instead describe two separate bivariate analyses (e.g., fluvial-coastal and pluvial-fluvial) as part of the same study that together consider three drivers.*

## 5.5) Review of Research Approaches

Across the database, the compound flood studies have tended to apply approaches that generally fall into two categories: (1) physical (process-based) numerical modelling, and/or (2) statistical modelling and analysis; similar findings to that of Tilloy, et al. [53]. The number of studies applying each approach are illustrated in Figure 6. In total, 96 (36%) studies used only numerical modelling approaches, 97 (36%) used only statistical approaches, and 76 (28%) studies applied hybrid methods involving a combination of numerical and statistical approaches. Within the main two approach classes are many different methods for investigating compound floods, each of which exhibiting their own benefits and limitations as discussed in Section 6. Lastly, 2 (<1%) studies used neither of these approaches, instead completing qualitative survey-based investigations related to the perception and understanding of compound flooding by disaster managers and the wider public [114,115].

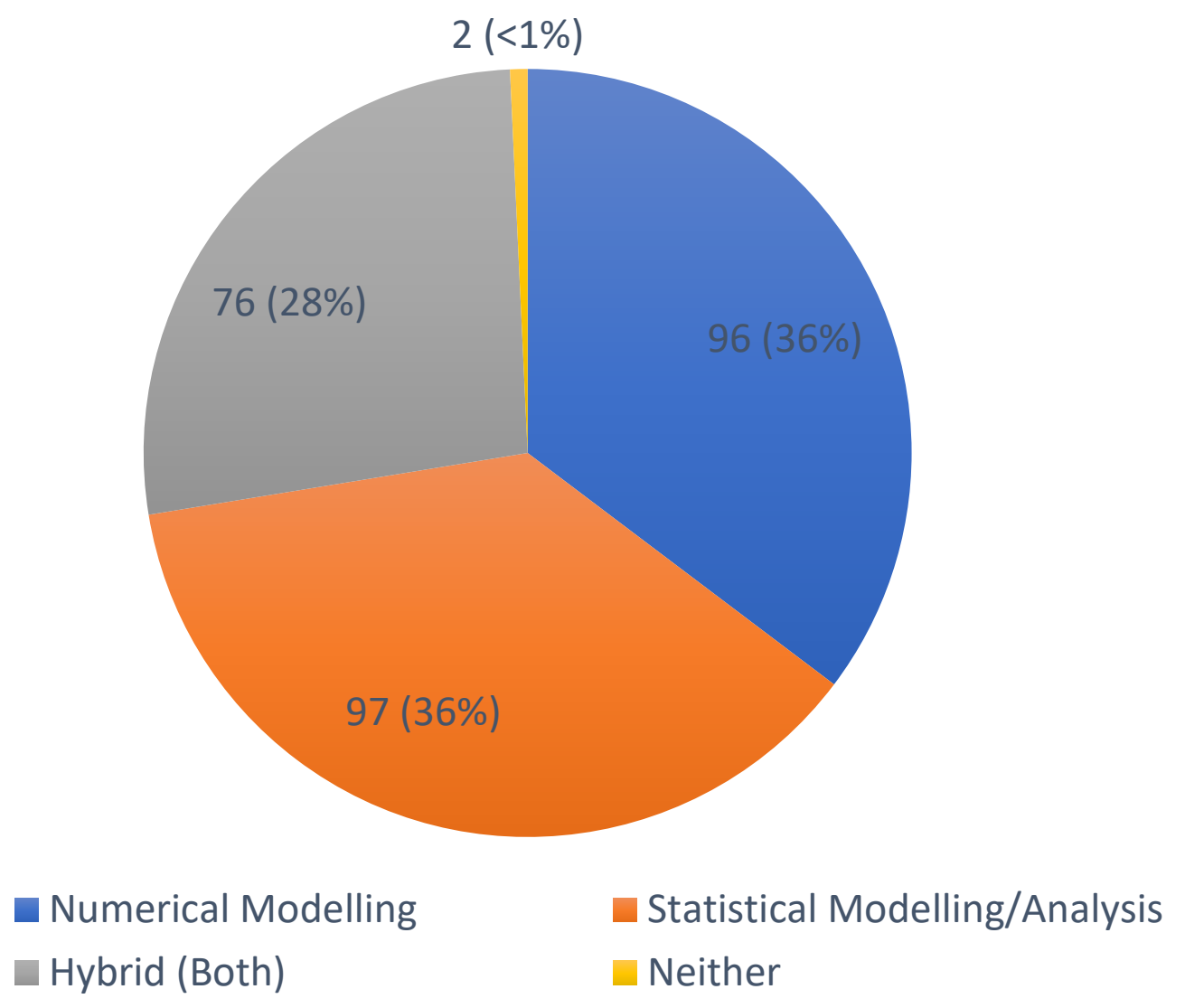

*Figure 6. Pie chart showing the proportion of compound flood literature review database studies that implement numerical modelling, statistical modelling/analysis, hybrid (both) methods, and neither of the two approaches.*

## 5.6) Review of Research Applications

Across the database, the compound flood studies have tended to relate to six main application themes, as illustrated in Figure 7. Assessing the individual research application categories nonexclusively, 129 (48%) studies consider Earth System Processes, 127 (47%) Risk Assessment, 12 (4.4%) Impact Assessment, 21 (7.7%) Forecasting, 29 (11%) Planning & Management, and 73 (27%)

Methodological Advancement (Figure 7). These applications are discussed in more detail in Section 6.7. Reflecting on the exclusive multi-classification of applications, the three most common classifications are ‘Earth System Processes’ (73, 27%), ‘Risk Assessment’ (49, 18%), and ‘Earth System Processes, Risk Assessment’ (30, 11%) which together account for over half of the literature database entries (Figure 7). This is to be expected as they are the broadest of application categories, but also the primary objective of most research. Other prominent research application classification categories include ‘Methodological Advancement’ (26, 9.6%); ‘Methodological Advancement, Risk Assessment’ (21, 7.7%); ‘Earth System Processes, Methodological Advancement’ (18, 6.6%); and ‘Planning & Management, Risk Assessment’ (12, 4.4%) (Figure 7).

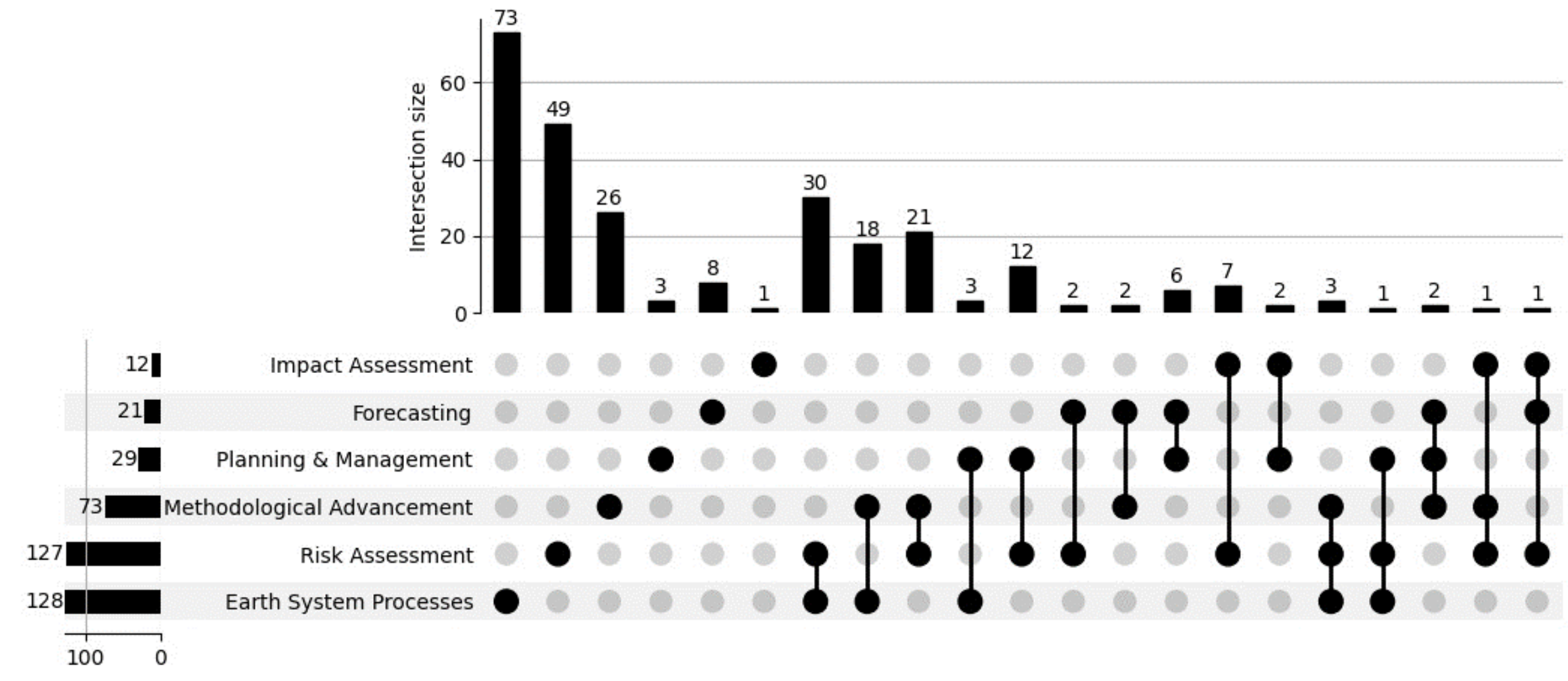

*Figure 7. UpSet plot [113] visualizing the combinations and frequency of application multi-classifications assigned across the literature. The vertical histogram presents the total count of studies considering each of the six application categorizes nonexclusively, while the horizontal histogram presents the total count for each application multi-classification combination exclusively. Instances of application classification are indicated by a non-linked circle.*

## 6) Discussion

Our fifth objective is to synthesize the key findings (e.g., dependence hotspots and driver dominance), considerations (e.g., uncertainty and climate change), and standard practices (e.g., application cases and analytical methods) of the compound flood research from across the database. First, we examine the global and regional hotspots of compound flooding, outlining where and when different driver pairs exhibit significant dependence (Section 6.1). Next, we discuss the tendency for certain drivers to dominate the compound flooding process and examine how this changes spatially as influenced by landscape characteristics (Section 6.2). We then consider compound flooding in the context of urban and coastal infrastructure and how these environments are particularly susceptible to the compounding drivers as it is a common consideration throughout the literature (Section 6.3). Next, we assess how climate change is expected to affect the frequency, variability, and severity of compound flooding in the future (Section 6.4). Then, we reflect on the different approaches that have been used in the literature to analyse compound flooding (Section 6.5). Finally, we investigate the range of different applications considered across the literature (Section 6.6).

### 6.1) Compound Flood Hotspots and Spatiotemporal Dependence Patterns

Our review highlights that knowledge of compound flooding hotspots, spatiotemporal patterns, and multivariate dependence characteristics has advanced considerably in recent years. However, the ways in which global meteorological and climate modulators affect the propensity of compound flooding in one region over another is not fully understood, and few studies consider the non-stationarity of multivariate flood variable dependence. Nonetheless, large-scale patterns in seasonal and interannual occurrence of compound events have become apparent in several regions [26,116-122].

Existing compound event literature has identified certain areas around the world that are especially prone to compound flooding, namely: Southern Asia, where monsoon floods and cyclones cause widespread damage; the Gulf and East Coasts of the United States, where hurricanes induce storm surge and heavy rainfall which exacerbate river flooding; global low-lying delta regions (e.g., Ganges, Irrawaddy, Mekong, Mississippi, Rhine, and Pearl) where riverine and coastal waters together induce severe flooding; northern and western Europe which are prone to river flooding plus extreme precipitation and surge from storm events; and coastal areas of East Asia, Southeast Asia, and Oceania, where TCs/ETCs drive joint fluvial and coastal flooding [118,123-128]. Below we further detail the spatiotemporal patterns in compound flooding and driver interdependence by region.

North America: The coasts of North America are the most studied in terms of compound flooding globally. Compound flooding predominantly occurs along the mid-eastern US coastline and the Gulf of Mexico due to TCs/ETCs that generate heavy rainfall and extreme sea levels [26,116,125,129]. Joint pluvial-fluvial extremes account for the majority of compound flood events and occur frequently with low return periods (<0.5 year) over the entire contiguous US, but particularly along the coasts [26]. Coastal-fluvial drivers too exhibit positive dependence at both coasts[26]. Dependence is also measured between flood drivers along Canada's coasts, albeit less frequent relative to the US [130]. Throughout the Great Lakes, consistent significant positive dependence is found between pluvial-coastal drivers. On the east coast, pluvial-fluvial extremes are frequent in late spring and early summer during the Atlantic hurricane season [26,131]. This region exhibits strong correlations between pluvial-coastal [118,132] and fluvial-coastal [133] drivers [125,131]. Lastly, the west coast features positive dependence for fluvial-coastal [134] and pluvial-coastal [118] pairs during the winter ETC season [131].

Central & South America: Current knowledge of compound flood events in Central and South America is lacking due to a void of localized research. Global studies on compound flooding indicate that fluvial-pluvial extremes are the most frequent cause of compound flooding in South America; and largely occur in the eastern half of the continent (particularly Brazil) during austral summer/late autumn [26]. Similarly, there is positive dependence between fluvial-coastal flood drivers on the southeast coast of Brazil, with large clustering in the highly populated states of São Paulo and Rio de Janeiro [26,126,134]. On the west coast, co-occurring fluvial-coastal extremes are located at the southern portion of Chile in austral summer [26,126].

Europe: Across Europe, large-scale low-pressure systems are a prominent modulator of compound floods [26], with most (~90%)[125] events occurring in the winter ETC season [26,116,118]. The main hotspots of compound flooding are the west coast of the UK, the northwest coast of the Iberian Peninsula, around the Strait of Gibraltar, coasts along the North Sea, and the eastern portion of the Baltic Sea [26,125,126,134]. Concomitant pluvial-fluvial and pluvial-coastal extremes are most prominent in western Europe [26,118,125,126]. In Ireland and the UK, joint occurrence of high skew surges and high river discharge are more common on the west and southwest coasts compared to the east coast [125,134-137]. Pluvial-fluvial drivers also show strong positive correlations in southern Italy, the east coast of Turkey, the eastern Mediterranean, the coasts along the North Sea, and parts of the Baltics. Compound rainfall and river discharge occur primarily in the early summer to late autumn. For fluvial-coastal and pluvial-coastal driver dependence, there are strong correlations along the Iberian coasts, the Strait of Gibraltar, and the UK west coast [118,125,134,137,138]. Lastly, positive pairwise dependence of temporally compounding pluvial-pluvial ("wet-wet") conditions are prominent along the coastal Mediterranean [139].

Africa: Research in Africa is sparse relative to the other continents; however, a few compound flood patterns have been ascertained along the northern, southern, and eastern coasts. Portions of northern Africa show significant positive pluvial-fluvial correlation along the southern Mediterranean and eastern Atlantic coasts including Libya, Tunisia, Algeria, and especially Morocco [125]. In fact, Morocco has the greatest compound flood potential in northern Africa as it also demonstrates strong dependence for coastal-pluvial [140] and coastal-fluvial extremes [125]. Analysis of rain gauges across northern Africa also reveals a select few sites in Algeria with pluvial-pluvial ("wet-wet") pairwise dependence [139]. In southern and eastern Africa, both South Africa and Mozambique experience compound flooding from seasonal TCs during austral summer [26,126,134,141,142]. As a result, this region has strong dependence relationships between the flood driver pairs coastal-fluvial, coastal-pluvial, and pluvial-fluvial [143-145]. Lastly, Madagascar has significant positive coastal-fluvial dependence [26,126] also due to its exposure to TCs [142].

Asia: Compound flood spatiotemporal distributions are highly varied throughout Asia but tend to be most frequent in the south, southeast, and east. Strong correlations for fluvial-coastal extremes are seen at the coasts of India and Bangladesh (Bay of Bengal), Indonesia (North Natuna Sea), Vietnam (East Sea), Philippines (West/East Philippine Seas), Malaysia, China, Taiwan, and Japan (Sea of Japan) [26,126,134]. Similarly, there is positive dependence for pluvial-fluvial drivers in India, Bangladesh, and Japan [26,142]. Co-occurring pluvial-coastal extremes are most prominent in east Asia (particularly China, Taiwan, and Japan)[117,118] and southeast Asia during the wet monsoon season [146]. Most compound flood events within Asia occur from summer to late autumn, corresponding with the TC/ETC seasonality in the western Pacific.

Oceania: Within Oceania, compound flood events have been primarily observed in Australia and to a lesser degree New Zealand. In Australia, the highest frequency of compound flood events is along the northern coastlines (bearing the brunt of TCs [142]) followed by the east and west coasts; all of which predominantly occur during TC season in austral summer. Examining dependence, these patterns are consistent for nearly all flood driver pair combinations, with strong positive correlation in all areas except the southern coast (particularly Victoria) for pluvial-coastal, fluvial-coastal, pluvial-fluvial, [26,117,118,121,126,134,147]. In New Zealand, compound flood events from pluvial-coastal and fluvial-coastal drivers have been observed as being substantial but are not strongly correlated [122]. Compound flooding likely affects small Pacific Island Nations; however they have been scarcely studied. To-date, there are only two localized studies [101,148] on co-occurring flood extremes for the entirety of Micronesia, Melanesia, and Polynesia. Habel, et al. [148] confirmed the occurrence of coastal-groundwater and pluvial-coastal flooding processes in Hawaii, and Chou [101] quantified the frequency of compound flooding from tide and storm surge along Saipan in the Mariana Islands.

### 6.2) Dominant Drivers of Compound Flooding

While compound flood events involve a combination of drivers, often one of the components contributes more than the other(s). Understanding how drivers dominate the flooding process and how these change with space and time is essential to improving compound flood forecasting and risk assessment. Most compound flood events highlighted in the literature contain regions that are pluvial-, fluvial-, coastal-, groundwater-, or compound-dominated in nature. Only a handful of studies examine driver dominance at a global scale [117,127], but those that do reveal general patterns that also tend be supported by more localized research. First, estuaries tend to have a mixture of dominant drivers. In a global assessment of 3,433 estuaries, Eilander, et al. [127] classified 19.7% as compound dominant, 69.2% as fluvial dominant, and 7.8% as coastal dominant. Next, coastal-only environments (i.e., coastal areas with little or no river interaction) have a much larger proportion of coastal-dominant compound floods due to the direct proximity of tide-surge processes and wave actions; and groundwater-dominated floods where sea level pushes the water table up. Excluding river processes, Lai, et al. [117] deduced that coastal (storm surge) and pluvial flooding contributed 65% and 35% to the global change in annual compound floods, respectively. Finally, urban coastal regions are expected to have greater amount of pluvial-dominated compound floods.

Flood driver dominance can depend on topography and channel morphology (i.e., depth, width, size, shape, volume, slope, friction, and damping) [86,127,149-151], spatial extent (i.e., location within hydrological network and distance to the coast) [33,86,152-159], elevation [33,160], ground-surface connectivity [71], and meteorologic modulator characteristics (i.e., storm event timing and intensity) [150,153]. Pluvial flooding is the least frequently reported dominating driver, and primarily only occurs in areas disconnected from the river network with no fluvial inundation [123,153,158] or at higher elevation [33,161]. Pluvial-dominated flooding is also prevalent in urban zones when the capacity of drainage systems is exceeded [162], areas with high antecedent soil moisture (e.g., Europe as a whole) and/or snow (rain-on-snow) (e.g., Scandinavia and northeast Europe) [161], and regions with strong connectivity of surface and groundwater networks [71]. Fluvial processes dominate inland flooding in watershed catchments from channelized freshwater in dynamic hydrological networks. Flooding can also be fluvial-dominant in coastal regions fed by steep mountainous rivers that respond quickly to rainfall and snowmelt (e.g., Zhejiang China) [160]. Within primarily coastal influenced regions, driver dominance can be further broken down into surge-, wave-, and tide-dominated. Which of the components of extreme sea level is the principal driver varies on continental to regional scale depending on meteorological modulators and characteristics of landmasses.

In the case of mixed fluvial and coastal flooding in estuaries and deltas, identifying the dominant driver is more challenging as it varies based on location and channel geomorphology. River-sea interactions are highly dynamic, and the sensitivities of flood components can fluctuate greatly within a single estuary [149]. Common methods of classifying regions of driver dominance usually involve using Flow Interaction Indices [31,154] and Compound Hazard Ratio Indices [31,154,163,164]. As might be expected, most researchers have found that the lower estuary is tide- or surge-dominated, the middle estuary transition zone may be considered compound-dominated, and the upper river region is discharge-dominated [33,86,152-159,165]. General patterns of driver dominance are different across estuaries depending on the properties of watershed drainage basins (i.e., topography and morphology) and behaviour of storm events (i.e., path, orientation, intensity, duration, and time lag between drivers). Numerous studies map out regions dominated by each of the different flood drivers [152,166-170], often zoned as coastal, hydrological (fluvial and/or pluvial), or transition/compound (combined drivers determine the max water levels) based on numerical model simulations using different scenarios. The exact scenario definitions however often vary between studies making it difficult to compare results. Compound-dominant floods usually have greater surge extremes and quicker discharge due in part to flatter topography [171]. Large rivers are usually fluvial-dominant, while smaller and less connected rivers are more likely to be influenced by precipitation at the coast [124]. Similarly, Familkhalili, et al. [151] found that increasing channel depth reduces the impact of fluvial processes while amplifying the effect of coastal drivers on total water level. Therefore, channel deepening pushes the compound-dominated region further upstream and shortens the length of fluvial-dominated estuary. Flood dominance can also be significantly affected by the magnitude and severity of storm events such that a single location can be dominated by different drivers from different return period storms. Gori, et al. [172] observed surge-dominated flooding at the coast for low return period events, but compound-dominated flooding for high (100-year) return periods.

Fewer studies have examined the role of timing on flood driver dominance. In the case of TC/ETC events there is a time lag such that it can be hypothesized that coastal areas are first inundated by storm-tide followed by river discharge from upstream rainfall. Thus, at the beginning of storm events flooding is likely coastal (and/or pluvial) dominated and later switches to being compound dominated and then finally fluvial (and/or pluvial) dominated. For instance, the 1991 cyclone that hit Chittagong Bangladesh had a 5-hour difference between peak surge and peak rainfall [150]. As a result, the flooding began as coastal-dominated and then shifted towards being pluvial-dominated. The importance of timing may also fluctuate depending on the size of the water bodies in question. Dykstra and Dzwonkowski [173] found that slowing of river propagation in larger watersheds (>5000 $km^2$) led to a greater time lag between storm surge and river discharge, indicating greater risk of fluvial-coastal compounding in smaller watersheds where discharge travels downstream faster. Likewise, differences observed in the UK's Humber and Dyfi estuaries explain why maximum flood depth from fluvial-coastal compounding is less sensitive to timing in the case of a larger estuary (Humber) subject to slow river discharge, compared with short intense discharge in a smaller estuary (Dyfi) [149].

## 6.3) Urban and Coastal Infrastructure

Urban areas are identified in the literature database to be especially vulnerable to compound flooding, as the built environment can exacerbate the effects of flooding, and the concentration of people and infrastructure can lead to significant losses. In the coastal environment, hazard modelling and risk assessment practices regularly consider the influence of flood defence structure (i.e., barriers, sea walls, groynes, breakwaters), however other aspects of human activity (e.g., coastal and floodplain development and modification, land use/land cover change) and urban infrastructure (e.g., sewer waste drainage systems, water management reservoirs) receive less attention. Furthermore, existing urban infrastructure planning and risk assessment practices generally do not consider the ramifications of compounding flood drivers and thus underperform or have greater chance of failure from compound flooding [66,129,174]. For instance, in Jasim, et al. [174], coastal earthen levees were simulated to experienced 8.7% and 18.6% reductions in the factor of safety for 2-year and 50-year recurrence intervals under compound pluvial-fluvial flood conditions compared to fluvial-only flooding. Similarly, Khanam, et al. [175] found that FEMA maps significantly underestimate risk at several power grid substations in coastal Connecticut by not accounting for compound flood interactions This section will discuss the ways in which compound floods influence the performance of urban and coastal infrastructure, and how infrastructure in these settings can either amplify or reduce the risks and impacts of compound floods.

It is well established that the risks and impacts of compound flooding can be elevated in coastal and urban settings. Private property and public utilities developed within floodplains and along shorelines are more likely to be exposed to multiple coinciding flood mechanisms. Over the past century, changes in land use/land cover have made the urban environment increasingly susceptible to flooding. Urban areas experience increased precipitation as unstable warm city air masses rise (i.e., urban heat island effect) and then cool, forming rainclouds. This rain falls onto impervious surfaces (i.e., asphalt and concrete) and compacted soils (from construction and agriculture) which prevent surface water from seeping into the ground and percolating down into underlying aquifers [176]. Instead, water finds its way into river channels and urban drainage networks which act as highways and rapidly deliver vast volumes of water to the coast. During TC events, rainfall and river discharge are more likely to temporally overlap with coastal storm surge due to the heightened mobility of water within the urban environment. It is this combination of urban land cover and storm-sewer drainage infrastructure that play a substantial part in amplifying the impacts of urban coastal compound flood [65]. It has been well demonstrated that elevated water levels at the coast from storm surge can significantly reduce the rates of urban drainage resulting in more severe flooding [140,162,177]. Accumulated surface runoff in cities is meant to flow into rivers and ultimately the ocean, but high tides or waves can either block or force this water back inland. It has also been shown that poorly maintained and leaking stormwater drainage systems can cause compound pluvial-groundwater and fluvial-groundwater flooding where seawater travels inland via drainage systems (known as 'drainage backflow' and 'seawater intrusion') and flood areas near (and sometimes far from) the coast [148,157,178,179]. Furthermore, human activity including coastal and riverine modifications (i.e., dredging and straightening) [180] in favour of water utilities (e.g., hydroelectric) and transportation (e.g., marine shipping) also may increase the risks and impacts of compound flooding. Changing the morphology of coastal channels as often seen in urban ports, can amplify fluvial-coastal and pluvial-coastal compound flooding due to of reduced dissipation of energy and thus increased extreme peaks. Lastly, urban environments also pose the rare but catastrophic potential of damming/dam failure related compound flooding. For instance, in 2013 a German dyke breach led to a compound pluvial-damming/dam failure flood that affected hundreds of households and caused major damages to transportation infrastructure [2].

Urban infrastructure can also reduce the risks and impacts of compound flooding if designed to be resilient and forward looking. Management and policy decisions regarding urban infrastructure investment, maintenance, and outreach can play a large role in shaping compound event risk through the lens of population exposure and vulnerability [56]. Well maintained and operated coastal urban infrastructure from flood defence (e.g., storm surge barriers, sea walls, levees, breakwaters, and groynes) to flow management systems (e.g., dams, stormwater sewers, sump pumps, dry wells) can act to minimize compound flood risk when the dependence of multiple drivers is adequately considered. Furthermore, sustainable urban drainage systems (e.g., swales, infiltration trenches, retention basins, green roofs, and permeable paving)[181] can reduce the likelihood of compound flooding as they can create a time lag between peak pluvial, groundwater, and coastal processes. Lastly, natural flood management practices (e.g., wetland/floodplain/lake restoration, riverbed material re-naturalisation, river re-meandering)[181], can also serve to spread out the duration and reduce acute impact of compounding involving fluvial and coastal drivers, advancing the resiliency of urban and coastal environments.

### 6.4) Compound Flooding and Changing Climate

Many studies in the database stress that future compound flood risk is likely to increase from changes in the variability, intensity, frequency, phasing, and seasonality of sea level, precipitation, river discharge, and temperature driven by climate change [19,149]. Under a changing climate the interrelationships and dependence between variables contributing to compound events are likely to change. These potential changes in dependence give rise to uncertainty around compound flood prevalence. Projected increasing rainfall and TCs/ETCs will pose higher risks of compound flooding in coastal and tropical regions [182]. Long-term increases in the frequency of compound coastal river flooding from intensifying precipitation has already been observed throughout the past century [173]. A warmer atmospheres will bring more frequent and extreme storm events in many parts of the world including Europe and the Mediterranean [183]. The UK is expected to see increased clustering and intensity of storms (particularly in the winter) such as those seen in 2013/14 [63,149]. In North America, coastal regions will be at further risk of compound flooding from changes in rainfall and storm surge [132]. A rise in the annual number of compound floods from rainfall and storm surge (1-4 per decade) has already been observed in northern Europe and the US east coast [117]. Increasing trends in concurrent extreme precipitation and storm surge events have been observed across most of the world [117]. SLR will likely pose the largest threat of compound flooding at the coast [86,149,184,185] with global mean sea level projected to increase 0.61-1.10m (RCP8.5) by 2100 (relative to 1986-2005) [186]. This is already drastically affecting island nations in Southeast Asia and the Pacific that are vulnerable to pluvial-coastal flooding from storm events. Furthermore, extreme sea level frequency will "very likely" increase over the century from the compounding of SLR, storm surge, and waves [12]. At a global scale (mid-latitudes especially), compound flooding will be increasingly driven by precipitation extremes and atmospheric driven storm surge.

In summary, across the studies reviewed, climate change is shown to be having a profound impact on the frequency and severity of compound flooding events [157]. The combination of heavy precipitation events, SLR, and changes in the frequency and intensity of storms and hurricanes are all contributing to the increased likelihood of these events.

### 6.5) Research Approaches

As highlighted in Section 5.4, we identified two main categories of approaches that have been used to assess compound flooding, namely, (1) physical (process-based) numerical modelling; (2)

and/or statistical modelling/analysis. In both approach classes we observed a diversity of methods, similarly to the findings of Tilloy, et al. [53]. Below, we discuss the use of computational numerical methods for compound flood modelling (Section 6.5.1), then provide an overview of the statistical and data science-based techniques for analysing compound flooding (Section 6.5.2), and finally reflect on the benefits of hybrid (numerical-statistical) approaches (Section 6.5.3).

### 6.5.1) Numerical Modelling

Compound flood events are often examined by numerically modelling the physics-based interactions of their processes and mechanisms. Through the simulation of historic and synthetic compound flood events, researchers can develop a better understanding of present and future inundation magnitude and extent. Given the highly complex nature of compound flooding, numerical modelling often requires a combination of hydrological, hydrodynamic, and atmospheric/climate models to represent all earth systems components contributing to compound flooding. A range of different numerical models are used in the literature, as we briefly discuss here. Further information on the hydrological, hydrodynamic, and atmospheric models, frameworks, systems, and toolsets used in the reviewed studies is provided in Appendix 2.

Hydrological models are used to simulate the movement, storage, and transformation of water within the hydrological cycle. These include land-atmosphere water exchange (precipitation and evapotranspiration), flow of water through the landscape (streamflow and rainfall-runoff), and the infiltration of water into the ground (groundwater recharge). Hydrodynamic models use a series of governing equations to simulate the flow of water in rivers, oceans, estuaries, and coastal areas. Coastal hydrodynamic models replicate the propagation and advection of water based on a combination of tide, surge, and waves. In the realm of compound flooding, hydrodynamic models are vital for simulating the effects of complex river-ocean interactions, storm surge, lake seiche, and flood infrastructure. Atmospheric models simulate various atmospheric processes based on primitive dynamic equations explaining radiation, convection, heat flux, gas exchange, kinematics of air masses, behaviour of water vapor (precipitation and clouds), and land/ocean-atmosphere interactions. In compound flood research, numerical atmospheric modelling is generally used to simulate synthetic or historical storm events (TCs/ETCs) and to generate meteorological inputs (e.g., precipitation, atmospheric pressure, and wind velocity) that force hydrological and hydrodynamic models.

Compound flood modelling often involves the use of coupled or linked models. Individually, hydrological and hydrodynamic models are unable to capture the full dynamic interactions between inland and coastal processes [187]. However, integrating the capabilities of both types of models can serve to better simulate the movement and transformation of water within a particular system as shortcomings of one model can be complemented by the strengths of another. Santiago-Collazo, et al. [36] define four techniques for linking different types of models: one-way coupled; two-way (or loosely) coupled; tightly-coupled; and fully-coupled. One-way coupling involves using the output of one model as the direct input for another model, such that data only transfers in one direction. Alternatively, two-way coupling describes a relationship in which the outputs of both models transfer information to each other iteratively, creating a two-way loop that influences behaviour of both. Tight coupling refers to the integration of two independent models into single model framework at the source code level. A common example of tight-coupling is the ADCIRC-SWAN model. SWAN sends simulated waves to ADCIRC, and ADCIRC sends water levels and wind velocities back to SWAN. Lastly, full coupling is the complete integration of all model components such that physical processes are calculated simultaneously under the same framework using the same governing equations. We observed that most of the existing compound flood indentation modelling implements simple one-way or two-way coupling approaches [36,37]. Fully coupled numerical models are rare in compound flood research, as most models only specialize in one or two earth systems (i.e., meteorology, climatology, hydrology, and oceanography).

### 6.5.2) Statistical Approaches and Dependence Analysis

Across the studies we have reviewed, a wide variety of statistical-based approaches have been employed to understand trends, patterns, and relationships using observed data, sometimes complemented by physically simulated data. This predominantly involves the use of statistical models as an indirect measure of compound flooding potential to better understand the dependence between different flood drivers and the likelihood of their joint occurrence.

There are several broad statistical techniques that are frequently used for compound flood research. Some of the most prominent methods include varying forms of spatial and temporal analysis, regression analysis, extreme value analysis, Bayesian probability, principal component analysis, index analysis, Markov chains, and machine learning (ML). Spatial and temporal analysis investigate correlations, covariance, trends, and patterns in where and when compound flood events occur. This can include identifying compound flood hotspots [26,116,117,119,125] and temporal clustering [24,125,188-190] or examining the underlying spatiotemporal preconditions and interactions of flood components [24,116]. Regression analysis involves using statistical functions to identify relationships between independent and dependent flood variables by fitting data to linear and higher order non-linear functions [65,86,109,117,185,191-199]. Extreme value analysis examines the tail distribution or threshold exceedances of extreme flood variables to better understand joint-probability, uncertainty, and severity [97,108,200-203]. Bayesian statistical approaches can iteratively recalculate the likelihood of an event based on new evidence. Bayesian frameworks are often used to update predictions about compound flood hazards based on new data and to understand the uncertainties associated with these hazards [86,109,153,192,204-207]. Principal component analysis is a method of reducing the dimensionality of data by selecting the most important variables and combining them into a smaller volume of composite variables. In compound flood research this approach can be used to reduce the complexity of compound flood data to identify the key factors contributing to compound flood hazards [116]. Index analysis is a method of data interpretation in which statistical indices simplify our understanding of the behaviour of multiple variables, a practice commonly used for flood risk and impact analysis [31,154,164,208-213]. Compound flood research takes this further using various indices that also consider the synergy of multiple flood drivers [154,164,208,210,211,213,214]. Markov chains use records of past variable states to describe the probability of future states. With this approach, flood variable data such as rainfall and river levels can be fit to stochastic models to simulate the probability of joint extreme states. Additionally, Monte Carlo Markov Chain (MCMC) approaches involving stochastic sampling of variables are sometimes also applied in compound flood research [87,139,184,214]. Lastly, in recent years ML models involving varying neural network structures have been trained using compound flood datasets to predict flood extremes or map inundation extents [194-196,204,209,215,216].

Understanding the dependence of compound flood variables is crucial as it tells us about their joint exceedance probability [37,134]. Failure to investigate driver dependence will lead to an underestimation of flood probabilities. Varying forms of the Joint Probability Method (JPM) [102,107,217], involving aspects of extreme value analysis, are commonly used to measure potential co-occurrence and dependence between compound flood drivers. Over time the analytical approaches have evolved, but generally involves three main steps for investigating dependence and frequency of cooccurring events. First, the flood variable event sets are sampled. The second step involves a simple calculation of varying correlation coefficients from the driver data. The third step consists of fitting a multivariate distribution function.

In preparation of the following steps, flood variables datasets are created by sampling events (according to varying compound scenarios, i.e., AND, OR, Kendall) via block-maxima or threshold-excess (peak-over-threshold, POT) methods. Block maxima sampling selects the maximum events within a given temporal block (annual, seasonal, daily), while the threshold-excess method selects events above a defined 'extreme' threshold value. Next, the correlation coefficient step typically implements different types of rank correlation coefficients and tail coefficients. Correlation coefficients such as Kendall's tau $\tau$ and Spearman's $\rho$ can reveal non-linear relationships between random variables based on their ordinal associations. Alternatively, the lower ($\lambda_L$) and upper ($\lambda_U$) tail coefficients help examine dependence between random variables at the extremes of their distributions. While random variables may appear to show no correlation, the co-movement of their tails may reveal dependence relationships that only occur at the extremes. The joint probability distribution is then constructed from the sampled variable event datasets as the probability of all possible pairs across each input variable. The joint probability distribution thus defines the probability of two or more simultaneous events, where the variables are at least partially dependent, and thus influence each other's occurrence.

In recent years copula have also been used to measure dependence, gaining considerable attention for their ability to simplify the analysis of highly stochastic multivariate processes. A total of 64 (24%) studies were observed using copula-based methods to assess dependence. Defined in Sklar's theorem [218], a copula is multivariate cumulative distribution made by joining or "coupling" the univariate marginal probability distributions of two or more individual variables. This can be done using several dependence structures, with common copula families being Elliptical and Archimedean. In addition to measuring dependence, copulas are used in compound flood research to assess the non-linear relationships and uncertainties between extreme flood variables [219,220]. By fitting copula functions to multivariate flood data, it is possible to understand the strength and nature of the dependence between these variables and to predict the likelihood of compound flood events. To date, the majority of compound flood research involves bivariate case studies. Nonetheless, several studies have implemented trivariate approaches to simultaneously analyse three partially dependent variables [71,130,159,164,221-227], and others have taken more complex procedures integrating copulas with MCMC [139,155,184,228] and Bayesian network [155,205,206,214] approaches. For further detail on copula-based multivariate flood analysis see Latif and Mustafa [229].

### 6.5.3) Hybrid Modelling and Analysis Approaches

Hybrid methods, involving linking numerical and statistical approaches off were commonly observed throughout the literature database, with around one-third of compound flood studies employing hybrid techniques (Figure 6). Hybrid approaches can complement each other or focus on multiple aspects of modelling in a way that would not be possible when using numerical or statistical approaches in isolation. For example, process-based numerical modelling of compound flood hazards may be ideal for physics-based inundation mapping and floodplain delineation, but can be very computationally expensive (this has pushed development of more computationally efficient models such as SFINCS [230]). Conversely, simplified statistical models are less computational expensive, but typically make general assumption about input data that do not fully consider the physical processes at play. In contrast, hybrid numerical-statistical approaches offer the benefit of computational efficiency of surrogate statistical modelling while still maintaining a realistic representation of the physical processes [196]. Additionally, numerical modelling can also be severely inhibited by historical data availability. Hydrodynamic modelling of astronomical tide and storm surge require atmospheric pressure and wind velocity forcing data, while past river level and rainfall data is dependent on the presence of in-situ tide and rain gauge monitors. If these datasets don't exist or have poor spatiotemporal coverage, numerical hydrodynamic models must rely on reanalysis data. Statistical approaches to compound flood analysis however can sometimes make do with limited data by interpolating or extrapolating extreme hazard probabilities and distributions. In the absence of historical data, one solution is to numerically simulate synthetic events that are physically capable of occurring, albeit not present in short term observations [196]. Many hybrid approach compound flood studies statistically simulate storm events that drive physical hydrodynamic and hydrological models [155,196].

## 6.6) Research Applications

As highlighted in Section 5.5, we identified that six main applications have been the focus of most compound flood studies in the database. Discussed in the following order, prominent case study applications include earth system processes (Section 6.6.1); risk assessment (Section 6.6.2); impact assessment (Section 6.6.3); forecasting (Section 6.6.4); planning and management (Section 6.6.5); and methodological advancement (Section 6.6.6). Note, many of the compound flood studies fall into multiple application categories.

### 6.6.1) Earth System Processes

From the 271 literature database entries, 128 (47%) seek to better understand the processes, interactions, and behaviour of earth systems associated with compound flooding. Research papers within the earth system processes application theme examine a variety of topics including the role of various dynamic earth systems on compound flooding, the environmental and landscape characteristics influencing flood drivers, the relationships between and relative significance of flood drivers, and the spatiotemporal distributions and frequency of compound flood events. Many of the papers discussed in Sections 6.1, 6.2, and 6.5 fall within this application category.

Focusing on flood drivers relationships, there is a plethora of research examining aspects of spatiotemporal distribution, correlation, covariance, dominance, and dependence structures as demonstrated in the US [131,154,170,231], UK [135-137,189,190], Europe [23,120,125,232], Australia [121,147,233,234], Canada [130,164], China [159,165,208], South Africa [143], India [24], Indonesia [156], New Zealand [122], Germany [108], and globally [26,118,126,134]. Many have simulated or projected how climate change (e.g., SLR and storm intensification) are expected to affect the future compounding interactions of flood drivers [86,132,183-185,202].

There is also notable insight into the large-scale meteorological and climatological modulators and underlying earth systems influencing the nature of compound flooding and behaviour of flood drivers. For instance, Camus, et al. [116], Hendry, et al. [135], and Rueda, et al. [212] identify the meteorological conditions associated with the compound occurrence of extreme flood drivers in the North Atlantic, the UK, and Spain respectively. Gori, et al. [14] and Gori, et al. [168] determine the type of TC events likely to cause compound pluvial-coastal flooding in North Carolina. Stephens and Wu [122] identify the weather types corresponding with both univariate and coincident pluvial, fluvial, and coastal extremes in New Zealand. Furthermore, Wu and Leonard [233] demonstrate how ENSO climate forcings impact the dependence between rainfall and storm surge extremes.

Other common focuses of earth system processes themed literature include characterizing the physical mechanics and environmental properties that shape the ways in which flood drivers interact. Several papers including Vongvisessomjai and Rojanakamthorn [105], Poulos, et al. [235], and Pietrafesa, et al. [236] evaluate the timing and mechanisms behind downstream blocking and dampening that often explain fluvial-coastal flooding. Similarly, Maymandi, et al. [170] measure the timing, extent, and intensity of storm surge, river discharge, and rainfall components to understand their relative importance. Likewise, Tanim and Goharian [150] observe how changes in tidal phase alter the depth and duration of urban compound pluvial-coastal flooding. Harrison, et al. [149] and Helaire, et al. [237] measure how estuary characteristics (e.g., shape, size, width) influence fluvial-coastal dynamics. Wolf [238] consider how wind-stress, bottom friction, depth, bathymetry, and ocean current refraction change co-occurring surge and wave extremes (coastal-coastal). Torres, et al. [239] and Gori, et al. [168] examine the influence of hurricane landfall location, angle of approach, and forward speed on compound rainfall-runoff and storm surge flooding (pluvial-coastal). Tao, et al. [208] explore compound fluvial-pluvial flood scenarios involving upstream and downstream water levels, and how intensity, timing, duration, and dependence change based on synoptic and topographic conditions.

Lastly, while the occurrence of compound flooding is well recognized in coastal, estuary, and delta environments, we note that emerging research has enhanced the understanding of compound flood processes in the context of coastal lake environments [164,188,207,240]. For example, Banfi and De Michele [188] determine that flooding of Italy's Lake Como is primarily (70%) from temporal compounding of rainfall (pluvial-pluvial). In Lake Erie, Saharia, et al. [240] analyses compound flooding involving river flow and lake seiche (fluvial-coastal), showing for the first time how seiches can combine with hydrological processes to exacerbate flooding. Finally, along Lake Ontario, Steinschneider [207] quantified the compounding nature and variability of storm surge and total water level (coastal-coastal).

### 6.6.2) Risk Assessment

The overarching goal of most compound flood research is to better understand risk, hence why 127 (46%) studies involve aspects of risk assessment. As defined by the UNDRR [48], risk assessment is an approach for determining the state of risk posed by a potential hazard taking into account conditions of exposure and vulnerability. Risk assessment inherently plays a key role in several of the reviews' other research application categories including hazard planning and management as well as impact assessment.

As the field of compound event sciences advances, it has become increasingly clear that conventional univariate analysis cannot accurately capture the synergistic and non-linear risk of compound processes [17,20,22,26,27,41,228]. A plethora of studies have concluded that traditional hazard analysis, in which flood variables dependence and synergy is not considered, underestimate the risk of compound extremes [33,135,171,241-243]. Jang and Chang [191] determine that by not considering the multivariate nature of pluvial-coastal flooding, Taiwan's flood risk would be severely misestimated causing incorrect warning alarms and inadequate protection. Khalil, et al. [13] assert that failing to consider the interactions of multiple flood drivers would reduce flood levels by 0.62m and 0.12m in Jidalee and Brisbane. Similarly, Santos, et al. [224] measured 15-35cm higher water levels for 1% annual exceedance probability events when considering dependence for trivariate fluvial-pluvial-coastal flooding in Sabine Lake, Texas.

There is a diversity of topics within the risk-themed compound flood literature, but many papers involve simple regional case studies or framework proposals [129,206,225,244]. Čepienė, et al. [245] examine risk associated with combined fluvial-coastal flooding and how it will change with SLR at the port city of Klaipėda. Bischiniotis, et al. [141] assess the influence of antecedent soil moisture on flood risk in sub-Saharan Africa, showing that precipitation alone cannot explain flood occurrence. Along the coasts of Mozambique, Eilander, et al. [144] demonstrate a globally applicable compound flood risk framework and Van Berchum, et al. [145] present the novel Flood Risk Reduction Evaluation and Screening (FLORES) model. Bass and Bedient [204] create joint pluvial-coastal flooding probabilistic risk models built upon TC risk products in Texas. A few studies examine the risk of Potential Loss of Life (PLL) such as De Bruijn, et al. [167] who present a Monte Carlo-based analysis framework for fluvial-coastal interactions in the Rhine-Meuse delta.

### 6.6.3) Impact Assessment

Impact assessment is the least common compound flood application with only 12 (4%) relevant studies. This may be because flood impact assessments have historically only been designed to address a single type of flooding at a time [246]. Additionally, flood loss modelling has largely targeted riverine floods, with less attention given to pluvial, coastal, or groundwater drivers [247]. This is slowly changing, and in recent years a small portion of research has been dedicated to analysing the impacts of compound flood events [148,211,213,246,247]. Impact assessment differs from risk assessment in that it looks at the realized or impending outcomes of flood events rather than simply the event likelihood as a product of exposure and vulnerability. This involves identifying and analysing the physical (e.g., building and infrastructure damage), social (e.g., loss of essential services, household displacement, and community cohesion), and economic (e.g., loss of income, damage to business and industry, and disruption of transportation and supply chain) impacts of flooding.

Physical parameters for quantifying the empirical impact of flooding in an affected area can include water depth, flow velocity, inundation duration, water quality (contamination), land use/land cover change, and infrastructure damage. For example, Habel, et al. [148] look at the influence of compound floods and SLR on urban infrastructure and identify the roadways, drainage inlets, and cesspools that would fail under compound extreme conditions.

Social and economic flood impacts are routinely measured using multifaceted indices and damage models. Preisser, et al. [211] and Tanir, et al. [213] assessed impacts of compound flooding with SVI (Social Vulnerability Index; 42 variables) and SOVI (Socio-Economic Vulnerability Index; 41 variables) respectively. Karamouz, et al. [248] apply a flood damage estimator (FDE) model to quantify pluvial-coastal flood damages to buildings structures in New York City. Similarly, Ming, et al. [225] calculate the average annual loss in value of residential buildings in the Thames River catchment from compound flooding. Lastly, Thieken, et al. [2] assessed the differing impacts and coping abilities (financial damage, psychological burden, and recovery) of residents following compound river-dyke breach (fluvial-damming/dam failure) and flash flood-surface saturation (pluvial-soil moisture) events.

### 6.6.4) Forecasting

A total of 21 (8%) compound flood studies in the database focus on flood forecasting. Flood forecasts are valuable emergency management tools that provide information on location, timing, magnitude, and potential impact of impending flood scenarios [249]. Together with monitoring and prediction, forecasts guide time sensitive early warning systems and disaster reduction strategies to help communities prepare for and respond to flooding. As compound event-based perspectives gain traction, there has been emerging development of flood forecast models that consider the compound interaction of multiple drivers.

Several studies have demonstrated the capabilities of integrated near-real-time observation-based hydrological river and hydrodynamic coastal flood models forced by already established meteorological forecasting systems [250-258]. For instance, the fluvial-coastal flood forecasting system Hydro-CoSMoS detailed in Tehranirad, et al. [258] can predict tidal river interactions in San Francisco Bay. Over the Korean peninsula, Park, et al. [256] design a model for real-time water level forecasting of pluvial-coastal inundation such as seen during Typhon Maemi.

Much of the existing compound flood forecasting research has focused on advances in the development of monitoring and early warning systems for the US East Coast and Gulf of Mexico. Blanton, et al. [250] feature development of the North Carolina Forecasting System (NCFS) which predicts fluvial-pluvial-coastal flood variables. Van Cooten, et al. [259] showcase the Coastal and Inland Flooding Observation and Warning (CI-FLOW) Project's 7-day total water levels forecasts and potential for near-real-time fluvial-pluvial-coastal flood prediction. Dresback, et al. [253] develop the coupled hydrological-hydrodynamic model ASGS-STORM for forecasting joint fluvial-coastal inundation. Multiple studies also concentrate on flood forecasting in the Chesapeake Bay and tidally-influenced Potomac River . Stamey, et al. [257] introduce the Chesapeake Bay Inundation Prediction System (CIPS), a prototype operational flood forecasting system for TC/ETC storm system induced fluvial-coastal flooding. This is followed by Mashriqui, et al. [254] and Mashriqui, et al. [255] who build a River-Estuary-Ocean (REO) forecast system to fill gaps in existing operational models.

Accurate forecast products are crucial to effective emergency management practices and reliable early warning systems. Ensemble modelling has been implemented in two compound forecasting studies as a means of minimizing uncertainty. Blanton, et al. [251] develop a hurricane ensemble hazard prediction framework and demonstrate the ability to forecast pluvial-coastal flooding with a 7-day lead simulation of Hurricane Isabel. Similarly, Saleh, et al. [260] showcase a 4-day advance operational ensemble forecasting framework for fluvial-coastal flooding in Newark Bay during Hurricanes Irene and Sandy.

A number of studies have also investigated the use-case of ML for forecasting compound flooding [194,204,209].. For instance, Sampurno, et al. [194] use a combined hydrodynamic and ML approach to forecast fluvial-pluvial-coastal flooding in Indonesia’s Kapuas River delta. Bass and Bedient [204] take peak inundation levels from a coupled hydrological-hydrodynamic model results to train an Artificial Neural Network (ANN) and Kriging ML model for rapid forecasting of TC-driven pluvial-coastal extremes in Houston, Texas as a result of Hurricanes Allison and Ike. Finally, Huang [209] constructs a Recurrent Neural Network (RNN) model that considers downstream geomorphological and hydrological characteristics to predict joint pluvial-coastal flooding in Taiwan.

### 6.6.5) Planning and Management

Within the literature database there are 29 (11%) papers that focus on different aspects of flood management from emergency response planning to risk mitigation strategies. The UNDRR [48] define disaster management as the organization, planning, and application of measures for disaster response and recovery. Subsequently, disaster risk management is described as the use of disaster risk reduction strategies and policies to prevent, reduce, and manage risk [48]. Flood management strategies might involve identifying areas for prioritized flood protection and building risk reduction structures such as building levees, dykes, barriers, and sea walls; or enacting changes in land use planning and zoning policy to minimize habitation and activity in floodplains.

Flood defence and water management structures have long been in use; however these features have predominantly been designed for responding to a single flood driver (e.g., storm surge) [157]. Several studies examine the effectiveness of flood defence structures protecting against compound events. Christian, et al. [261] investigate the feasibility of a proposed storm surge barrier for mitigating pluvial-coastal flooding in the Houston Shipping Channel. Findings on the magnitude of reductions in surface height and floodplain area help guide project development decision making by coastal and port authorities. Del-Rosal-Salido, et al. [152] develop management maps to support decision making and long-term climate and SLR adaptation planning in Spain's Guadalete estuary, identifying sites for potential flood barriers.

During extreme flood events, unpredictable impacts to utility and transportation infrastructure can exacerbate loss. Thus, another key component of flood management is flexible emergency response planning. Several articles address these elements of response planning, identify evacuation areas, routes, and emergency shelters in the event of compound flooding. In their analysis of urban infrastructure failure from compound flooding in Hawaii, Habel, et al. [148] locate road networks and urban spaces that are likely to be impassable and estimate the effects of traffic on resident evacuation. In the event of Typhon landfall in the Korean peninsula, Park, et al. [256] design an early warning system for pluvial-coastal flooding that supports decision making and response from local officials by identifying areas to evacuate. Blanton, et al. [251] also address emergency planning, developing a hurricane-driven inundation evacuation model that dynamically accounts for interactions of compound drivers.

Effective communication and outreach are additional critical components of flood hazard planning and mitigation. This includes educating the public about the types and considerations of flooding, collaborating with hazard managers and policy makers to address challenges in flood management, and timely dissemination of information on flood risk, evacuation routes, and emergency shelters. In a unique narrative paper, Curtis, et al. [115] interview emergency managers and planners on compound flood risk perceptions and challenges and reveal inadequacies in communication mediums and the ability to convey compound flood severity to the public. Similarly, Thieken, et al. [2] survey German residents affected by two compound flood events on their understanding of compounding drivers and the communication medium through which they learned about the events. Modrakowski, et al. [114] centres on the use of precautionary risk management strategies in the Netherlands, and how perception of compound flood events in-part shapes the flood management practices of local authorities. Interestingly, both Curtis, et al. [115] and Thieken, et al. [2] discovered a greater perception of risk from fluvial and coastal dominant flooding as opposed to pluvial inundation. Conversely, Modrakowski, et al. [114] found that pluvial flooding (specifically heavy rainfall from cloudbursts) had a larger perceived risk, being equal if not greater than fluvial and coastal. These findings on compound flood communication and perception help hazard managers determine how to approach emergency response and risk mitigation planning.

### 6.6.6) Methodological Advancement

The third most common application category is methodological advancement with 73 (27%) of the 271 studies aimed at testing and developing methodologies for research on compound floods. Methodological advancement is a broad application category, but most often describes research studies that investigate either new setups and frameworks for running numerical model simulations, or novel statistical modelling and analysis techniques for quantifying the likelihood of compounding extremes or behaviour of interacting drivers. Papers classified as methodological advancement seek to better understand and showcase the feasibility, development, and/or performance of compound flood research methods. Here forward see Appendix 2 for full model names and descriptions.

In relation to advancements in numerical-based methodologies, many papers explicitly state their primary research objective is the development of a compound flood modelling system itself, such as Chen and Liu [262] and Lee, et al. [263], who test whether their respective SELFE and HEC-HMS + Delft3D-FLOW model frameworks can sufficiently replicate the fluvial-coastal flood conditions observed during historical storm events. Bates, et al. [264] showcase a sophisticated 30m resolution large-scale LISFLOOD-FP centric model of the contiguous US that incorporates pluvial, fluvial, and coastal processes under the same methodological framework. Numerous papers focus on assessing the performance of specific computational software applications for simulating compound flooding. These primarily seek to provide insight for future development and use case application. For instance, Bush, et al. [265] examine the benefits and drawbacks between ADCIRC and combined ADCIRC + HEC-RAS simulations of fluvial-coastal flooding. Bilskie, et al. [166] demonstrate a new approach for delineating coastal floodplains and simulating water level using ADCIRCs "rain-on-mesh" modules forced by antecedent rainfall, TC-driven rainfall, and storm surge. Ye, et al. [187] use SCHISM to develop a 3D model that incorporate the baroclinic effects of storm surge and compare its performance against 3D barotropic and 2D models alternatives. Numerous studies incorporate sensitivity assessments, experimenting with model parameters and settings, and examining how they influence performance and uncertainty [13,230,266-271]. For example, Khalil, et al. [13] investigate how model mesh resolution affects flood discharge rates, revealing that finer meshes best replicate peak flows. Some studies introduce newly developed numerical models, such as Olbert, et al. [271], who present the first instance of a dynamically linked and nested POM + MSN_Flood framework for fluvial-pluvial-coastal flooding. Others focus on the computational efficiency of compound flood frameworks, for instance Leijnse, et al. [230] assess the reduced-physical solver SFINCS's ability to accurately simulate fluvial-pluvial-coastal interactions with less computational resources.

Many of the literature database studies showcase innovations in statistical approaches to compound flood research. Sampurno, et al. [194] assess the operational viability and performance of three ML algorithms for compound flood forecasting system. Similarly, Muñoz, et al. [216] examine the capability of ML and data fusion-based approaches for post-event mapping of compound floods from satellite imagery. Muñoz, et al. [272] demonstrate techniques for employing data assimilation to reduce uncertainty in compound flood modelling. Wu, et al. [273] experiment with three methods of compound flood frequency analysis and discuss the advantages and disadvantages of each approach. Phillips, et al. [274] examine combinations of varying copula structure and statistical fitting frameworks to further approaches for measuring driver dependence. Thompson and Frazier [275] test out different means of deterministic and probabilistic modelling for quantifying compound flood risk. Lastly, some studies expand on existing methodologies to overcome known limitations, such as Gouldby, et al. [276] who develop a method of full multivariate probability analysis that overcomes drawbacks of the prevalent joint probability contours (JPC) method by directly quantifying response variable extremes.

## 7) Knowledge Gaps and Improvements for Future Research

Our final objective is to reflect on the knowledge gaps in compound flood research and suggest potential directions for research going forward. Based on our detailed review we have five main recommendations moving forward, as follows:

**Recommendation 1 - Adopt consistent definitions, terminology, and approaches:** Definitions and use-cases of compound event, compound hazard, multi-hazard, and associated terminology (Table 1) are highly inconsistent throughout the literature [53,55,277]. This is well recognized in Tilloy, et al. [53], who refer to the variety of terms as a “fragmentation of [the] literature.” Similarly, Pescaroli and Alexander [57] draw attention to trends in “superficial” and “ambiguous” use of hazard terms by academics and practitioners. This tendency to use differing concepts synonymously is blurring the state of compound flood research (something we observed ourselves while completing this review). They warn of potential confusion and duplication of research as a result of overlapping definitions. In summary, compound event and related terms have a wide range of overlapping and interlinked definitions, and there is a considerable need for clarity. Recent preliminary efforts by the collaborative MYRIAD-EU project to develop a multi-hazard and multi-risk definitions handbook appear promising for fostering a common understanding of hazard concepts across disciplines [278].

**Recommendation 2 - Expand the geographic coverage of research:** Geographically, much of the existing compound flood research is too narrowly focused on a select few regions (i.e., North America, Europe, Southeast Asia, UK, China, the Netherlands, Australia) (Figure 3b). To date there are no English-language studies, to our knowledge, on compound flooding in any parts of South America, Central America, or the Middle East. South America regularly experiences catastrophic flooding from both long-term heavy rainfall and extreme river discharge (e.g., 2015/16 [279] and 2016/17 [280] South American floods), however existing research in these regions has not considered their combined interactions. Furthermore, there are very few compound flood papers within the African subcontinent [139,141,143,145] (a region deserving of greater attention given the projected extreme coastal hazard exposure as a result of SLR, population growth, and coastal urbanization [281]) due to a lack of data. Thus, for much of the world, knowledge on the interactions and dependence of flood variables is missing. Future compound flood research must be dedicated to improving our understanding of these neglected regions and developing methodologies for assessing compound flooding in data sparse areas.

**Recommendation 3 - Pursue more inter-comparison and collaborative compound flood projects:** Current methodologies for analysing compound flooding are highly diverse, inhibiting quantitative comparisons between studies. Considerable subjectivity is observed in compound event mechanism and variable selection, temporal and spatial bounds, hazard scenario design, conditional and joint probability, and dependence measurement [19]. Standard approaches for compound flood risk analysis have yet to be established [55,157]. Furthermore, methods for analysing compound events vary across scientific communities [53,236]. Discussions between emergency manager and stakeholder have revealed the leading barrier to the use of multi-hazard and multi-risk approaches was a lack of common methodologies and data [52]. Further highlighting this point, Tilloy, et al. [53] identified a staggering 79 unique uses of 19 different methods for analysing compound events. There is a substantial need for a standardized framework that addresses assorted analytical methods and considerations [157] including flood variable choice and pairing, flood threshold definition, case study hazard design, spatiotemporal scales and resolutions, statistical model assumptions, and numerical parameter choice. Future water management practices and coastal hazard mitigation strategies must better reflect the perspectives of compound events. To aid this we would recommend that the community create a compound flood inter-comparison project, similar to that set up for the wave and coastal modelling communities (i.e., COWCLIP [282] and CoastMIP [283]).

**Recommendation 4 - Develop modelling frameworks that holistically represent dynamic earth systems:** While there have been substantial advancements in compound flood research over the past decade, the overall ability to identify, model, quantify, and forecast compound flood events remains a substantial challenge. These difficulties stem from the highly complex and chaotic nature of hydrological, meteorological, and oceanographic systems [157]. Connections between flood modulators and drivers are spatiotemporally dynamic, and how those relationships are affected by the changing climate is uncertain and everchanging. Stand-alone numerical models generally lack the ability to holistically simulate the dynamic interconnected systems necessary to explain compound flooding (especially in the coastal setting). The skill of compound flood forecasting systems and numerical models have improved but still largely remains inadequate [236,255]. Going forward, we recommend adoption of standardized modelling interfaces (e.g., Basic Model Interface [284]) to facilitate coupling between numerical models to develop holistic modelling frameworks that better disentangle the complex earth system processes driving compound floods. Compound flood research also serves to greatly benefit from the use of hybrid modelling frameworks that couple numerical and statistical models. While this review discovered many studies that employed hybrid numerical-statistical methods, few explicitly outlined a standardized frameworks for linking the models. Thus, we additionally recommend further evaluation of hybrid frameworks as the linking of statistical and numerical models has considerable room for improvement.

**Recommendation 5 – Plan and design urban and coastal infrastructure with compound flooding in mind:** We advise reshaping the planning, design, and operation of urban and coastal infrastructure to fully recognize the dependence and synergetic extremes of interacting flood drivers. As we look to a future of increasing flood frequency, proactive flood management is vital to lowering the vulnerability and exposure of urban and coastal communities. This can include investing in long-term resilient infrastructure (i.e., >100-year extremes), supporting development blue-green and natural flood management (e.g., wetland protection, riverbank restoration, and leaky dams), enacting operational early warning systems and emergency response measures, and educating the public about the risks of inhabiting coastal floodplains.

## 8) Conclusions

We have long known that high-impact hazard events involve a combination of drivers, however existing research has largely been limited to single-factor or univariate analysis of climate extremes due to technical or methodological constraints. Such is the case with flooding, as standard flood hazard assessment practices have traditionally accounted for the effects of the different drivers of flooding independently. Only in recent years has flood research more closely examined the non-linear combination of these variables through the lens of compound events.

This paper has presented a systematic review of the existing literature on compound flooding in coastal regions. Analysis of 271 studies up to 2022 has revealed significantly increased attention to compound flood research in recent years. This review identified different definitions and terminologies of compound flood events, categories of compound flood drivers, numerical modelling frameworks, and statistical analysis techniques. Furthermore, several compound flood hotspots have been identified throughout the world including the US East Coast and Gulf of Mexico, Northern

Europe, East Asia, Southern Asia, Southeast Asia, Northern Australia, and global low-lying deltas and estuaries. Research has shown that compound floods are likely to have increasing frequency and severity in the future as a result of climate change, and that societal risks of extreme climate hazards are underestimated when the compound effects of climatic processes are not considered in combination. Compound flood research thus requires a more holistic and integrated approach to risk analysis that reflects on the complex interactions and nonstationary of Earth systems. We must recognize the threats posed by the interactions between hazard drivers for accurate risk assessment. Further research must also focus on identifying the dominant drivers of flooding, the precursors that make certain regions particularly susceptible to compound flooding, the dependence relationships between flood drivers, and investigate how all these aspects change spatiotemporally. Going forward, an improved understanding of compound flooding processes and precursors is vital to coastal management, hazard risk reduction, and community resilience in the face of changing climates.

**Acknowledgements**

We thank Kate Davis[1] for creation of schematic diagram (Figure 1). JG time was supported by the UKRI Natural Environmental Research Council (NERC) grant NE/S007210/1. ID time was supported by NERC grant NE/S010262/1. JN time was supported by the NERC grant NE/S015639/1. T.W. acknowledges support by the National Science Foundation (grant numbers 1929382 and 2103754) and the USACE Climate Preparedness and Resilience Community of Practice and Programs.

**CRediT (Contributor Roles Taxonomy)**

Joshua Green: – Conceptualization, Writing - Original Draft Preparation, Writing - Review & Editing, Data Curation, Methodology, Formal analysis

Ivan D. Haigh: – Conceptualization, Writing - Review & Editing

Niall Quinn: Writing - Review & Editing

Jeff Neal: Writing - Review & Editing

Thomas Wahl: Writing - Review & Editing

Melissa Wood: Writing - Review & Editing

Dirk Eilander: Writing - Review & Editing

Marleen de Ruiter: Writing - Review & Editing

Philip Ward: Writing - Review & Editing

Paula Camus: Writing - Review & Editing

## Supplementary Material

*Appendix 1. Overview of the literature database containing 271 compound flood research publications. Note: Numerical models without defined names are given simple descriptions. Statistical methods are defined as explicitly stated in the literature and then simplified for brevity.*

| Author | Geographic Region | Scenario / Event | Application | Compound Drivers | Numerical | Statistical | Numerical & Statistical | Numerical Models | Statistical Methods / Tools |
|---|---|---|---|---|---|---|---|---|---|
| Acreman 1994 | UK (River Roding) | Varying climate change scenarios, Varying return period scenarios | Planning & Management, Risk Assessment | Fluvial, Coastal | TRUE | TRUE | TRUE | ONDA | Joint Probability Method (JPM) |
| Ai et al. 2018 | China (Jiangsu) | - | Risk Assessment | Pluvial, Coastal | FALSE | TRUE | FALSE | - | Joint Probability Method (JPM), Copula |
| Apel et al. 2016 | Vietnam (Can Tho, Mekong Delta) | - | Risk Assessment | Fluvial, Pluvial | TRUE | TRUE | TRUE | 2D Hydrodynamic Model | Joint Probability Method (JPM), Copula, Peak-over-Threshold (POT) |
| Archetti et al. 2011 | Italy (Rimini) | - | Planning & Management, Risk Assessment | Pluvial, Coastal | TRUE | TRUE | TRUE | 1D Hydrodynamic Drainage Model (InfoWorks CS) | Joint Probability Method (JPM), Copula |
| Bacopoulos et al. 2017 | US (Florida) | Tropical Storm Fay | Risk Assessment | Fluvial, Coastal | TRUE | FALSE | FALSE | ADCIRC, SWAT | - |
| Bakhtyar et al. 2020 | US (Delaware, Delaware Bay Estuary) | - | Forecasting | Fluvial, Coastal | TRUE | FALSE | FALSE | ADCIRC, D-FLOW FM, HEC-RAS, NWM, WW3 | - |
| Banfi and Michele 2022 | Italy (Lake Como) | Lake Flood Events (1980 - 2020) | Earth System Processes | Pluvial | FALSE | TRUE | FALSE | - | Temporal Analysis (Clustering), Peak-over-Threshold (POT) |
| Bao et al. 2022 | US (North Carolina, Cape Fear River Basin) | - | Methodological Advancement | Pluvial, Coastal | TRUE | FALSE | FALSE | COAWST | - |
| Bass and Bedient 2018 | US (Texas) | Tropical Storm Allison (2001), Hurricane Ike (2008) | Forecasting, Risk Assessment | Pluvial, Coastal | TRUE | TRUE | TRUE | ADCIRC, HEC-HMS, HEC-RAS, SWAN | Machine Learning (Artificial Neural Networks (ANN)), Storm Surge Statistical Emulator (Kriging/Gaussian Process Regression (GPR)), Principal Components Analysis, Bayesian Regularization Algorithm |
| Bates et al. 2021 | US (CONUS) | Varying climate change scenarios | Methodological Advancement | Fluvial, Pluvial, Coastal | TRUE | FALSE | FALSE | LISFLOOD-FP | - |
| Beardsley et al. 2013 | US (Massachusetts) | 2010 Nor'easter Storm | Forecasting | Fluvial, Coastal | TRUE | FALSE | FALSE | FVCOM | - |
| Benestad and Haugen 2007 | Norway | - | Earth System Processes | Pluvial, Temp/Heat, Snow | FALSE | TRUE | FALSE | ECHAM4, HIRHAM | Joint Probability Method (JPM), Monte Carlo Simulation |
| Bermúdez et al. 2019 | Spain (Betanzos, Mandeo River) | - | Earth System Processes, Methodological Advancement | Fluvial, Coastal | TRUE | TRUE | TRUE | Iber | Least Square Support Vector Machine (LS-SVM) Regression |
| Bermúdez et al. 2021 | Spain (Betanzos, Mandeo River) | Varying climate change scenarios | Earth System Processes, Methodological Advancement | Fluvial, Pluvial, Coastal, Temp/Heat | TRUE | TRUE | TRUE | Iber, MISDc | Machine Learning (Artificial Neural Networks (ANN)), Least Square Support Vector Machine (LS-SVM) Regression, Bayesian Regularization Algorithm |
| Bevacqua et al. 2017 | Italy (Ravenna) | February 2015 Flood Event | Risk Assessment | Fluvial, Coastal | FALSE | TRUE | FALSE | - | Joint Probability Method (JPM), Copula, Linear Gaussian Autoregressive Model |
| Bevacqua et al. 2019 | Europe | Varying climate change scenarios, Varying return period scenarios | Earth System Processes, Risk Assessment | Pluvial, Coastal | FALSE | TRUE | FALSE | - | Joint Probability Method (JPM), Copula |
| Bevacqua et al. 2020a | Global | Varying climate change scenarios | Earth System Processes | Pluvial, Coastal | TRUE | TRUE | TRUE | Delft3D-Flow | Joint Probability Method (JPM), Copula |
| Bevacqua et al. 2020b | Global | Varying return period scenarios | Risk Assessment | Fluvial, Pluvial | FALSE | TRUE | FALSE | - | Joint Probability Method (JPM), Copula |
| Bevacqua et al. 2022 | Australia (Perth, Swan River Estuary) | - | Risk Assessment | Fluvial, Coastal | FALSE | TRUE | FALSE | - | Multivariate Non-linear Regression, Copula, Temporal Analysis, Kendall's Correlation Coefficient tau (τ), Tail Dependence Coefficient (λ), Block Maxima |
| Bilskie et al. 2021 | US (Louisiana, Barataria and Lake Maurepas Watersheds) | 21 Tropical Cyclone Events (1948–2008) | Methodological Advancement | Pluvial, Coastal | TRUE | FALSE | FALSE | ADCIRC | - |
| Bischiniotis et al. 2018 | Africa (Sub-Saharan Region) | 501 Flood Events (1980 - 2010) | Forecasting, Risk Assessment | Pluvial, Soil Moisture | FALSE | TRUE | FALSE | - | Temporal Analysis, Risk Ratio (RR) |
| Blanton et al. 2012 | US (North Carolina) | Hurricane Irene (2011) | Forecasting, Planning & Management | Pluvial, Coastal | TRUE | FALSE | FALSE | ADCIRC, HL-RDHM | - |
| Blanton et al. 2018 | US (North Carolina) | Hurricane Isabel (2003) | Forecasting, Planning & Management | Pluvial, Coastal | TRUE | FALSE | FALSE | ADCIRC, CREST, WRF | - |
| Bliskie and Hagen, 2018 | US (Louisiana) | Hurricane Gustav (2008) and 2016 Louisiana Flood | Risk Assessment | Pluvial, Coastal | TRUE | FALSE | FALSE | ADCIRC | - |
| Brown et al. 2007 | UK (Canvey Island) | - | Methodological Advancement | Coastal | TRUE | FALSE | FALSE | Delft-FLS, SWAN | - |
| Bunya et al. 2010 | US (Louisiana and Mississippi) | - | Methodological Advancement | Fluvial, Pluvial, Coastal | TRUE | FALSE | FALSE | ADCIRC, ECWAM, H*WIND, IOKA, STWAVE, | - |
| Bush et al. 2022 | US (North Carolina) | - | Methodological Advancement | Fluvial, Pluvial, Coastal | TRUE | FALSE | FALSE | ADCIRC, HEC-RAS | - |
| Camus et al. 2021 | Europe | - | Earth System Processes | Fluvial, Pluvial, Coastal | FALSE | TRUE | FALSE | - | Joint Probability Method (JPM), Spatial Analysis, Correlation Coefficients (Kendall's tau (τ), Spearman's rho (ρ)), Block Maxima, Peak-over-Threshold (POT) |
| Camus et al. 2022 | Global (US and Europe, North Atlantic) | Flood Events (1980-2014) | Earth System Processes | Fluvial, Coastal | TRUE | TRUE | TRUE | CaMa-Flood, GTSM | Joint Occurrence Method, Spatial Anaylsis (Clustering K-Means Algorithm (KMA)), Principal Component Analysis (PCA), Temporal Analysis, Kendall's Correlation Coefficient tau (τ), Peak-over-Threshold (POT) |
| Cannon et al. 2008 | US (Colorado and California) | - | Earth System Processes | Pluvial, Fire | FALSE | TRUE | FALSE | - | Spatial Analysis, Temporal Analysis |
| Čepienė et al. 2022 | Lithuania (Klaipėda) | - | Risk Assessment | Fluvial, Coastal | TRUE | FALSE | FALSE | HEC-RAS | - |
| Chen and Liu 2014 | Taiwan (Tainan City, Tsengwen River basin) | Typhoon Krosa (2007), Kalmegei (2008), Morakot (2009), and Haiyan (2013) | Methodological Advancement | Fluvial, Coastal | TRUE | FALSE | FALSE | SELFE | - |
| Chen and Liu, 2016 | Taiwan (Kaohsiung City, Gaoping River) | Typhoon Kalmegei (2008), Morakot (2009), Fanapi (2010), Nanmadol (2011), and Talim (2012), Varying return period scenarios | Risk Assessment | Fluvial, Coastal | TRUE | FALSE | FALSE | SELFE | - |
| Chen et al. 2010 | UK (Bradford, Keighley, River Aire) | Varying return period scenarios | Planning & Management, Risk Assessment | Fluvial, Pluvial | TRUE | FALSE | FALSE | SIPSON, UIM | - |
| Chen et al. 2013 | Taiwan (Tainan City) | Typhone Haitang (2005) and Kalmaegi (2008), Varying return period scenarios | Risk Assessment | Pluvial, Coastal | TRUE | FALSE | FALSE | SELFE | - |
| Chou 1989 | Saipan (West Coast) | 168 Synthetic Typhoon Events, Varying return period scenarios | Risk Assessment | Coastal | TRUE | TRUE | TRUE | SHAWLWV, WIFM | Joint Probability Method (JPM), Frequency Analysis |
| Christian et al. 2015 | US (Texas, Galveston Bay) | Hurricane Ike (2008) | Planning & Management | Pluvial, Coastal | TRUE | FALSE | FALSE | ADCIRC, HEC-RAS, Vflo | - |
| Cifelli et al. 2021 | US (California, San Francisco) | - | Forecasting | Fluvial, Coastal | TRUE | FALSE | FALSE | Hydro-CoSMoS | - |
| Coles and Tawn 1994 | UK (Cornwall) | - | Methodological Advancement, Risk Assessment | Coastal | FALSE | TRUE | FALSE | - | Joint Probability Method (JPM), Chi Squared Test (χ2) |
| Coles et al. 1999 | UK (Southwest Coast) | - | Methodological Advancement, Risk Assessment | Pluvial, Coastal | FALSE | TRUE | FALSE | - | Joint Probability Method (JPM), Copula, Chi Squared Test (χ2) |
| Comer et al. 2017 | Ireland (Cork City) | 2009 Flood Event | Methodological Advancement | Fluvial, Coastal | TRUE | FALSE | FALSE | MSN_Flood, POM | - |
| Couasnon et al. 2018 | US (Texas) | - | Methodological Advancement, Risk Assessment | Fluvial, Coastal | TRUE | TRUE | TRUE | 1D Hydrodynamic Model | Bayesian Network (BN), Copula |
| Couasnon et al. 2020 | Global | - | Earth System Processes | Fluvial, Coastal | FALSE | TRUE | FALSE | - | Joint Probability Method (JPM), Spatial Analysis, Temporal Analysis, Spearman's Correlation Coefficient rho (ρ) |
| Curtis et al. (2022) | US (North Carolina) | - | Risk Assessment | Fluvial, Coastal | FALSE | FALSE | FALSE | - | - |
| Daoued et al. 2021 | France (Le Havre) | Varying return period scenarios | Risk Assessment | Pluvial, Coastal | FALSE | TRUE | FALSE | - | Joint Probability Method (JPM), Probabilistic Flood Hazard Assessment (PFHA), Belief Functions, Peak-over-Threshold (POT) |
| De Bruijn et al. 2014 | Netherlands (Rhine-Meuse Delta) | - | Risk Assessment | Fluvial, Coastal | FALSE | TRUE | FALSE | - | Copula, FN-Curve, Potential Loss of Life (PLL), Monte Carlo Simulation |
| De Michele et al. 2020 | Global (Europe and North Africa) | - | Earth System Processes | Pluvial | FALSE | TRUE | FALSE | - | Copula, Binary Markov Chain Network, Monte Carlo Simulation |
| Deidda et al. 2021 | UK | - | Earth System Processes | Fluvial, Pluvial | FALSE | TRUE | FALSE | - | Joint Occurrence Method, Spatial Analysis, Kendall's Correlation Coefficient tau (τ), Block Maxima |
| Del-Rosal-Salido et al. 2021 | Europe (Iberian Peninsula, Guadalete Estuary) | Varying climate change scenarios, Varying return period scenarios | Forecasting, Planning & Management | Fluvial, Coastal | TRUE | TRUE | TRUE | Delft3D | Spatial Analysis (Vector Autoregressive (VAR) Model), Block Maxima, Peak-over-Threshold (POT), |
| Dietrich et al. 2010 | US (Louisiana and Mississippi) | Hurricane Katrina (2005) and Rita (2005) | Methodological Advancement | Fluvial, Coastal | TRUE | FALSE | FALSE | ADCIRC, IOKA, H*WIND, STWAVE, WAM | - |
| Dixon and Tawn 1994 | UK | - | Earth System Processes | Coastal | FALSE | TRUE | FALSE | - | Joint Probability Method (JPM), Extreme Value Analysis, Chi Squared Test (χ2) |
| Dresback et al. 2013 | US (North Carolina) | Hurricane Irene (2011) | Forecasting | Fluvial, Coastal | TRUE | FALSE | FALSE | ASGS-STORM, ADCIRC, Holland Wind Model, HL-RDHM, SWAN | - |
| Dykstra et al. 2021 | US (Gulf Coast; Ascagoula, Tombigbee-Alabama River, and Apalachicola watersheds) | - | Earth System Processes | Fluvial, Coastal | FALSE | TRUE | FALSE | - | Joint Probability Method (JPM), Kendall's Correlation Coefficient tau (τ), Frequency Analysis, Temporal Analysis (Pettitt Test), Wavelet Transformations (Mortlet-type Wave), Peak-over-Threshold (POT), Bootstrap Method |
| Eilander 2022 | Global | - | Earth System Processes, Risk Assessment | Fluvial, Coastal | TRUE | FALSE | FALSE | HydroMT | - |
| Eilander et al. 2020 | Global | - | Earth System Processes, Risk Assessment | Fluvial, Coastal | TRUE | FALSE | FALSE | CaMa-Flood, FES2012, GTSM | - |
| Eilander et al. 2022 | Mozambique (Sofala) | Varying return period scenarios | Impact Assessment, Risk Assessment | Fluvial, Pluvial, Coastal | TRUE | TRUE | TRUE | CaMa-Flood, Delft-FIAT, SFINCS | Copula, Block Maxima |
| Erikson et al. 2018 | US (California, San Francisco) | Varying climate change scenarios, Varying return period scenarios | Impact Assessment, Risk Assessment | Fluvial, Coastal | TRUE | FALSE | FALSE | CoSMoS | - |
| Familkhalili et al. 2022 | US (North Carolina, Cape Fear Estuary) | Hurricane Irene (2011) | Earth System Processes | Fluvial, Coastal | TRUE | FALSE | FALSE | 1D Hydrodynamic Model | - |
| Fang et al. 2021 | China | Varying climate change scenarios, Varying return period scenarios | Earth System Processes | Pluvial, Coastal | FALSE | TRUE | FALSE | - | Kendall's Correlation Coefficient tau (τ), Temporal Analysis, Peak-over-Threshold (POT) |
| Feng and Brubaker, 2016 | US (Washington DC) | Varying climate change scenarios, Varying return period scenarios | Impact Assessment, Risk Assessment | Pluvial, Coastal | TRUE | FALSE | FALSE | HEC-RAS | - |
| Ferrarin et al. 2022 | Italy (Venice, Adriatic Sea) | November 2019 Flood Event | Earth System Processes, Risk Assessment | Coastal | FALSE | TRUE | FALSE | - | Joint Probability Method (JPM), Copula, Kendall's Correlation Coefficient tau (τ), Temporal Analysis, Mann-Whitney U Test |
| Flick 1991 | US (California, San Francisco) | - | Risk Assessment | Coastal | FALSE | TRUE | FALSE | - | Joint Probability Method (JPM) |
| Galiatsatou and Prinos 2016 | Greece (Aegean Sea) | - | Earth System Processes | Coastal | TRUE | TRUE | TRUE | RegCM3, SWAN | Joint Probability Method (JPM), Copula, Block Maxima |
| Ganguli and Merz 2019a | Europe (Northwest) | - | Earth System Processes | Fluvial, Coastal | FALSE | TRUE | FALSE | - | Spatial Analysis, Compound Hazard Ratio (CHR) Index, Kendall's Correlation Coefficient tau (τ) |
| Ganguli and Merz 2019b | Europe (Northwest) | Flood Events (1970-2014) | Earth System Processes | Fluvial, Coastal | FALSE | TRUE | FALSE | - | Spatial Analysis, Frequency Analysis, Compound Hazard Ratio (CHR) Index, Kendall's Correlation Coefficient tau (τ) |
| Ganguli et al. 2020 | Europe (Northwest) | Varying climate change scenarios | Earth System Processes | Fluvial, Coastal | TRUE | TRUE | TRUE | Delft3D-FLOW, WGHM | Copula, Markov Chain, Monte Carlo Simulation |
| Georgas et al. 2016 | US (New York and New Jersey) | Winter Storm Jonas (2016) | Forecasting | Fluvial, Coastal | TRUE | FALSE | FALSE | ESTOFS, ETSS, sECOM, SFAS, NAM, NYHOPS | - |
| Ghanbari et al. 2021 | US (CONUS) | Varying return period scenarios, Varying climate change scenarios | Earth System Processes, Risk Assessment | Fluvial, Coastal | FALSE | TRUE | FALSE | - | Joint Probability Method (JPM), Copula, Quantile Regression, Kendall's Correlation Coefficient tau (τ), Peak-over-Threshold (POT) |
| Gori and Lin 2022 | US (North Carolina, Cape Fear River) | Varying climate change scenarios | Risk Assessment | Fluvial, Pluvial, Coastal | TRUE | TRUE | TRUE | ADCIRC, HEC-HMS, HEC-RAS | Joint Probability Method Optimal Sampling Bayesian Quadrature Optimization (JPM-OS-BQ) |
| Gori et al. 2020a | US (North Carolina, Cape Fear River) | Varying return period scenarios | Earth System Processes | Pluvial, Coastal | TRUE | FALSE | FALSE | ADCIRC, HEC-HMS, HEC-RAS | - |
| Gori et al. 2020b | US (North Carolina, Cape Fear River) | Tropical Cyclone Fran (1996), Floyd (1999), and Matthew (2016), Varying return period scenarios | Earth System Processes | Fluvial, Pluvial, Coastal | TRUE | TRUE | TRUE | ADCIRC, HEC-HMS, HEC-RAS | Joint Probability Method (JPM), Copula |
| Gori et al. 2022 | US (East Coast and Gulf of Mexico) | Varying climate change scenarios, Varying return period scenarios | Risk Assessment | Pluvial, Coastal | TRUE | TRUE | TRUE | ADCIRC | Joint Probability Method (JPM), Kendall's Correlation Coefficient tau (τ), Statistical-Deterministic TC Model, Spatial Analysis, Temporal Analysis, Bootstrap Method |
| Gouldby et al. 2017 | UK (South Coast) | Varying return period scenarios | Methodological Advancement | Coastal | TRUE | TRUE | TRUE | SWAN, WW3 | Joint Probability Method (JPM), Wave Transformation Model Emulator, Monte Carlo Simulation |
| Gutenson et al. 2022 | US (Texas, Galveston Bay) | Hurricane Harvey (2017) | Impact Assessment, Methodological Advancement, Risk Assessment | Fluvial, Pluvial, Coastal | TRUE | TRUE | TRUE | AutoRoute, HEC-RAS, LISFLOOD-FP | Spatial Analysis |
| Habel et al. 2020 | US (Hawaii, Honolulu) | Varying climate change scenarios, Varying return period scenarios | Impact Assessment, Planning & Management | Coastal, Groundwater | TRUE | TRUE | TRUE | MODFLOW | Frequency Analysis, Bayesian Hierarchical Model, Spatial Analysis |
| Haigh et al. 2016 | UK | 2013-2014 Winter Storm Season | Earth System Processes | Coastal | FALSE | TRUE | FALSE | - | Joint Probability Method (JPM), Spatial Analysis, Temporal Analysis, Peak-over-Threshold (POT) |
| Harrison et al. 2022 | UK (Humber and Dyfi Estuaries) | - | Earth System Processes | Fluvial, Pluvial, Coastal | TRUE | FALSE | FALSE | 2D Hydrodynamic Model | - |
| Hawkes 2003 | UK | - | Earth System Processes, Methodological Advancement | Fluvial, Coastal | FALSE | TRUE | FALSE | - | Joint Probability Method (JPM), JOIN-SEA Model |
| Hawkes 2006 | UK | - | Planning & Management, Risk Assessment | Fluvial, Pluvial, Coastal | FALSE | TRUE | FALSE | - | Joint Probability Method (JPM), JOIN-SEA Model, Chi Squared Test (χ2) |
| Hawkes 2008 | UK (South Coast) | - | Methodological Advancement, Risk Assessment | Coastal | FALSE | TRUE | FALSE | - | Joint Probability Method (JPM), JOIN-SEA Model, Temporal Analysis, Monte Carlo Simulation |
| Hawkes and Svensson 2003 | UK | - | Earth System Processes, Risk Assessment | Fluvial, Pluvial, Coastal | FALSE | TRUE | FALSE | - | Joint Probability Method (JPM), JOIN-SEA Model, Monte Carlo Simulation |
| Hawkes et al. 2002 | UK (England and Wales) | Varying return period scenarios | Risk Assessment | Coastal | FALSE | TRUE | FALSE | - | Joint Probability Method (JPM), Monte Carlo Simulation |
| Helaire et al. 2020 | US (Washington, Portland-Vancouver, Columbia River Estuary) | Varying climate change scenarios | Earth System Processes | Fluvial, Coastal | TRUE | FALSE | FALSE | Delft3D | - |
| Hendry et al. 2019 | UK | - | Earth System Processes | Fluvial, Coastal | FALSE | TRUE | FALSE | - | Joint Occurrence Method, Kendall's Correlation Coefficient tau (τ), Temporal Analysis, Block Maxima, Peak-over-Threshold (POT) |
| Herdman et al. 2018 | US (California, San Francisco) | - | Forecasting | Fluvial, Coastal | TRUE | FALSE | FALSE | Delft3D-FM | - |
| Ho and Myers 1975 | US (Florida, St. George Sound, Apalachicola Bay) | Varying return period scenarios | Methodological Advancement, Risk Assessment | Coastal | TRUE | TRUE | TRUE | SPLASH, 2D Hydrodynamic Bay-Ocean Model (Overland 1975) | Joint Probability Method (JPM), Frequency Analysis |
| Hsiao et al. 2021 | Taiwan | Typhoon Megi (2016), Low-Pressure Rainstorm (2018), Varying climate change scenarios | Risk Assessment | Fluvial, Pluvial, Coastal | TRUE | TRUE | TRUE | SCHISM, COS-Flow, 39 General Circulation Models (GCM) | Index Method (2 Hazard Indices, 4 Exposure Indices, 6 Vulnerability Indices) |
| Huang 2022 | Taiwan (Touqian and Fengshan Rivers) | Hurricane Harvey (2017) | Forecasting | Pluvial, Coastal | TRUE | TRUE | TRUE | ADCIRC | Machine Learning (Recurrent Neural Network (RNN)), Topographic Wetness Index (TWI) |
| Huang et al. 2021 | US (Texas, Galveston Bay) | - | Earth System Processes | Fluvial, Pluvial, Coastal | TRUE | TRUE | TRUE | SCHISM | Compound Ratio (CR), Spatial Analysis |
| Ikeuchi et al. 2017 | Bangladesh (Ganges-Brahmaputra-Meghna Delta) | Cyclone Sidr (2007) | Methodological Advancement | Fluvial, Coastal | TRUE | FALSE | FALSE | CaMa-Flood, MATSIRO-GW | - |
| Jalili Pirani and Reza Najafi 2020 | Canada | - | Earth System Processes | Fluvial, Pluvial, Coastal | FALSE | TRUE | FALSE | - | Spatial Analysis, Temporal Analysis (Mann-Kendall Test), Probability Space (PS) Index, Correlation Coefficients (Kendall's tau (τ), Spearman's rho (ρ)) |
| Jalili Pirani and Reza Najafi 2022 | Canada (East and West Coast, Great Lakes) | Varying return period scenarios | Earth System Processes | Fluvial, Pluvial, Coastal | FALSE | TRUE | FALSE | - | Joint Probability Method (JPM), Compound Hazard Ratio (CHR) Index, Copula, Kendall's Correlation tau (τ) |
| Jane et al. 2020 | US (Florida) | - | Earth System Processes | Pluvial, Coastal, Groundwater | FALSE | TRUE | FALSE | - | Joint Probability Method (JPM), Copula, Kendall's Correlation Coefficient tau (τ) |
| Jane et al. 2022 | US (Texas, Sabine and Brazos River Basins) | Varying return period scenarios | Planning & Management, Risk Assessment | Fluvial, Coastal | FALSE | TRUE | FALSE | - | Joint Probability Method (JPM), Copula, Peak-over-Threshold (POT) |
| Jang and Chang 2022 | Taiwan (Chiayi) | Varying return period scenarios | Risk Assessment | Pluvial, Coastal | TRUE | TRUE | TRUE | COS-Flow | Joint Probability Method (JPM), Copula, Monte Carlo Simulation |
| Jasim et al. 2020 | US (California, Sherman Island) | Varying return period scenarios | Planning & Management, Risk Assessment | Fluvial, Pluvial | TRUE | TRUE | TRUE | RS3 | Joint Probability Method (JPM), Frequency Analysis, Copula |
| Jones 1998 | UK (Thames Estuary) | - | Earth System Processes, Methodological Advancement | Fluvial, Coastal | FALSE | TRUE | FALSE | - | Joint Probability Method (JPM), Temporal Analysis, Historical Emulation Model |
| Jong-Levinger et al. 2022 | US (California) | Varying climate change scenarios, Varying return period scenarios | Earth System Processes | Pluvial, Fire | FALSE | TRUE | FALSE | - | Markov Chain Monte Carlo (MCMC) Algorithm |
| Joyce et al. 2018 | US (Florida) | Varying climate change scenarios | Methodological Advancement, Risk Assessment | Pluvial, Coastal | TRUE | FALSE | FALSE | ADCIRC, SWAN, ICPR | - |
| Juárez et al. 2022 | US (Florida, Jacksonville, Lower St. Johns River) | Hurricane Irma (2017), Varying climate change scenarios | Earth System Processes, Methodological Advancement | Fluvial, Coastal | FALSE | TRUE | FALSE | - | Flow Interaction Index (μ), Temporal Analysis |
| Karamouz et al. 2014 | US (New York, New York City) | Varying return period scenarios, Varying climate change scenarios | Planning & Management | Pluvial, Coastal | TRUE | TRUE | TRUE | HEC-RAS, GSSHA, SWMM | Machine Learning (Multilayer Perceptron (MLP) Feedforward Neural Network (FNN)), Markov Chain Monte Carlo (MCMC) Algorithm, DREAM_ZS, Max Relevance Min Redundancy (MRMR) Algorithm |
| Karamouz et al. 2017 | US (New York, New York City) | Hurricane Irenne (2011) and Sandy (2012), Varying future climate change flood scenarios, Varying return period scenarios | Methodological Advancement | Pluvial, Coastal | TRUE | TRUE | TRUE | GSSHA | Joint Probability Method (JPM), Frequency Analysis, Copula |
| Karamouz et al. 2017 | US (New York, New York City) | Varying return period scenarios | Impact Assessment, Risk Assessment | Pluvial, Coastal | TRUE | TRUE | TRUE | GSSHA | Joint Probability Method (JPM), Frequency Analysis, Flood Damage Estimator (FDE) Model, Copula, Correlation Coefficients (Kendall's tau (τ), Pearson's (r), Spearman's rho (ρ)) |
| Kerr et al. 2013 | US (Louisiana and New Orleans, Mississippi River) | Hurricane Betsy (1965), Camille (1969), Andrew (1992), Katrina (2005), Rita (2005), Gustav (2008), Ike (2008), 15 Synthetic Storm Events | Earth System Processes | Fluvial, Coastal | TRUE | TRUE | TRUE | ADCIRC, H*WIND, SWAN | Joint Probability Method (JPM) with Optimal Sampling (JPM-OS), Frequency Analysis |
| Kew et al. 2013 | Netherlands (Rhine Delta) | Varying return period scenarios, Varying climate change scenarios | Earth System Processes | Fluvial, Coastal | TRUE | TRUE | TRUE | ECHAM5, MPI-OM | Joint Probability Method (JPM), Extreme Value Analysis, Peak-over-Threshold (POT) |
| Khalil et al. 2022 | Australia (Brisbane, Brisbane River and Moreton Bay) | Flood Events (2006, 2011, 2013) | Earth System Processes, Methodological Advancement | Fluvial, Coastal | TRUE | FALSE | FALSE | MIKE21 | - |
| Khanal et al. 2019 | Europe (Rhine River Basin) | - | Earth System Processes | Fluvial, Coastal | TRUE | TRUE | TRUE | DCSM, HBV, RACMO2, SPHY, WAQUA | Joint Probability Method (JPM), Temporal Analysis |
| Khanam et al. 2021 | US (Connecticut) | Varying climate change scenarios | Impact Assessment, Risk Assessment | Pluvial, Coastal | TRUE | FALSE | FALSE | CREST-SVAS, HEC-RAS, WRF | - |
| Khatun et al. 2022 | India (Upper Mahanadi River basin) | Varying return period scenarios, Varying climate change scenarios | Earth System Processes | Fluvial, Pluvial | TRUE | TRUE | TRUE | MIKE11, NAM | Bivariate Hazard Ratio (BHR) Index, Copula, Kendall's Correlation Coefficient tau (τ), Peak-over-Threshold (POT) |
| Kim et al. 2022 | US (Texas, Houston, Dickinson Bayou Watershed) | Hurricane Harvey (2017) | Earth System Processes | Pluvial, Coastal | FALSE | TRUE | FALSE | - | Copula, Kendall's Correlation Coefficient tau (τ), Peak-over-Threshold (POT) |
| Kirkpatrick and Olbert 2020 | Ireland (Cork City) | Varying climate change scenarios, Varying return period scenarios | Earth System Processes, Risk Assessment | Fluvial, Coastal | TRUE | FALSE | FALSE | - | - |
| Klerk et al. 2015 | Netherlands (Hoek van Holland and Lobith, Rhine-Meuse Delta) | Varying climate change scenarios, Varying return period scenarios | Earth System Processes | Fluvial, Coastal | FALSE | TRUE | FALSE | CKF, Delft3D-FLOW, DCSM, HBV-96 | Temporal Analysis, Chi Squared Test (χ2), Peak-over-Threshold (POT) |
| Kowalik and Proshutinsky 2010 | US (Alaska, Cook Inlet) | - | Earth System Processes | Coastal, Tsunami | TRUE | FALSE | FALSE | 1D/2D Hydrodynamic Models | - |
| Kudryavtseva et al. 2020 | Europe (Baltic Sea) | - | Risk Assessment | Coastal | TRUE | TRUE | TRUE | NEMO, WAM | Joint Probability Method (JPM), Copula |
| Kumbier et al. 2018 | Australia (New South Wales, Nowra, Shoalhaven River) | 2016 Cyclone | Earth System Processes, Risk Assessment | Fluvial, Coastal | TRUE | FALSE | FALSE | Delft3D-FLOW | - |
| Kupfer et al. 2022 | South Africa (Breede Estuary) | Varying return period scenarios | Earth System Processes, Risk Assessment | Fluvial, Coastal | TRUE | FALSE | FALSE | Delft3D-FLOW, Delft3D-WAVE | - |
| Lai et al. 2021a | Global | - | Earth System Processes, Risk Assessment | Pluvial, Coastal | FALSE | TRUE | FALSE | - | Joint Probability Method (JPM), Copula, Kendall's Correlation Coefficient tau (τ), Peak-over-Threshold (POT) |
| Lai et al. 2021b | Global | Varying climate change scenarios, Varying return period scenarios, Flood Events (1948–2014, 1979–2014) | Earth System Processes | Pluvial, Coastal | FALSE | TRUE | FALSE | - | Frequency Analysis, Spatial Analysis, Temporal Analysis (Mann-Kendall Test), Multivariate Regression, Peak-over-Threshold (POT) |
| Láng-Ritter et al. 2022 | Spain | - | Forecasting, Impact Assessment, Risk Assessment | Fluvial, Pluvial | TRUE | FALSE | FALSE | EFAS, ReAFFIRM | - |
| Latif and Simonovic 2022a | Canada (West Coast) | - | Methodological Advancement, Risk Assessment | Fluvial, Pluvial, Coastal | FALSE | TRUE | FALSE | - | Joint Probability Method (JPM), Copula |
| Latif and Simonovic 2022b | Canada (West Coast) | - | Methodological Advancement, Risk Assessment | Fluvial, Pluvial, Coastal | FALSE | TRUE | FALSE | - | Joint Probability Method (JPM), Copula |
| Lawrence et al. 2014 | Norway | Varying return period scenarios | Risk Assessment | Pluvial, Snow | TRUE | TRUE | TRUE | HBV, PQRUT | Stochastic Probability (SCHADEX Probabilistic Method, GRADEX Probabilistic Method) |
| Lee et al. 2019 | South Korea | Typhone Maemi (2003) | Methodological Advancement, Risk Assessment | Fluvial, Coastal | TRUE | FALSE | FALSE | Delft3D, HEC-HMS | - |
| Lee et al. 2020 | South Korea (Busan, Marine City) | - | Methodological Advancement | Pluvial, Coastal | TRUE | FALSE | FALSE | ADCIRC, FLOW-3D, SWAN, XPSWMM | - |
| Leijnse et al. 2021 | US (Florida, Jacksonville) and Phillippines | Hurricane Irma (2017) and Typhoon Haiyan (2013) | Methodological Advancement | Fluvial, Pluvial, Coastal | TRUE | FALSE | FALSE | ADCIRC | - |
| Li and Jun 2020 | South Korea (Han River) | - | Earth System Processes, Risk Assessment | Fluvial, Coastal | TRUE | FALSE | FALSE | 1D Hydrodynamic Model | - |
| Li et al. 2022 | Hong Kong (Hong Kong-Zhuhai-Macao Bridge) | - | Risk Assessment | Pluvial, Coastal | TRUE | TRUE | TRUE | MIKE+ | Joint Probability Method (JPM), Temporal Analysis, Damage Curves |
| Lian et al. 2013 | China (Fuzhou City) | Typhoon Longwang (2005), Varying return period scenarios | Risk Assessment | Pluvial, Coastal | TRUE | TRUE | TRUE | HEC-RAS, SWAT | Joint Probability Method (JPM), Copula, Peak-over-Threshold (POT) |
| Lian et al. 2017 | China (Hainan Province, Haikou) | - | Planning & Management, Risk Assessment | Pluvial, Coastal | TRUE | TRUE | TRUE | HEC-RAS, SWMM | Disaster Reduction Analysis, Cost-Benefit Analysis (CBA) |
| Liang and Zhou 2022 | China (Zhejiang, Qiantang River) | Typhoon Lekima (2019) | Methodological Advancement, Risk Assessment | Fluvial, Coastal | TRUE | FALSE | FALSE | CaMa-Flood, MIKE21 | - |
| Lin et al. 2010 | US (East Coast, Chesapeake Bay) | - | Earth System Processes | Pluvial, Coastal | TRUE | FALSE | FALSE | ADCIRC, WRF | - |
| Liu et al. 2022 | China (Haikou City) | - | Risk Assessment | Pluvial, Coastal | TRUE | FALSE | FALSE | Delft3D | - |
| Loganathan et al. 1987 | US (Virginia, Rappahannock River) | - | Earth System Processes, Risk Assessment | Fluvial, Coastal | FALSE | TRUE | FALSE | - | Joint Probability Method (JPM), Box-Cox Transformation, Chi Squared Test (χ2) |
| Loveland et al. 2021 | US (Texas, Lower Neches River) | Hurricane Harvey (2017) | Methodological Advancement | Fluvial, Coastal | TRUE | FALSE | FALSE | ADCIRC, HEC-RAS | - |
| Lu et al. 2022 | China (Southeast) | - | Risk Assessment | Fluvial, Coastal | FALSE | TRUE | FALSE | - | Joint Probability Method (JPM), Copula, Multivariate Copula Analysis Toolbox (MvCAT), Kendall's Correlation Coefficient tau (τ) |
| Lucey et al. 2022 | US (California, Los Angeles, Huntington Beach, San Diego) | Varying return period scenarios | Risk Assessment | Pluvial, Coastal | FALSE | TRUE | FALSE | - | Copula, Correlation Coefficients (Kendall's tau (τ), Pearson's (r), Spearman's rho (ρ)) |
| Lyddon et al. 2022 | UK | - | Earth System Processes, Methodological Advancement | Coastal | FALSE | TRUE | FALSE | - | Frequency Analysis, Temporal Analysis, Spatial Analysis, Kendall's Correlation Coefficient tau (τ), Annual Mean Compound Event Measure, Block Maxima, Peak-over-Threshold (POT) |
| Manoj et al. 2022 | India | - | Earth System Processes | Pluvial, Soil Moisture | FALSE | TRUE | FALSE | - | Event Coincidence Analysis (ECA), Chi Squared Test (χ2), Spatial Analysis, Temporal Analysis |
| Mantz and Wakeling 1979 | UK (Norfolk, Yare Basin) | Varying return period scenarios | Planning & Management, Risk Assessment | Pluvial, Coastal | FALSE | TRUE | FALSE | - | Joint Probability Method (JPM), Extreme Value Analysis |
| Martyr et al. 2013 | US (Louisiana) | Hurricane Gustave (2008) | Methodological Advancement | Fluvial, Coastal | TRUE | FALSE | FALSE | ADCIRC | - |
| Mashriqui et al. 2010 | US (Washington DC) | 1996 Flood, Hurricane Isabel (2003) | Forecasting, Methodological Advancement, Planning & Management | Fluvial, Coastal | TRUE | FALSE | FALSE | HEC-RAS | - |
| Mashriqui et al. 2014 | US (Washington DC) | Hurricane Isabel (2003) | Forecasting, Methodological Advancement, Planning & Management | Fluvial, Coastal | TRUE | FALSE | FALSE | HEC-RAS | - |
| Masina et al. 2015 | Italy (Ravenna) | - | Risk Assessment | Coastal | FALSE | TRUE | FALSE | - | Joint Probability Method (JPM), Copula, Correlation Coefficients (Kendall's tau (τ), Pearson's (r), Spearman's rho (ρ)) |
| Maskell et al. 2014 | UK (England) | Varying return period scenarios | Earth System Processes | Fluvial, Coastal | TRUE | FALSE | FALSE | FVCOM, LISFLOOD-FP | - |
| Maymandi et al. 2022 | US (Texas, Sabine-Neches Estuary) | Hurricane Rita (2005), Ike (2008), and Harvey (2017) | Earth System Processes | Fluvial, Pluvial, Coastal | TRUE | FALSE | FALSE | ADCIRC, Delft3D | - |
| Mazas et al. 2014 | France (Brest) | Varying return period scenarios | Methodological Advancement, Risk Assessment | Coastal | FALSE | TRUE | FALSE | - | Revised Joint Probability Method (RJPM), Chi Squared Test (χ2), Peak-over-Threshold (POT) |
| McInnes et al. 2002 | Australia (Queensland, Gold Coast Broadwater) | Tropical Cyclones (1989 and 1974) | Earth System Processes, Methodological Advancement | Fluvial, Coastal | TRUE | FALSE | FALSE | GCOM2D, RAMS, WAM | - |
| Meyers et al. 2021 | US (Florida) | Hurricane Hermine (2017), 79 Sanitary Sewer Overflow Events (1996 - 2017), Varying climate change scenarios | Risk Assessment | Pluvial, Coastal | FALSE | TRUE | FALSE | - | Logistic Regression Model (LRM), Temporal Analysis |
| Ming et al. 2022 | UK (London, Thames Estuary) | Varying return period scenarios, 27 Flood Scenarios | Risk Assessment | Fluvial, Pluvial, Coastal | TRUE | TRUE | TRUE | HiPIMS | Joint Probability Method (JPM), Copula, Correlation Coefficients (Kendall's tau (τ), Spearman's rho (ρ)), Peak-over-Threshold (POT), |
| Modrakowski et al. 2022 | Netherlands (Odense, Hvidovre, Vejle) | - | Planning & Management, Risk Assessment | Fluvial, Pluvial, Coastal, Soil Moisture | FALSE | FALSE | FALSE | - | - |
| Moftakhari et al. 2017 | US (Philadelphia, Pennsylvania; San Francisco, California; and Washington DC) | Varying climate change scenarios | Earth System Processes, Risk Assessment | Fluvial, Coastal | FALSE | TRUE | FALSE | - | Copula, Kendall's Correlation Coefficient tau (τ), Block Maxima |
| Moftakhari et al. 2019 | US (California, Newport Bay) | - | Methodological Advancement, Risk Assessment | Fluvial, Coastal | TRUE | TRUE | TRUE | BreZo | Joint Probability Method (JPM), Copula, Correlation Coefficients (Kendall's tau (τ), Spearman's rho (ρ)) |
| Mohammadi et al. 2021 | US (Idaho, Clearwater River; Montana, Yellowstone River; New Jersey, Delaware River) | - | Earth System Processes, Risk Assessment | Fluvial, Pluvial, Coastal, Snow | FALSE | TRUE | FALSE | - | Copula, Bayesian Network (BN), Storm Surge Statistical Emulator (Kriging/Gaussian Process Regression (GPR) |
| Mohor et al.  2020 | Germany | Flood Events (2002-2013) | Impact Assessment | Fluvial, Pluvial, Groundwater, Damming/Dam Failure | FALSE | TRUE | FALSE | - | Multivariate Ordinary Least Squares (OLS) Regression, Building Loss Ratio, Chi Squared Test (χ2), Univariate Normality and Variance (Levene's Test, Box's M Test, Kruskal-Wallis Test, Dunn's Test), Bootstrap Method |
| Muñoz et al. 2020 | US (Georgia, Savannah, Savannah River Delta) | Hurricane Matthew (2016), Varying return period scenarios | Earth System Processes | Fluvial, Coastal | TRUE | TRUE | TRUE | Delft3D-FM | Spatial Analysis, Copula, Multi-hazard Scenario Analysis Toolbox (MhAST), Correlation Coefficients (Kendall's tau (τ), Spearman's rho (ρ)) |
| Muñoz et al. 2021 | US (Southeast Coast; Savannah River Estuary, Florida, Georgia, South Carolina, and North Carolina) | Hurricane Matthew (2016) | Methodological Advancement | Fluvial, Coastal | TRUE | TRUE | TRUE | Delft3D-FM | Machine Learning (Convolutional Neural Network (CNN)), Data Fusion (DF) |
| Muñoz et al. 2022 | US (Alabama, Mobile Bay) | Varying climate change scenarios | Earth System Processes, Planning & Management, Risk Assessment | Fluvial, Pluvial, Coastal | TRUE | TRUE | TRUE | Delft3D-FM | Joint Probability Method (JPM), Copula, Multi-hazard Scenario Analysis Toolbox (MhAST), Peak-over-Threshold (POT) |
| Muñoz et al. 2022 | US (Texas, Galveston Bay; Delaware, Delaware Bay) | Hurricane Harvey (2017), Hurricane Sandy (2012) | Methodological Advancement | Fluvial, Pluvial, Coastal | TRUE | TRUE | TRUE | Delft3D-FM | Baysesian Data Assimilation (DA), Ensemble Kalman Filter (EnKF) |
| Myers 1970 | US (New Jersey, Atlantic City, Long Beach Island) | - | Methodological Advancement, Risk Assessment | Coastal | FALSE | TRUE | FALSE | - | Joint Probability Method (JPM), Frequency Analysis |
| Najafi et al. 2021 | Saint Lucia | Hurricane Matthew (2016) | Risk Assessment | Fluvial, Pluvial, Coastal | TRUE | TRUE | TRUE | HyMOD, LISFLOOD-FP | Stongest Path Method (SPM) Network Risk Analysis, Risklogik Platform, Monte Carlo Simulation |
| Naseri and Hummel 2022 | US (CONUS) | Varying return period scenarios | Risk Assessment | Pluvial, Coastal | FALSE | TRUE | FALSE | - | Copula, Kendall's Correlation Coefficient tau (τ), Spatial Analysis, Temporal Analysis (Mann-Kendall Test), Markov Chain Monte Carlo (MCMC) Algorithm |
| Nash et al. 2018 | Ireland (Cork City) | November 2009 Flood | Earth System Processes | Fluvial, Coastal | TRUE | FALSE | FALSE | MSN_Flood, POM | - |
| Nasr et al. 2021 | US (CONUS) | - | Earth System Processes, Methodological Advancement | Fluvial, Pluvial, Coastal | FALSE | TRUE | FALSE | - | Temporal Analysis, Spatial Analysis, Kendall's Correlation Coefficient tau (τ), Tail Dependence Measure chi (χ), Bootstrap Method |
| Olbert et al. 2013 | Ireland | 48 Storm Events (1959-2005), Varying return period scenarios | Earth System Processes, Risk Assessment | Coastal | FALSE | TRUE | FALSE | - | Joint Probability Method (JPM) |
| Olbert et al. 2017 | Ireland (Cork City) | 2009 Flood Event | Methodological Advancement, Risk Assessment | Fluvial, Pluvial, Coastal | TRUE | FALSE | FALSE | MSN_Flood, POM | - |
| Orton et al. 2012 | US (New York) | - | Methodological Advancement | Fluvial, Pluvial, Coastal | TRUE | FALSE | FALSE | sECOM, WRF | - |
| Orton et al. 2015 | US (New York) | 533 Synthetic Tropical Cyclones, 76 Flood Events | Risk Assessment | Fluvial, Coastal | TRUE | TRUE | TRUE | sECOM, SELFE | Bayesian Simultaneous Quantile Regression, Markov Chain Monte Carlo (MCMC) Algorithm |
| Orton et al. 2016 | US (New York, New York Harbor) | Hurricane Irene (2011), Northeaster Storm (2010), 42 Storm Events (1950-2013), 606 Synthetic Storms, Varying return period scenarios | Risk Assessment | Coastal | TRUE | TRUE | TRUE | NYHOPS, sECOM, Holland Wind Model | Hall Stochastic TC Life Cycle Model (Hall and Jewson 2007; Hall and Yonekura 2013), Extreme Value Analysis, Markov Chain Monte Carlo (MCMC) Algorithm, Bootstrap Method |
| Orton et al. 2018 | US (New York, Hudson River) | 76 Storm Events (1900–2010) | Risk Assessment | Fluvial, Coastal | TRUE | TRUE | TRUE | sECOM | Hall Stochastic TC Life Cycle Model, Bayesian Simultaneous Quantile Regression, Extreme Value Analysis |
| Pandey et al. 2021 | India (Mahanadi River) | Cyclone Odisha (1999) and Phailin (2013) | Earth System Processes, Methodological Advancement | Fluvial, Pluvial, Coastal | TRUE | FALSE | FALSE | ADCIRC, HEC-RAS | - |
| Paprotny et al. 2020 | Europe (Northwest) | - | Earth System Processes | Fluvial, Pluvial, Coastal | TRUE | TRUE | TRUE | EFAS, Delft3D, LISFLOOD-FP | Tail Dependence Coefficient (λ), Correlation Coefficients (Kendall's tau (τ), Spearman's rho (ρ)), Peak-over-Threshold (POT) |
| Park et al. 2011 | South Korea | Typhoon Meami (2003) | Forecasting, Planning & Management | Pluvial, Coastal | TRUE | FALSE | FALSE | Holland Wind Model, Hydrodynamic Model (MATLAB) | - |
| Pasquier et al. 2019 | UK (East Coast) | Varying climate change scenarios, Varying return period scenarios | Earth System Processes, Risk Assessment | Fluvial, Pluvial, Coastal | TRUE | FALSE | FALSE | HEC-RAS | Extreme Value Analyis, Peak-over-Threshold (POT) |
| Peña et al. 2022 | US (Florida, Arch Creek Basin) | - | Earth System Processes, Methodological Advancement, Risk Assessment | Fluvial, Pluvial, Coastal, Groundwater | TRUE | FALSE | FALSE | FLO-2D, MODFLOW-2005 | - |
| Petroliagkis et al. 2016 | Europe | - | Earth System Processes | Fluvial, Coastal | TRUE | TRUE | TRUE | Delft3D-Flow, ECWAM, LISFLOOD, | Joint Probability Method (JPM), Tail Dependence Measure chi (χ), Peak-over-Threshold (POT) |
| Petroliagkis et al. 2018 | Europe (Rhine River) | Top 80 Compound Events at 32 Rivers Each | Earth System Processes | Coastal | FALSE | TRUE | FALSE | Delft3D-FLOW, ECWAM | Joint Probability Method (JPM), Tail Dependence Measure chi (χ), Peak-over-Threshold (POT) |
| Phillips et al. 2022 | US (Southeast Coast; Florida, Georgia, and South Carolina) | - | Earth System Processes | Pluvial, Coastal | FALSE | TRUE | FALSE | - | Copula, Locally Weighted Scatterplot Smoothing (LOWESS) Autoregressive Moving Average (ARMA) Model |
| Piecuch et al. 2022 | US (West Coast; California, Oregon, and Washington) | Atmospheric Rivers Events (1980-2016) | Earth System Processes | Coastal | FALSE | TRUE | FALSE | - | Temporal Analysis, Regression Analysis, Peak-over-Threshold (POT), Bootstrap Method |
| Pietrafesa et al. 2019 | US (North Carolina) | Hurricanes Dennis and Floyd (1999) | Earth System Processes, Methodological Advancement | Pluvial, Coastal | TRUE | FALSE | FALSE | POM | - |
| Poulos et al. 2022 | Greece (Thrace, Evros River Delta) | 8 Flood Events (2005–2018) | Earth System Processes, Risk Assessment | Fluvial, Pluvial | FALSE | TRUE | FALSE | - | Temporal Analysis, Spatial Analysis, Spearman's Correlation Coefficient rho (ρ) |
| Prandle and Wolf (1978) | UK (East Coast, North Sea, River Thames) | - | Earth System Processes | Coastal | TRUE | FALSE | FALSE | 1D Hydrodynamic Model (Prandle 1975) | - |
| Preisser et al. 2022 | US (Texas, Austin) | 2015 Memorial Day Flood | Impact Assessment, Risk Assessment | Fluvial, Pluvial | TRUE | TRUE | TRUE | GeoFlood, GeoNet, ProMaIDes | Social Vulnerability Index (SVI), Principal Component Analysis (PCA), Spatial Analysis |
| Qiang et al. 2021 | Hong Kong (Tseung Kwan O Town Centre) | Typhoon Mangkhut (2018) | Risk Assessment | Pluvial, Coastal | TRUE | FALSE | FALSE | FLO-2D, SWMM | - |
| Qiu et al. 2022 | China (Guangdong, Pearl River Delta) | 76 Tropical Cyclone Events (1957-2018), Varying climate change scenarios | Earth System Processes | Fluvial, Coastal | TRUE | FALSE | FALSE | ADCIRC | - |
| Quagliolo et al. 2021 | Italy (Liguria) | - | Methodological Advancement, Risk Assessment | Pluvial, Coastal | TRUE | FALSE | FALSE | InVEST-UFRM | - |
| Rahimi et al. 2020 | US (California, Oakland Flatlands) | - | Methodological Advancement, Risk Assessment | Pluvial, Coastal, Groundwater | TRUE | FALSE | FALSE | HEC-RAS | - |
| Ray et al. 2011 | US (Texas, Galveston Bay) | Hurricane Ike (2008) | Earth System Processes | Pluvial, Coastal | TRUE | FALSE | FALSE | HEC-HMS, HEC-RAS | - |
| Razmi et al. 2022 | US (New York, New York City) | Hurricane Sandy (2012), Hurricane Irene (2011), Varying return period scenarios | Earth System Processes, Methodological Advancement | Pluvial, Coastal | FALSE | TRUE | FALSE | - | Joint Probability Method (JPM), Copula, Kendall's Correlation Coefficient tau (τ), Temporal Analysis (Mann-Kendall Test) |
| Ridder et al. 2018 | Netherlands | - | Earth System Processes | Pluvial, Coastal | TRUE | FALSE | FALSE | WAQUA | - |
| Ridder et al. 2020 | Global | 27 Hazard Pairs (1980–2014), Spatial analysis | Earth System Processes | Pluvial, Coastal, Drought, Soil Moisture | FALSE | TRUE | FALSE | - | Joint Probability Method (JPM), Spatial Analysis, Likelihood Multiplication Factor (LMF) |
| Robins et al. 2011 | UK (Dyfi Estuary) | Varying climate change scenarios | Earth System Processes, Planning & Management | Pluvial, Coastal | TRUE | FALSE | FALSE | TELEMAC | - |
| Robins et al. 2021 | UK (Humber and Dyfi Estuaries) | 56 Flood Events | Earth System Processes | Fluvial, Coastal | FALSE | TRUE | FALSE | - | Linear Regression, Temporal Analysis, Cross-correlation Analysis, Correlation Coefficients (Kendall's tau (τ), Spearman's rho (ρ)), Chi Squared Test (χ2) |
| Rodríguez et al. 1999 | Spain (Northwest Coast) | - | Risk Assessment | Coastal | FALSE | TRUE | FALSE | - | Joint Probability Method (JPM) |
| Rueda et al. 2016 | Spain (Santander) | - | Earth System Processes | Coastal | FALSE | TRUE | FALSE | - | Joint Probability Method (JPM), Copula, Climate-based Extremal Index (Θ), Extreme Value Analysis, Monte Carlo Simulation |
| Ruggiero et al. 2019 | US (Washington, Grays Harbor) | Varying climate change scenarios, Varying return period scenarios | Planning & Management, Risk Assessment | Fluvial, Coastal | TRUE | TRUE | TRUE | ADCIRC, HEC-RAS, SWAN | Managing Uncertainty in Complex Models (MUCM) Hydrodynamic Emulator, Temporal Analysis |
| Sadegh et al. 2018 | US (Washington DC, Potomac River) | Varying return period scenarios | Methodological Advancement, Risk Assessment | Fluvial, Coastal | FALSE | TRUE | FALSE | - | Copula, Correlation Coefficients (Kendall's tau (τ), Pearson's (r), Spearman's rho (ρ)), Block Maxima |
| Saharia et al. 2021 | US (New York, Buffalo River & Lake Erie) | Varying return period scenarios | Earth System Processes, Risk Assessment | Fluvial, Coastal | TRUE | TRUE | TRUE | HEC-RAS | Joint Probability Method (JPM), Copula, Kendall's Correlation Coefficient tau (τ) |
| Saleh et al. 2017 | US (New Jersey, Newark Bay) | Hurricane Irene (2011) and Sandy (2012) | Forecasting | Pluvial, Coastal | TRUE | FALSE | FALSE | HEC-HMS, HEC-RAS, sECOM, NYHOPS | - |
| Sampurno et al. 2022a | Indonesia (Pontianak, Kapuas River Delta) | December 2018 Flood Event | Forecasting, Methodological Advancement | Fluvial, Pluvial, Coastal | TRUE | TRUE | TRUE | SLIM, SWAT | Machine Learning (Random Forest (RF), Multiple Linear Regression (MLR), Support Vector Machine (SVM)) |
| Sampurno et al. 2022b | Indonesia (Pontianak, Kapuas River Delta) | - | Earth System Processes | Fluvial, Coastal | TRUE | FALSE | FALSE | SLIM | - |
| Samuels and Burt 2002 | UK (Wales, Pontypridd, Taff River, Ely River) | Varying return period scenarios, Varying climate change scenarios | Planning & Management, Risk Assessment | Fluvial, Coastal | TRUE | TRUE | TRUE | Flood Modeller/ISIS | Joint Probability Method (JPM), JOIN-SEA Model, Monte Carlo Simulation |
| Sangsefidi et al. 2022 | US (California, Imperial Beach) | - | Risk Assessment | Pluvial, Coastal, Groundwater | TRUE | FALSE | FALSE | PCSWMM | - |
| Santiago-Collazo et al. 2021 | US (Mississippi, Mississippi River Delta) | - | Earth System Processes, Risk Assessment | Pluvial, Coastal | TRUE | FALSE | FALSE | ADCIRC | - |
| Santos et al. 2017 | UK | 92 Extreme Wave Events (2002-2016), Varying return period scenarios | Earth System Processes | Coastal | FALSE | TRUE | FALSE | - | Spatial Analysis, Temporal Analysis, Extreme Value Analysis, Kendall's Correlation tau (τ), Peak-over-Threshold (POT) |
| Santos et al. 2021a | US (Texas, Sabine Lake) | - | Earth System Processes, Methodological Advancement, Risk Assessment | Fluvial, Pluvial, Coastal | FALSE | TRUE | FALSE | - | Copula, Multiple Linear Regression (MLR), Extreme Value Analysis, Kendall's Correlation tau (τ), Peak-over-Threshold (POT) |
| Santos et al. 2021b | Netherlands | Varying return period scenarios | Earth System Processes, Methodological Advancement | Pluvial, Coastal | TRUE | TRUE | TRUE | RTC-Tools | Joint Probability Method (JPM), Copula, Machine Learning (Artifical Neural Network (ANN), Multiple Linear Regression (MLR), Random Forest (RF)), Kendall's Correlation Coefficient tau (τ), Block Maxima |
| Serafin and Ruggiero 2014 | US (Oregon) | Varying return period scenarios | Earth System Processes, Risk Assessment | Coastal | FALSE | TRUE | FALSE | - | Total Water Level Full Simulation Model (TWL-FSM), Temporal Analysis (Declustering), Extreme Value Analysis, Monte Carlo Simulation, Peak-over-Threshold (POT) |
| Serafin et al. 2019 | US (Washington) | Varying return period scenarios | Earth System Processes | Fluvial, Coastal | TRUE | TRUE | TRUE | ADCIRC, HEC-RAS, SWAN | Total Water Level Full Simulation Model (TWL-FSM), Extreme Value Analysis, Temporal Analysis, Spatial Analysis, Monte Carlo Simulation |
| Shahapure et al. 2010 | India (Maharashtra, Navi Mumbai) | 5 Rainfall Events | Methodological Advancement | Pluvial, Coastal | TRUE | FALSE | FALSE | 1D Hydrodynamic Model (GIS-based) | - |
| Shen et al. 2019 | US (Virginia, Norfolk) | Varying return period scenarios | Planning & Management, Risk Assessment | Pluvial, Coastal | TRUE | FALSE | FALSE | ESTRY, TUFLOW | Transition Zone Index (TZI), Spatial Analysis, Temporal Analysis |
| Sheng et al. 2022 | US (Florida) | Varying Tropical Cyclone events, Varying climate change scenarios, Varying return period scenarios | Earth System Processes, Risk Assessment | Pluvial, Coastal | TRUE | TRUE | TRUE | ADCIRC, CAM, CESM, CH3D, HiRAM, RFMS, SWAN | Joint Probability Method with Optimal Sampling (JPM-OS), Monte Carlo Life-Cycle (MCLC) Simulation, Peak-over-Threshold (POT) |
| Shi et al. 2022 | China (Zhejiang, Xiangshan) | Typhoons Haikui (2012) and Fitow (2013) | Earth System Processes, Planning & Management | Pluvial, Coastal | TRUE | FALSE | FALSE | ADCIRC, SWMM | - |
| Silva-Araya et al. 2018 | US (Puerto Rico) | Hurricane Georges (1998) | Methodological Advancement | Pluvial, Coastal | TRUE | FALSE | FALSE | ADCIRC, GSSHA, SWAN | - |
| Skinner et al. 2015 | UK (Humber Estuary) | 2013 Storm Event | Methodological Advancement, Risk Assessment | Coastal | TRUE | FALSE | FALSE | CAESAR-LISFLOOD, LISFLOOD-FP | - |
| Sopelana et al. 2018 | Spain (Betanzos) | 40 Flood Events | Methodological Advancement | Fluvial, Coastal | TRUE | FALSE | FALSE | Iber | - |
| Stamey et al. 2007 | US (Maryland and Virginia) | Hurricane Isabel (2003), Tropical Storm Ernesto (2006), and 2006 Nor'easter Storm | Forecasting, Planning & Management | Fluvial, Coastal | TRUE | FALSE | FALSE | AHPS, ELCIRC, RAMS, ROMS, UnTRIM, WRF | - |
| Steinschneider 2021 | Canada (Ontario, Lake Ontario) | - | Earth System Processes, Risk Assessment | Coastal | TRUE | TRUE | TRUE | LOOFS | Bayesian Hierarchical Model, Monte Carlo Simulation, Spatial Analysis, Chi Squared Test ($\chi 2$) |
| Stephens and Wu 2022 | New Zealand | - | Earth System Processes | Fluvial, Pluvial, Coastal | FALSE | TRUE | FALSE | - | Joint Probability Method (JPM), Kendall's Correlation Coefficient tau (τ), Spatial Analysis, Temporal Analysis, Peak-over-Threshold (POT) |
| Sui and Koehler 2001 | Germany | Varying return period scenarios | Earth System Processes | Pluvial, Snow | FALSE | TRUE | FALSE | - | Extreme Value Analysis, Spatial Analysis, Temporal Analysis |
| Svensson and Jones 2002 | UK (East Coast) | - | Earth System Processes | Fluvial, Pluvial, Coastal | FALSE | TRUE | FALSE | - | Dependence Measure chi (χ), Temporal Analysis, Spatial Analysis, Peak-over-Threshold (POT), Bootstrap Method |
| Svensson and Jones 2004 | UK (South and West Coast) | - | Earth System Processes | Fluvial, Pluvial, Coastal | FALSE | TRUE | FALSE | - | Dependence Measure chi (χ), Temporal Analysis, Spatial Analysis, Peak-over-Threshold (POT), Bootstrap Method |
| Tahvildari et al. 2022 | US (Virginia) | Hurricane Irene (2011) | Planning & Management | Pluvial, Coastal | TRUE | TRUE | TRUE | Delft3D-FLOW, TUFLOW | Spatial Analysis (Traffic Network Analysis) |
| Tanim and Goharian 2021 | Bangladesh (Chittagong) | - | Earth System Processes, Methodological Advancement | Pluvial, Coastal | TRUE | TRUE | TRUE | Delft3D-FLOW, SWAN, SWMM | Joint Probability Method (JPM), Copula, Spearman's Correlation Coefficient rho (ρ), Spatial Analysis, Temporal Analysis |
| Tanir et al. 2021 | US (Washington DC, Potomac River) | - | Impact Assessment, Risk Assessment | Fluvial, Pluvial, Coastal | TRUE | TRUE | TRUE | HEC-RAS | Socio-Economic Vulnerability Index (SOVI), Exposure Index (EI), Flood Socio-Economic Vulnerability Index (FSOVI), HAZUS-MH Damage Assessment Tool, Principal Component Analysis (PCA), Spatial Analysis |
| Tao et al. 2022 | China (Wuhan, Yangtze River) | Compound Events (1980 - 2020) | Earth System Processes, Risk Assessment | Fluvial, Pluvial | FALSE | TRUE | FALSE | - | Compound Intensity Index (CII), Joint Probability Method (JPM), Copula, Multivariate Copula Analysis Toolbox (MvCAT), Correlation Coefficients (Kendall's tau (τ), Pearson's (r), Spearman's rho (ρ)), Temporal Analysis (Mann-Kendall Test) |
| Tawn 1992 | UK | - | Methodological Advancement, Risk Assessment | Coastal | FALSE | TRUE | FALSE | - | Joint Probability Method (JPM), Revised Joint Probability Method (RJPM), Extreme Value Analysis |
| Tehranirad et al. 2020 | US (California, San Francisco Bay) | February 2019 Storm Event | Forecasting, Planning & Management | Fluvial, Pluvial | TRUE | FALSE | FALSE | Hydro-CoSMoS | - |
| Thieken et al. 2022 | Germany | 2013 and 2016 Flood Events | Impact Assessment, Planning & Management | Pluvial, Damming/Dam Failure | FALSE | TRUE | FALSE | - | Socioeconomic Metrics, Mann-Whitney U Test, Chi Squared (χ2) Value, Spatial Analysis |
| Thompson and Frazier, 2014 | US (Florida, Sarasota County) | Varying climate change scenarios | Methodological Advancement, Risk Assessment | Pluvial, Coastal | TRUE | TRUE | TRUE | ICPR, SLOSH | Spatial Analysis (Geographic Weighted Regression (GWR), Moran's I, Linear Probability Model (LPM)) |
| Torres et al. 2015 | US (Texas, Galveston Bay) | Hurricane Katrina (2005), Ike (2008), and Isaac (2012) | Earth System Processes, Planning & Management | Pluvial, Coastal | TRUE | FALSE | FALSE | ADCIRC, HEC-RAS, SWAN, Vflo | - |
| Tromble et al. 2010 | US (North Carolina, Tar and Neuse River) | Tropical Storm Alberto (2006) | Methodological Advancement | Pluvial, Coastal | TRUE | FALSE | FALSE | ADCIRC, HL-RDHM, Vflo | - |
| Tu et al. 2018 | China (Xixiang Basin) | - | Risk Assessment | Pluvial, Coastal | FALSE | TRUE | FALSE | - | Joint Probability Method (JPM), Copula, Kendall's Correlation Coefficient tau (τ), Block Maxima, Peak-over-Threshold (POT) |
| Valle-Levinson et al. 2020 | US (Texas, Houston, Galveston Bay) | Hurricane Harvey (2017) | Earth System Processes | Pluvial, Coastal | TRUE | TRUE | TRUE | ROMS | Flow Interaction Index (μ), Temporal Analysis |
| Van Berchum et al. 2020 | Mozambique (Beira) | - | Risk Assessment | Pluvial, Coastal | TRUE | FALSE | FALSE | FLORES | - |
| Van Cooten et al. 2011 | US (North Carolina) | Hurricane Isabelle (2003), Earl (2010) and Irene (2011), Tropical Storm Nicole (2010) | Forecasting, Methodological Advancement | Fluvial, Pluvial, Coastal | TRUE | FALSE | FALSE | ADCIRC, CI-FLOW, HL-RDHM, RUC | - |
| Van Den Hurk et al. 2015 | Netherlands | January 2012 Near Flood, 800-Year Climate Simulation | Risk Assessment | Pluvial, Coastal | FALSE | TRUE | FALSE | EC-Earth, RACMO2, RTC-Tools | Joint Probability Method (JPM), Spatial Analysis, Temporal Analysis |
| Vitousek et al. 2017 | Global | Varying climate change scenarios | Earth System Processes, Risk Assessment | Coastal | FALSE | TRUE | FALSE | - | Extreme Value Analysis, Monte Carlo Simulation |
| Vongvisessomjai and Rojanakamthorn 1989 | Thailand (Chao Phraya River) | - | Earth System Processes | Fluvial, Coastal | TRUE | TRUE | TRUE | 1D Hydrodynamic Model | Analytical Perturbation Method, Harmonic Analysis, Temporal Analysis |
| Wadey et al. 2015 | UK (Sefton and Suffolk) | Cyclone Xaver (2013), Varying return period scenarios | Earth System Processes, Risk Assessment | Coastal | FALSE | TRUE | FALSE | - | Joint Probability Method (JPM), Temporal Analysis (Clustering) |
| Wahl et al. 2015 | US (CONUS) | - | Earth System Processes, Risk Assessment | Pluvial, Coastal | FALSE | TRUE | FALSE | - | Joint Probability Method (JPM), Copula, Temporal Analysis, Kendall's Correlation Coefficient tau (τ) |
| Walden et al. (1982) | UK (South Coast) | - | Earth System Processes, Methodological Advancement | Coastal | FALSE | TRUE | FALSE | - | Joint Probability Method (JPM), Temporal Analysis |
| Wang et al. 2014 | US (New York, New York City) | Hurricane Sandy (2012) | Methodological Advancement | Coastal | TRUE | FALSE | FALSE | SELFE, RAMS, UnTRIM | - |
| Wang et al. 2015 | US (Washington DC, Potomac River) | Hurricane Isabel (2003) | Methodological Advancement | Fluvial, Coastal | TRUE | FALSE | FALSE | UnTRIM | - |
| Wang et al. 2021 | Canada (Newfoundland and Labrador) | Varying return period scenarios, Varying climate change scenarios | Earth System Processes, Risk Assessment | Fluvial, Coastal | TRUE | FALSE | FALSE | HEC-HMS, HEC-RAS, WRF | - |
| Ward et al. 2018 | Global | - | Earth System Processes | Fluvial, Coastal | FALSE | TRUE | FALSE | - | Joint Probability Method (JPM), Copula, Kendall's Correlation Coefficient tau (τ), Spatial Analysis, Block Maxima, Peak-over-Threshold (POT) |
| Webster et al. 2014 | Canada (Nova Scotia, Bridgewater, LaHave River estuary) | Varying climate change scenarios, Varying return period scenarios | Risk Assessment | Fluvial, Coastal | TRUE | TRUE | TRUE | MIKE11, MIKE21 | Joint Probability Method (JPM), Extreme Value Analysis |
| White 2007 | UK (East Sussex, Lewes, Ouse River) | October 2000 Flood Event | Earth System Processes, Methodological Advancement, Risk Assessment | Fluvial, Coastal | TRUE | TRUE | TRUE | - | Joint Probability Method (JPM), Dependence Measure chi (χ), Block Maxima, Peak-over-Threshold (POT) |
| Williams et al. 2016 | Europe (UK, US, Netherlands, and Ireland) | - | Earth System Processes | Coastal | FALSE | TRUE | FALSE | - | Joint Probability Method (JPM), Kendall's Correlation Coefficient tau (τ), Temporal Analysis |
| Wolf 2009 | Myanmar (Irrawaddy River Delta) | May 2008 Flood Event | Earth System Processes | Coastal | TRUE | FALSE | FALSE | ADCIRC, SWAN | - |
| Wu and Leonard 2019 | Australia | - | Earth System Processes | Pluvial, Coastal | TRUE | TRUE | TRUE | ROMS | Joint Probability Method (JPM), Kendall's Correlation tau (τ), Spatial Analysis, Peak-over-Threshold (POT) |
| Wu et al. 2018 | Australia | - | Earth System Processes | Pluvial, Coastal | TRUE | TRUE | TRUE | ROMS | Extreme Value Analysis, Temporal Analysis, Spatial Analysis, Pearson's Correlation Coefficient (r), Peak-over-Threshold (POT) |
| Wu et al. 2021 | Australia (Swan River) | Varying return period scenarios | Methodological Advancement, Risk Assessment | Fluvial, Pluvial, Coastal | TRUE | TRUE | TRUE | MIKE21 | Joint Probability Method (JPM), Frequency Analysis, Peak-over-Threshold (POT) |
| Xiao et al. 2021 | US (Delaware, Delaware Bay Estuary) | Hurricane Irene (2011), Isabel (2003), Sandy (2012); and Tropical Storm Lee (2011) | Earth System Processes | Fluvial, Coastal | TRUE | TRUE | TRUE | FVCOM | Temporal Analysis (Complex Demodulation, Singular Spectral Analysis (SSA)) |
| Xu et al. 2014 | China (Fuzhou City) | - | Risk Assessment | Pluvial, Coastal | FALSE | TRUE | FALSE | - | Joint Probability Method (JPM), Copula, Temporal Analysis (Mann-Kendall U Test, Pettitt Test) |
| Xu et al. 2019 | China (Haikou City) | - | Risk Assessment | Pluvial, Coastal | FALSE | TRUE | FALSE | - | Joint Probability Method (JPM), Copula |
| Xu et al. 2022 | China (Shanghai) | Tropical Cyclones and Peak Water Level Events (1961-2018) | Risk Assessment | Pluvial, Coastal | TRUE | TRUE | TRUE | D-Flow FM | Copula, Correlation Coefficients (Kendall's tau (τ), Spearman's rho (ρ)) |
| Xu et al. 2022 | China (Hainan, Haikou) | - | Earth System Processes, Risk Assessment | Pluvial, Coastal | TRUE | TRUE | TRUE | PCSWMM | Joint Probability Method (JPM), Copula, Monte Carlo Simulation, Kendall's Correlation Coefficient tau (τ) |
| Yang and Qian 2019 | China (Shenzhen, Pearl River) | - | Earth System Processes, Methodological Advancement | Pluvial, Coastal | FALSE | TRUE | FALSE | - | Joint Probability Method (JPM), Copula, Particle Swarm Optimization (PSO) |
| Yang et al. 2020 | China (Jiangsu Province, Lianyungang, Yancheng and Nantong) | - | Earth System Processes | Pluvial, Coastal | FALSE | TRUE | FALSE | - | Joint Probability Method (JPM), Copula, Particle Swarm Optimization (PSO) |
| Ye et al. 2020 | US (East Coast and Gulf of Mexico, Deleware Bay) | Hurricane Irene (2011) | Methodological Advancement | Fluvial, Pluvial, Coastal | TRUE | FALSE | FALSE | NWM, SCHISM, 3D Baroclinic Atmospheric Model | - |
| Ye et al. 2021 | US (Southeast Coast, North Carolina & South Carolina) | Hurricane Florence (2018) | Earth System Processes, Methodological Advancement | Fluvial, Pluvial, Coastal | TRUE | FALSE | FALSE | HYCOM, NWM, SCHISM, SMS | - |
| Yeh et al. 2006 | Taiwan (Longdong, Hualien, Chiku, and Eluanbi) | 30 Typhoon Events (2001-2005), Varying return period scenarios | Risk Assessment | Coastal | FALSE | TRUE | FALSE | - | Joint Probability Method (JPM), Frequency Analysis |
| Zellou and Rahali 2019 | Morocco (Bouregreg River) | Varying return period scenarios | Risk Assessment | Pluvial, Coastal | TRUE | TRUE | TRUE | CAESAR-LISFLOOD | Joint Probability Method (JPM), Copula, Kendall's Correlation Coefficient tau (τ), Tail Dependence Coefficient (λ) |
| Zhang and Chen 2022 | China | - | Earth System Processes, Risk Assessment | Fluvial, Pluvial, Coastal | FALSE | TRUE | FALSE | - | Joint Probability Method (JPM), Copula, Kendall's Correlation Coefficient tau (τ), Spatial Analysis, Temporal Analysis, Peak-over-Threshold (POT), Block Maxima |
| Zhang and Najafi 2020 | Saint Lucia | Hurricane Mathew (2016) | Risk Assessment | Fluvial, Pluvial, Coastal | TRUE | FALSE | FALSE | HYMOD, LISFLOOD-FP | - |
| Zhang et al. 2011 | US (Alaska, Prince William Sound) | 1964 Alaska Tsunami | Earth System Processes | Coastal, Tsunami | TRUE | FALSE | FALSE | SELFE | - |
| Zhang et al. 2020 | US (Delaware, Delaware Bay) | Hurricane Irene (2011) | Earth System Processes, Methodological Advancement | Pluvial, Coastal | TRUE | FALSE | FALSE | SCHISM | - |
| Zhang et al. 2022 | China (Zhejiang, Ling River Basin) | Typhoon Lekima (2019) and Wiph (2007) | Earth System Processes | Fluvial, Pluvial, Coastal | TRUE | FALSE | FALSE | 1D/2D Coupled Hydrodynamic Model | - |
| Zheng et al. 2013 | Australia | - | Earth System Processes | Pluvial, Coastal | FALSE | TRUE | FALSE | - | Joint Probability Method (JPM), Extreme Value Analysis, Dependence Measure chi (χ), Spatial Analysis, Temporal Analysis, Peak-over-Threshold (POT) |
| Zheng et al. 2014 | Australia (Sydney, Hawkesbury-Nepean Catchmen) | - | Earth System Processes, Risk Assessment | Pluvial, Coastal | FALSE | TRUE | FALSE | - | Joint Probability Method (JPM), Extreme Value Analysis, Block Maxima, Peak-over-Threshold (POT) |
| Zhong et al. 2013 | Netherlands (Lower Rhine Delta) | Varying climate change scenarios | Risk Assessment | Fluvial, Coastal | TRUE | TRUE | TRUE | 1D Hydrodynamic Model | Joint Probability Method (JPM), Copula, Temporal Analysis (Mann-Kendall Test), Monte Carlo Simulation, Correlation Coefficient (Kendall's tau (τ), Spearman's rho (ρ)), Chi Squared Test (χ2), |

*Appendix 2. Table of numerical models, frameworks, systems, and toolsets observed in literature database studies for simulating hydrologic, hydrodynamic, oceanographic, and atmospheric systems that contribute to compound flooding.*

| Model Acronym | Full Names | Model Type |
|---|---|---|
| ADCIRC | Advanced CIRCulation | Hydrodynamic Model |
| ADCIRC-SWAN | | Coupled Hydrodynamic Model System of ADCIRC and SWAN |
| AHPS | Advanced Hydrologic Prediction Service | Coupled Atmospheric & Hydrological Model System |
| ASGS | ADCIRC Surge Guidance System | Hydrodynamic Model System |
| ASGS-STORM | ASGS-Scalable, Terrestrial, Ocean, River, Meteorology | Coupled Model System of ASGS, SWAN, HL-RDHM, DAH, and NAM |
| AutoRoute | - | Hydrological Model |
| BreZo | - | Hydrodynamic Model |
| CAESAR-Lisflood | - | Coupled Model System of Lisflood-FP and CAESAR |
| CAM | Community Atmosphere Model | Atmospheric Model |
| CaMa-Flood | Catchment-based Macro-scale Floodplain | Hydrodynamic Model |
| CESM | Community Earth System Model | Atmospheric Model |
| CH3D | Curvilinear-grid Hydrodynamics 3D Model | Hydrodynamic Model |
| CI-FLOW | Coastal and Inland Flooding Observation and Warning Project | Hydrological Model |
| CKF | Climate Knowledge Facility System | Coupled Hydrological & Hydrodynamic Model System |
| COAWST | Coupled-Ocean-Atmosphere-Wave-Sediment Transport Modeling System | Coupled Hydrodynamic & Atmospheric Model System |
| COS-Flow | Coupled Overland-Sewer Flow model | Hydrodynamic Model |
| CoSMoS | Coastal Storm Modeling System | Atmospheric Model |
| CREST | Coupled Routing and Excess Storage | Hydrological Model |
| CREST-SVAS | Coupled Routing and Excess Storage-Soil-Vegetation-Atmosphere-Snow | Hydrological Model |
| D-Flow FM | D-Flow Flexible Mesh | Hydrodynamic Model |
| DCSM | Dutch Continental Shelf Model | Hydrodynamic Model |
| Delft3D-FM | Delft 3D Flexible Mesh Suite | Toolset |
| Delft3D-FLOW | - | Hydrodynamic Model |
| Delft3D-WAVE | - | Coupled Hydrodynamic Model of Delft3D and SWAN |
| Delft-FIAT | Flood Impact Analysis Tool | Toolset |
| Delft-FLS | DELFT FLooding System | Hydrodynamic Model |
| EC-Earth | European community Earth System Model | Atmospheric, Hydrological, & Hydrodynamic Model System |
| ECHAM5 | ECMWF Hamburg Model Version 5 | Atmospheric Model |
| ECWAM | ECMWF Ocean Wave Model | Hydrodynamic Model |
| EFAS | European Flood Awareness System | Hydrological Model |
| ELCIRC | Eulerian-Lagrangian CIRCulation | Hydrodynamic Model |
| ESTRY | - | Hydrodynamic Model |
| ESTOFS | Extra Tropical Storm and Tide Operational Forecast System | Hydrodynamic Model |
| ETSS | Extratropical Storm Surge model | Hydrodynamic Model |
| FES2012 | Finite Element Solution Model | Hydrodynamic Model |
| FLO-2D | - | Hydrodynamic Model |
| Flood Modeller/ISIS | - | Hydrodynamic Model |
| FLORES | Flood risk Reduction Evaluation and Screening | Hydrodynamic Model |
| FLOW-3D | - | Hydrodynamic Model |
| FVCOM | Finite Volume Community Ocean Model | Hydrodynamic Model |
| GCOM2D | Global Environmental Modelling Systems (GEMS) 2D Coastal Ocean Model | Hydrodynamic Model |
| GeoFlood | - | Hydrological Model |
| GeoNet | - | Toolset |
| GSSHA | Gridded Surface Subsurface Hydrologic Analysis | Hydrological Model |
| GTSM | Global Tide and Surge Model | Hydrodynamic Model |
| H*WIND | Hurricane Wind Analysis System | Atmospheric Model |
| HADGEM | HADley Centre Global Environment Model | Coupled Atmospheric & Hydrodynamic Model System |
| HBV | Hydrologiska Byråns Vattenbalansavdelning | Hydrological Model |
| HEC-HMS | Hydrologic Engineering Centre's - Hydrologic Modeling System | Hydrological Model |
| HEC-RAS | Hydrologic Engineering Centre's - River Analysis System | Hydrological Model |
| HiPIMS | High-Performance Integrated Hydrodynamic Modelling Software | Hydrological & Hydrodynamic Model |
| HiRHAM | High Resolution Atmospheric Model | Atmospheric Model |
| HL-RDHM | Hydrology Laboratory - Research Distributed Hydrologic Model | Hydrological Model |
| Holland Wind Model | Holland Wind Model | Atmospheric Model |
| HYCOM | HYbrid Coordinate Ocean Model | Hydrodynamic Model |
| Hydro-CoSMoS | Hydro-Coastal Storm Modeling System | Hydrodynamic Model |
| HydroMT | Hydro Model Tools | Toolset |
| HyMOD | HYdrological MODel | Hydrological Model |
| Iber | Iberaula | Hydrodynamic Model |
| ICRP | Interconnected Channel and Pond Routing Model | Hydrological & Hydrodynamic Model |
| InVEST-UFRM | Integrated Valuation of Ecosystem Services and Tradeoffs - Urban Flood Risk Mitigation model | Toolset |
| IOKA | Oceanweather's Interactive Kinematic Objective Analysis System | Atmospheric Model |
| LISFLOOD-FP | - | Hydrological & Hydrodynamic Mode |
| LOOFS | Lake Ontario Operational Forecast System | Coupled Hydrodynamic Model System of FVCOM and CICE |
| MATSIRO-GW | Minimal Advanced Treatments of Surface Integration and RunOff - Groundwater | Hydrological Model |
| MIKE+ | - | Hydrological & Hydrodynamic Model |
| MIKE11 | - | Hydrodynamic Model |
| MIKE21 | - | Hydrodynamic Model |
| MISDc | Modello Idrologico SemiDistribuito in continuo | Hydrological Model |
| MODFLOW | Modular Hydrologic Model | Hydrological Model |
| Mog2D | | Hydrodynamic Model |
| MPI-OM | Max Planck Institute - Ocean/Sea-Ice Model | Hydrodynamic Model |
| MRI-CGCM2 | Meteorological Research Institute coupled General Circulation Model Version 2 | Coupled Atmospheric & Hydrodynamic Model |
| MSN_Flood | - | Hydrodynamic Model |
| NAM | Nedbor-Afstromnings Model | Hydrological Model |
| NAM | North American Mesoscale Forecast System | Atmospheric Model |
| NEMO | Nucleus for European Modelling of the Ocean | Hydrodynamic Model |
| NWM | National Water Model | Hydrological Model |
| NYHOPS | New York Harbor Observing and Prediction System | Hydrodynamic Model |
| ONDA | - | Hydrodynamic Model |
| PCSWMM | Personal Computer Storm Water Management Model | Hydrological & Hydrodynamic Model System |
| POM | Princeton Ocean Model | Hydrodynamic Model |
| PQRUT | - | Hydrological Model |
| ProMaIDes | Protection Measures against Inundation Decision Support | Hydrodynamic Model & Toolset |
| RACMO2 | Regional Atmospheric Climate Model Version 2 | Atmospheric Model |
| RAMS | Regional Atmospheric Modelling System | Atmospheric Model |
| ReAFFIRM | Real-time Assessment of Flash Flood Impacts Framework | Hydrological Model |
| RegCM3 | Regional Climate Model Version 3 | Atmospheric Model |
| RFMS | Rapid Forecasting and Mapping System | Coupled Hydrodynamic Model System of SLOSH and CH3D |
| ROMS | Regional Ocean Modelling System | Hydrodynamic Model |
| RS3 | Rocscience 3D Finite Element Analysis | Toolset |
| RTC-Tools | - | Hydrological Model & Toolset |
| RUC | Rapid Update Cycle | Atmospheric Model |
| SCHISM | Semi-implicit Cross-scale Hydroscience Integrated System Model | Hydrodynamic Model |
| sECOM | Stevens Estuarine and Coastal Ocean Model | Hydrodynamic Model |
| sECOM-NYHOPS | - | Coupled Hydrodynamic Model System of sECOM and NYHOPS |
| SELFE | Semi-Implicit Finite-Element/Volume Eulerian-Lagrangian Algorithm | Hydrodynamic Model |
| SFAS | Stevens Flood Advisory System | Coupled Hydrologic & Hydrodynamic Model System |
| SFINCS | Super-Fast Inundation of CoastS | Hydrodynamic Model |
| SHAWLWV | Model for Simulation of Shallow Water Wave Growth, Propagation, and Decay | Hydrodynamic Model |
| SIPSON | Simulation of Interaction between Pipe flow and Surface Overland flow in Networks | Hydrodynamic Model |
| SLIM | Second-generation Louvain-la-Neuve Ice-ocean Model | Hydrodynamic Model |
| SLOSH | Sea, Lake, and Overland Surges from Hurricanes | Hydrodynamic Model |
| SMS | Surface-water Modeling System | Toolset |
| SNAP | Stevens Northwest Atlantic Prediction Model | Hydrodynamic Model |
| SPHY | Spatial Processes in HYdrology | Hydrological Model |
| SPLASH | Special Program to List Amplitudes of Surges From Hurricanes | Atmospheric and Hydrodynamic Model System |
| STWAVE | Steady State Spectral Wave | Hydrodynamic Model |
| SWAN | Simulating Waves Nearshore | Hydrodynamic Model |
| SWAT | Soil & Water Assessment Tool | Toolset |
| SWMM | Storm Water Management Model | Hydrological Model |
| TELEMAC | TELEMAC-MASCARET | Hydrodynamic Model |
| TUFLOW | - | Hydrodynamic Model |
| UIM | Urban Inundation Model | Hydrodynamic Model |
| UnTRIM | - | Hydrodynamic Model |
| Vflo | Vieux FLOod | Hydrological Model |
| WAM | Wave Model | Hydrodynamic Model |
| WAQUA | WAter movement and water QUAlity modelling | Hydrodynamic Model |
| WGHM | WaterGAP Global Hydrology Model | Hydrological Model |
| WIFM | WES Implicit Flooding Model | Hydrodynamic Model |
| WRF | Weather Research and Forecast Model | Atmospheric Model |
| WW3/WaveWatch III | WAVE-height, WATer depth and Current Hindcasting Version 3 | Hydrodynamic Model Framework |
| XPSWMM | XP Solutions Storm Water Management Model | Hydrological & Hydrodynamic Model |